\documentclass[aps,prx,nofootinbib,longbibliography,amsmath,amssymb,amsfonts,superscriptaddress,notitlepage,floatfix,reprint]{revtex4-2}

\usepackage{mathtools}
\usepackage{physics}
\usepackage{bbm}
\usepackage{bm}
\usepackage{wasysym} 

\usepackage[bookmarksnumbered,pdfpagelabels=true,plainpages=false,colorlinks=true,linkcolor=blue,citecolor=magenta,urlcolor=blue]{hyperref}
\usepackage[capitalize]{cleveref}

\usepackage[dvipsnames]{xcolor}
\usepackage{graphicx}
\usepackage[percent]{overpic}
\usepackage[usestackEOL]{stackengine}
\usepackage{rotating}

\usepackage{tikz}
\usepackage{tabularray}

\definecolor{cblue}{RGB}{55,126,184}
\hypersetup{colorlinks=true,linkcolor=cblue,citecolor=cblue,urlcolor=cblue}

\definecolor{cL}{RGB}{59, 83, 140}
\definecolor{cM}{RGB}{33, 145, 141}
\definecolor{cH}{RGB}{95, 202, 98}
\definecolor{cG}{RGB}{204,204,204}

\def\be{\begin{align}}
\def\ee{\end{align}}

\begin{document}

\title{Unconventional Spin Dynamics and Supersolid Excitations in the Triangular-Lattice XXZ Model}
\author{Rafael Flores-Calder\'{o}n}
\affiliation{Max Planck Institute for the Physics of Complex Systems, N\"othnitzer Strasse 38, D-01187 Dresden, Germany}

\author{Roderich Moessner}
\affiliation{Max Planck Institute for the Physics of Complex Systems, N\"othnitzer Strasse 38, D-01187 Dresden, Germany}

\author{Frank Pollmann}
\affiliation{Department of Physics, Technical University of Munich, 85748 Garching, Germany}

\begin{abstract}

Motivated by recent experiments, we investigate the  spin-1/2 XXZ model on the triangular lattice with strong Ising anisotropy, combining large-scale numerical simulations and analytical methods to uncover unconventional spin dynamics { at $T=0$}.
First, we compute the dynamical spin structure factor using density matrix renormalization group (DMRG) simulations and find excellent agreement with inelastic neutron scattering data on the layered compound $\text{K}_2\text{Co}(\text{SeO}_3)_2$. The low-energy spectrum reveals a roton-like minimum at the $M$ point—absent in linear spin-wave theory—accompanied by peak intensity and a broad continuum above it. Near the $\Gamma$ point, we observe an approximately linear dispersion with vanishing spectral weight.
Second, we compare two analytical frameworks that reproduce the observed features. The first is a hard-core boson approach, which includes: (i) an effective staggered boson model (ESBM) at zero magnetic field, (ii) perturbation theory applied to the one-third magnetization plateau, and (iii) a self-consistent mean-field Schwinger boson theory (SBT). The second framework is based on a variational supersolid quantum dimer model (QDM) ansatz, combined with a single-mode approximation. The SBT captures the broad continuum, the $M$-point minimum, and linear dispersion at $\Gamma$, whereas the QDM reproduces the roton minimum and linear dispersion at finite momentum near $\Gamma$. Remarkably, both the QDM wavefunction and the DMRG ground state exhibit nearly identical structure factors with pronounced transverse photon-like excitations.
Together, our comprehensive theoretical and numerical analysis elucidates the microscopic origin of supersolid excitations in the XXZ triangular lattice model and their proximity to a spin liquid phase observed experimentally.
\end{abstract}
\maketitle

\tableofcontents

\section{Introduction}

In recent years, the study of frustrated quantum magnets has revealed various exotic phases that challenge our conventional understanding of magnetic order. Among these, the supersolid phase—where superfluidity and crystalline order coexist \cite{LeggettSupersolid}—has garnered considerable attention, both experimentally and theoretically. Originally, supersolid order was predicted for $^4\text{He}$ \cite{Kim2004,Balibar2010,Kim2012,Boninsegni2012,Nyeki2017} where the concept was defined as a phase of matter which is simultaneously superfluid and a crystal, breaking in this way {the global $U(1)$ phase rotation symmetry of the bosons} and translational symmetry. Subsequent studies extended this idea to cold atom systems \cite{Li2017,Leonard2017,Tanzi2019,Norcia2021} and to unfrustrated hard-core boson models, where numerical methods are not hindered by the sign problem \cite{Melko2005,Wessel2005,HeidarianDamle2005,Boninsegni2005,HeidarianDamle2005,Jiang2009,HeidarianParamekanti2010}. These works primarily focused on ground state properties.

Earlier theoretical studies had already identified a supersolid phase of spin-1/2 particles on the triangular lattice \cite{Wessel2005,Melko2005,HeidarianDamle2005}. However, realizing such a two-dimensional lattice model in solid-state materials remained out of reach until recently. This became possible with the synthesis of compounds featuring weak interlayer coupling. In particular, the experiments reported in \cite{chen_phase_2024} and \cite{zhu_continuum_2024} successfully realized such a model. Both experiments demonstrate that the layered compound $\text{K}_2\text{Co}(\text{SeO}_3)_2$  is well described by spin-1/2 moments that form a triangular lattice of the $\text{Co}^{+2}$ ions. Due to the strong anisotropy, there is a dominant antiferromagnetic Ising exchange and antiferromagnetic XX anisotropy. Through neutron scattering experiments, the authors conjecture that the ground state is in a supersolid phase close to the Ising point. Moreover, the experiments show evidence for unconventional excitations with substantial differences with respect to predictions of linear spin wave theory, as we will review in the next sections. 

Although the interplay of quantum fluctuations and geometric frustration in the triangular lattice has been known to host a plethora of interesting magnetic ground states, already years ago \cite{moessner_ising_2001}, it is only recently that large-scale numerical methods can give insight into the nature of the excitations. Indeed, signatures of a Dirac spin liquid (DSL) have been identified in the dynamical structure factor of the $J_1$–$J_2$ triangular Heisenberg model \cite{drescher_dynamical_2023, Sherman_2023}. For the XXZ model, some features of the dynamical structure factor—such as a minimum at the $M$ point—have been theoretically predicted \cite{zhu_wannier_2024,chen_phase_2024,ulaga_easy-axis_2024,ulaga_finite-temperature_2024}. While connections to a supersolid ground state have been proposed, it remains uncertain whether these features can be linked to a nearby Dirac spin liquid (DSL) phase \cite{xiang_giant_2024}. Developing a coherent understanding of the supersolid phase, its excitations, and its potential relation to quantum spin liquids is essential for interpreting future experiments and exploring possible applications. One notable example is the recently proposed giant magnetocaloric effect relevant for millikelvin cooling \cite{Jia2024}.

\begin{figure*}[t]
    \centering
    \includegraphics[scale=0.31]{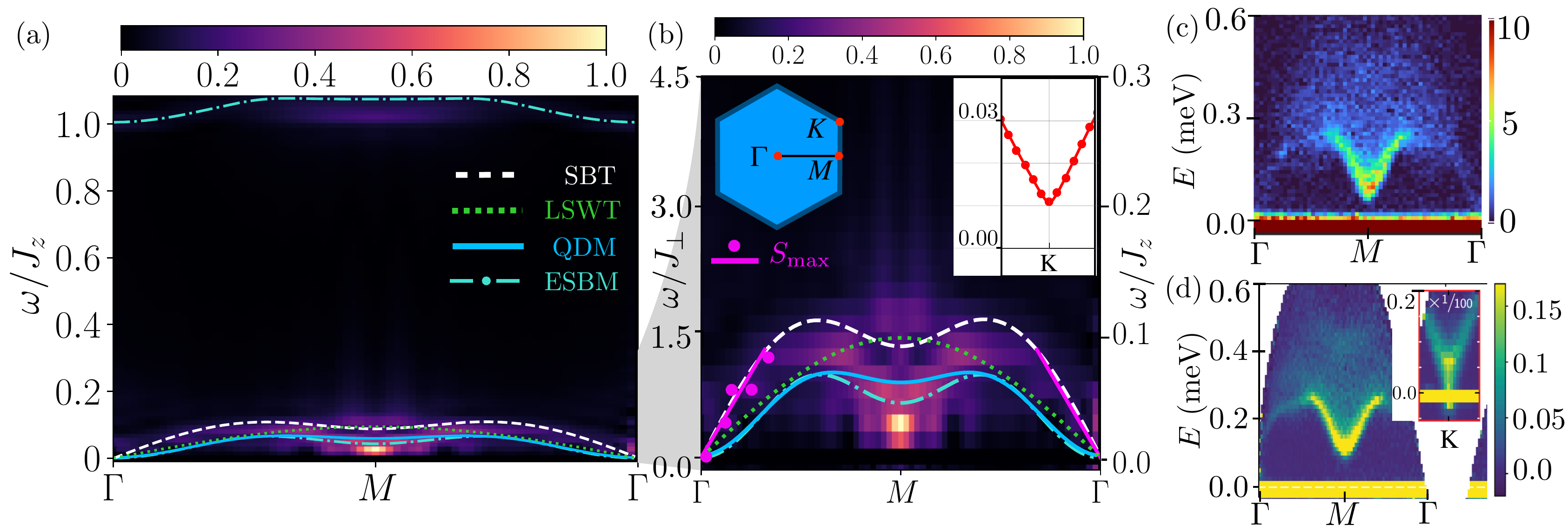}
    \caption{Dynamical structure factor for $J_z = 3\  \text{meV}$,  $J_\perp = 0.2 \ \text{meV}$ of the spin-1/2 XXZ model \cite{zenodo_data}. (a) DMRG results shown as a density plot for a $24\times 6$ cylinder geometry with circumference $L = 6$ and bond dimensions up to $\chi = 800$. The lines overlaid on the heatmap show predictions from Schwinger boson theory (SBT), linear spin-wave theory (LSWT), the variational quantum dimer model wavefunction (QDM), the effective staggered boson model (ESBM) {and in pink a fit $\omega = 0.0849 \abs{\textbf{k}}$  to the maximum dynamical structure factor $S_{\text{max}}$ computed from DMRG with representative points. }. (b) A zoom into the low-energy region with the same theoretical curves overlaid. The left inset shows the Brillouin zone with the relevant momenta and the path used for the plots. The right inset shows the energy bound near the $K$ point computed using the single-mode approximation of the QDM. Panels (c) and (d) display inelastic neutron scattering data for $\text{K}_2\text{Co}(\text{SeO}_3)_2$, taken from Refs. \cite{zhu_continuum_2024} and \cite{chen_phase_2024}, respectively. Note that the experimental plots have been cropped and rescaled to match the energy scales of the DMRG data shown in (b).}
    \label{fig:DSF_exp}
\end{figure*}

In our study, we combine large-scale density matrix renormalization group (DMRG) simulations \cite{white1992dmrg, hauschild_efficient_2018} with analytical approaches to investigate the dynamics and low-energy excitations of the frustrated spin-$1/2$ triangular lattice XXZ model {at $T=0$.}
Our main results show both quantitative and qualitative agreement with neutron scattering data for the low-energy excitations near the roton-like minimum at the $M$ point, as shown in Fig.~\ref{fig:DSF_exp}.
Moreover, we find strong evidence for gapless, photon-like transverse excitations near the $\Gamma$ point, as shown in Fig.~\ref{fig:DMRG_QDM_dimer}.
Our comprehensive approach, using complementary methods, allows us to confidently establish these features in the excitation spectrum while providing an intuitive understanding of their microscopic origin.

\section{Model} \label{sec:model}

We consider the Hamiltonian for the spin-1/2 XXZ model on the triangular lattice in the frustrated Ising limit {$J_z \gg J_\perp > 0$}

\begin{align}
        H= J_z \sum_{\expval{i,j}} S^{z}_iS^{z}_j+J_\perp \sum_{\expval{i,j}} (S^{x}_iS^{x}_j+S^{y}_iS^{y}_j). \label{XXZmodel}
\end{align}

The Hamiltonian has a $U(1)\times \mathbb{Z}_2$ symmetry, which is the remnant after anisotropy is introduced in the full $SU(2)$ Heisenberg point. The $\mathbb{Z}_2$ symmetry corresponds to flipping the $S_z$ operator, while the $U(1)$ symmetry rotates the spins in the $XY$ plane.
{It has been argued according to Refs.\cite{sellmann_phase_2015,HeidarianParamekanti2010,melko_supersolid_2005} that in the parameter regime $0<J_\perp/J_z<1$ the long-range order of the $XY$ type persists together with the translational symmetry breaking, showcasing three sublattice order. If a perpendicular magnetic field is included, $-h\sum_i S^z_i$, then a $1/3$-Neel state appears for larger values of the field beyond a critical value of $h^*=3(J_z + J_\perp /2)$ and below the fully polarized state. We will study the excitations of this Neel state in section \ref{UUDexcit} and compare them to the zero-field case. Focusing on this last one, we will reveal the microscopic origin of the low-energy excitations and their relation to the ground state supersolid order.  } 

\section{Summary of main results}

\begin{figure*}[t]
    \centering
    \includegraphics[scale=0.25]{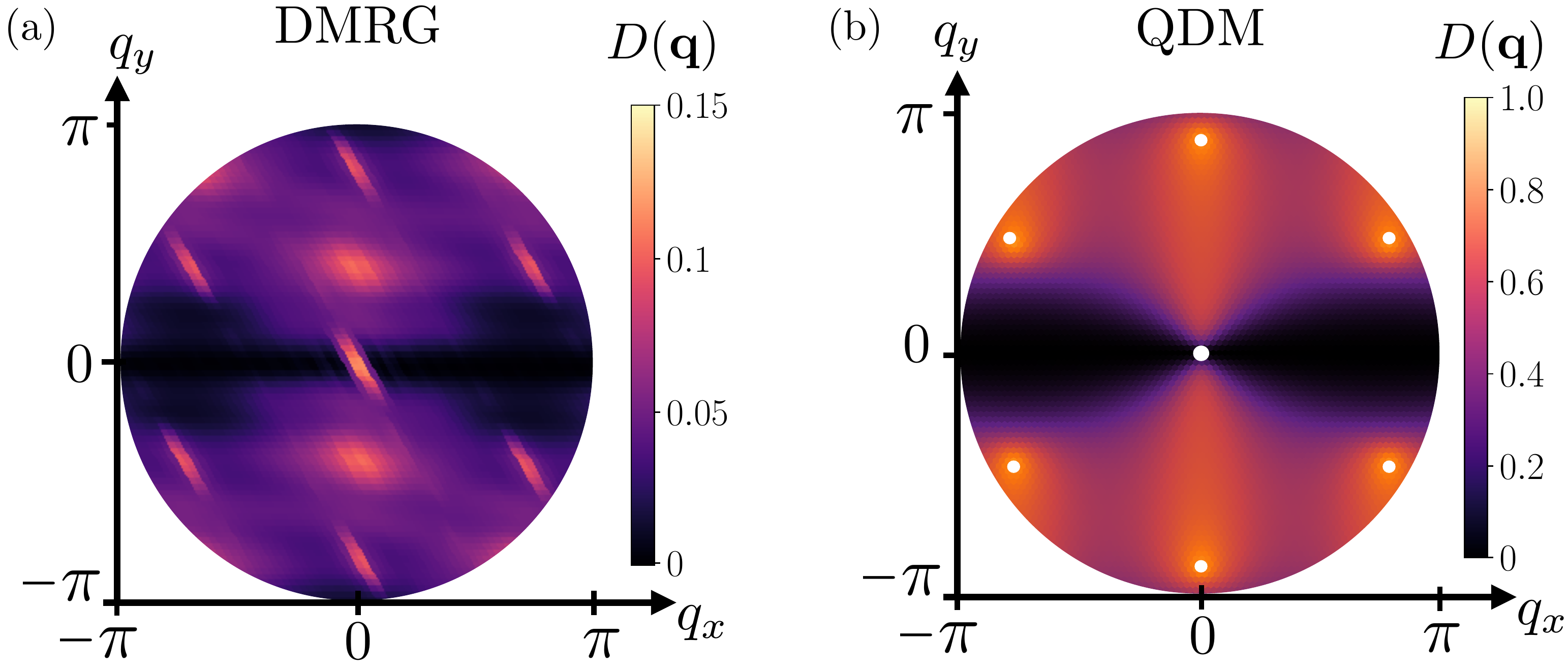}
    \caption{ Static horizontal dimer structure factor calculated (a) for the ground state of the XXZ model Eq. \eqref{XXZmodel} from DMRG simulations of a simulate a $36\times 6$ cylinder geometry with circumference $L = 6$, with bond dimension $\chi=800$, normalized and interpolated in momentum space. (b)  Dimer structure factor calculated exactly from the supersolid variational wavefunction of \cite{sen_variational_2008} with $z=0.925$ and the mapping to free Kasteleyn fermions. We have added in white the points where the structure factor diverges with system size. }
    \label{fig:DMRG_QDM_dimer}
\end{figure*}

 In this section, we briefly discuss our main results shown in Figs. \ref{fig:DSF_exp},\ref{fig:DMRG_QDM_dimer}. 

Our numerical simulations are carried out within the matrix product state (MPS) formalism and the DMRG algorithm. The dynamical structure factor (DSF) as a function of momentum and frequency is shown as a heat plot in Fig.~\ref{fig:DSF_exp}; a more exhaustive discussion of the numerics and experiments is presented in Sec.~\ref{sec:mps}. The different superimposed lines correspond to the different analytical treatments we implement, as we will discuss shortly. It is worth mentioning that we choose the parameter values $J_z = 3\  \text{meV}$,  $J_\perp = 0.2 \ \text{meV}$ relevant to the experimental data of Ref.~\cite{chen_phase_2024}.

Our first main result is the excellent qualitative agreement between our DMRG simulations of the XXZ model close to the Ising point and the neutron scattering experiments. Specifically, the low-energy branch displays a roton-like minimum of the dispersion at the $M$ point, with two maxima at $\omega \approx 2 J_\perp$ appearing close to the mid-point between $M$ and $\Gamma$.  The DMRG simulations reveal a strong maximum of the DSF at the M point, with broad spectral weight close to the M point that persists for frequencies up to $0.2J_z$, as seen in Fig.~\ref{fig:DSF_exp}b, which occurs for very small frequencies. Moreover, the dispersion close to the $\Gamma$ point is approximately linear, although the spectral weight diminishes strongly as the $\Gamma$ point is approached. Experimentally, the diminishing weight upon approaching $\Gamma$ is observed, but the resolution only allows one to see a linear momentum dependence at finite momentum. Finally, we numerically observe a higher-energy Ising excitation branch at around $\omega \approx J_z$ with a very flat dispersion of bandwidth close to $J_\perp$, as shown in Fig.~\ref{fig:DSF_exp}a, concordant with the branch found in Ref.~ \cite{chen_phase_2024}. 

Our second main result concerns the analytical approaches we implement to understand the physical origins of such dynamical signatures. The lines shown Fig.~\ref{fig:DSF_exp} are calculated using the Schwinger boson theory (SBT), the variational quantum dimer model wavefunction (QDM), the effective staggered boson model (ESBM), and linear spin wave theory (LSWT). Except for LSWT, all our methods capture the roton-like minimum of the dispersion at the $M$ point. We find that the two methods that have the best agreement with the DMRG simulations are the QDM and SBT.  { The Schwinger boson formalism decomposes an S=1 quasiparticle into two $S=1/2$ excitations. In linear spin-wave theory (LSWT), a spin flip directly creates a single, $S=1$ quasiparticle with a well-defined dispersion. In contrast, within the Schwinger boson theory (SBT), a flipped spin does not correspond to an elementary excitation but to a composite object made of two $S=1/2$  spinons. As a result, the momentum and energy of a simple spin flip can get redistributed into both constituents, creating a continuum of scattering. } In turn, the SBT shows a linear dispersion near the $\Gamma$ point, and the magnitudes of the energies are shifted. The QDM is instead based on a mapping of the Ising ground states onto dimer coverings of the honeycomb lattice.  {As explained in detail in Sec.~\ref{sec:qdm}, a dimer is placed at the midpoint of each frustrated bond of the original triangular lattice; the endpoints of the dimer lie at the centers of the triangles and form an auxiliary honeycomb lattice, see Fig. \ref{fig:dimers}. The Ising ground states satisfy one frustrated bond per triangle, and are in this way mapped to hard-core dimer coverings of the honeycomb lattice}. This approach can access the low-energy branch via a single-mode approximation as a form of dimer density waves. Indeed, these quasiparticles, which correspond to photon-like excitations, have also been known to display vanishing spectral weight in the context of quantum spin ice \cite{seeing_light}. The QDM exhibits a linear dispersion for finite momenta, but gives no information at the $\Gamma$ point due to the vanishing of the dimer structure factor for the supersolid variational wavefunction presented in Sec.~\ref{sec:qdm}. Nevertheless, together with knowledge of the symmetry-breaking pattern, this can be used as a starting point to derive an effective field theory, which we propose in Sec. \ref{eff_fieldtheory}. 

Another important result comes from comparing our variational supersolid wave function with the DMRG static simulations. In particular, we calculate the dimer-dimer structure factor, as shown in Fig.~\ref{fig:DMRG_QDM_dimer} for horizontal dimers. {The dimer-dimer structure factor $D(\textbf{q})$ is defined as the Fourier transform of the dimer density correlation function $\expval{n^d_{(ab)}n^d_{(cd)}}$, with $n^d_{(ab)}$ equal to one (zero) if a dimer is present (absent) in the $(ab)$ bond with $a,b$ sites of the auxiliary honeycomb lattice. In terms of the original spins, this quantity corresponds to a four-point correlation function of the form $\expval{S_a^zS_b^zS_c^zS_d^z}$.}  There is a remarkable agreement between both frameworks, specifically along the dimer direction, here $q_x$. Both frameworks show a vanishing weight along this line, consistent with there being only transverse photon-like excitations, as we will discuss in Sec.~\ref{sec:SMA}. The maximum weight displayed at the $K,K'$ points is also well captured by both approaches, indicative of the three-sublattice order in the supersolid phase. The QDM approach additionally gives an energy dispersion near $K,K'$ points closely resembling the experimental one, as shown when comparing the inset of Fig.~\ref{fig:DSF_exp}c with Fig.~\ref{fig:SMA}. Finally, we find a higher-energy excitation branch within the DMRG simulations in Fig.~\ref{fig:DSF_exp}a at around $\omega \approx J_z$ with a very flat dispersion of bandwidth close to $J_\perp$, which is captured analytically by the ESBM.


 The remainder of the paper delves deeper into the details of each method. We detail our DMRG approach, along with comparisons to relevant experimental measurements, in Sec.~\ref{sec:mps}, where we emphasize how the static and dynamical spin structure factors provide striking signatures of the supersolid phase. Next, in Sec.~\ref{sec:hardcore}, we turn to an analytical hard-core boson framework that captures the interplay of hole and particle excitations at finite field, before proposing an effective staggered boson hopping model that reproduces the major qualitative features of our full DMRG results. Subsequently, we develop a Schwinger boson mean-field description in Sec.~\ref{sec:schwinger}, highlighting how fractionalized excitations can yield a very similar dynamical structure factor as our DMRG methods. In Sec.~\ref{sec:qdm}, we derive a quantum dimer model (QDM) formulation of the supersolid ground state and demonstrate in Sec.~\ref{sec:SMA}, via a single-mode approximation, that both the dynamical spectra and the dimer structure factors compare favorably with numerical simulations, emphasizing the ``photon-like'' excitations that underpin the supersolid’s emergent gauge structure. Finally, in Sec.~\ref{sec:conclusions}, we summarize our main conclusions.

\section{Matrix Product State methods and experimental comparison}\label{sec:mps}

We use large-scale DMRG simulations \cite{white1992dmrg,white1993}, performed with the TeNPy library \cite{tenpy2024,hauschild_efficient_2018}, to first calculate the ground state of Eq.~\eqref{XXZmodel}, and then, using the time-dependent variational principle (TDVP) \cite{haegeman2011,haegeman2016,vanderstraeten2019} to evolve our state.
We use periodic boundary conditions along the circumference $L_y$, resulting in a cylinder geometry for the triangular lattice \cite{gohlke_dynamics_2017,white_review_dmrg_2d,drescher_dynamical_2023,Sherman_2023}. We choose to have the maximum resolution across the $\Gamma-M-\Gamma$ line which we accomplish by identifying sites $\textbf{r}$ and $\textbf{r}+L_y \textbf{a}_2+n \textbf{a}_1$   with $\textbf{a}_1=(\sqrt{3}/2,1/2),\textbf{a}_2=(1,0)$ are the primitive vectors of the Bravais lattice and $n$ is an integer, more on the details of our DMRG methods can be found in appendix \ref{appendix:DMRG}. We find a DMRG ground-state approximation using a combination of both infinite DMRG (iDMRG) and finite DMRG, which we first characterize in terms of the static spin structure factor.

\subsection{Static spin structure factor}
\begin{figure}[t]
    \centering
    \includegraphics[scale=0.3]{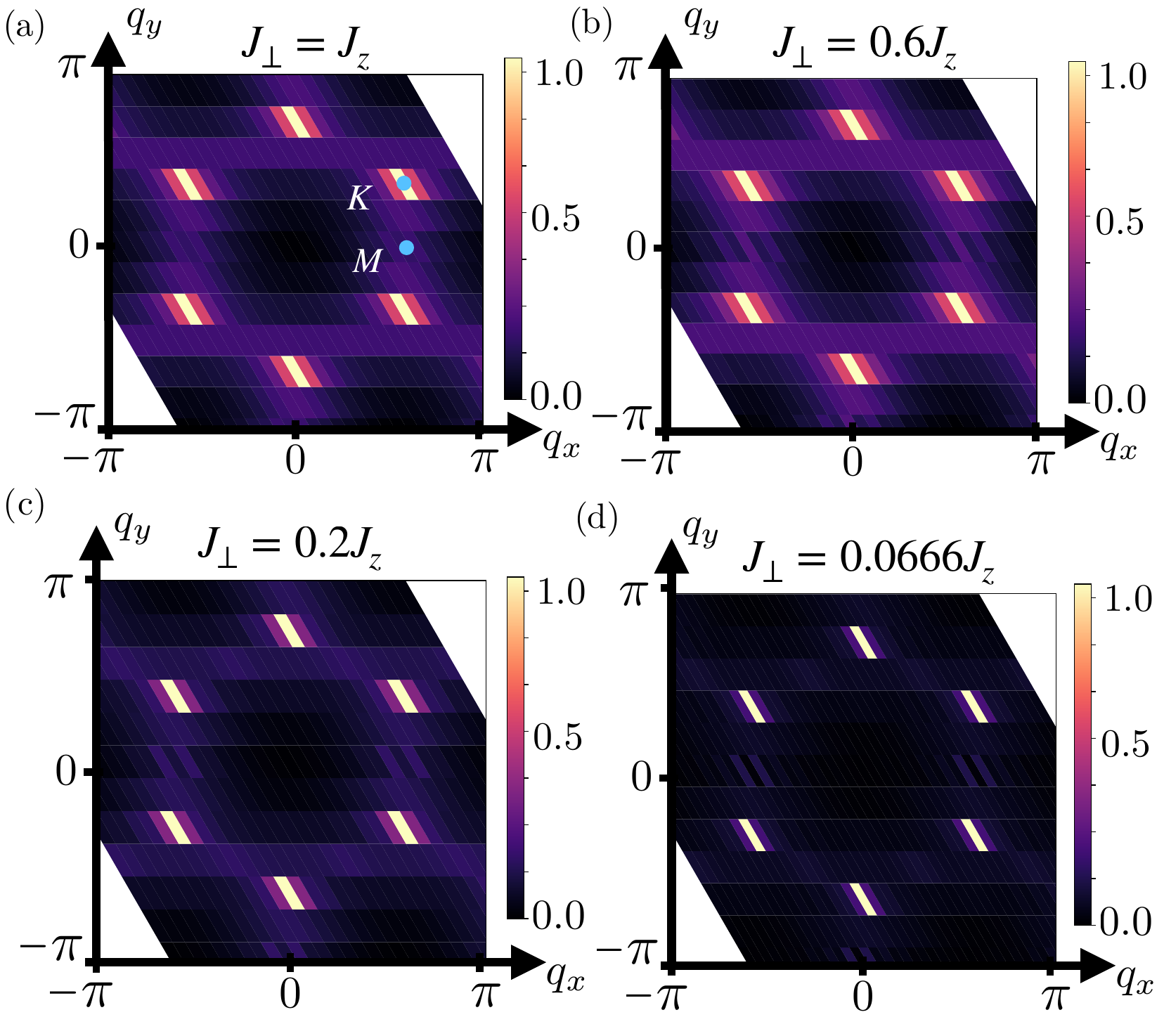}
    \caption{ Static spin structure factor obtained from DMRG for different anisotropy values. We simulate a $18 \times 6$ cylinder geometry with circumference $L = 6$, with bond dimensions up to $\chi = 512$ for (a),(b) (c) and circumference $L = 6$ length $24$ for (d) with bond dimension $\chi=1000$.}
    \label{fig:dmrg_ssf}
\end{figure}

\begin{figure}[t]
    \centering
    \includegraphics[scale=0.18]{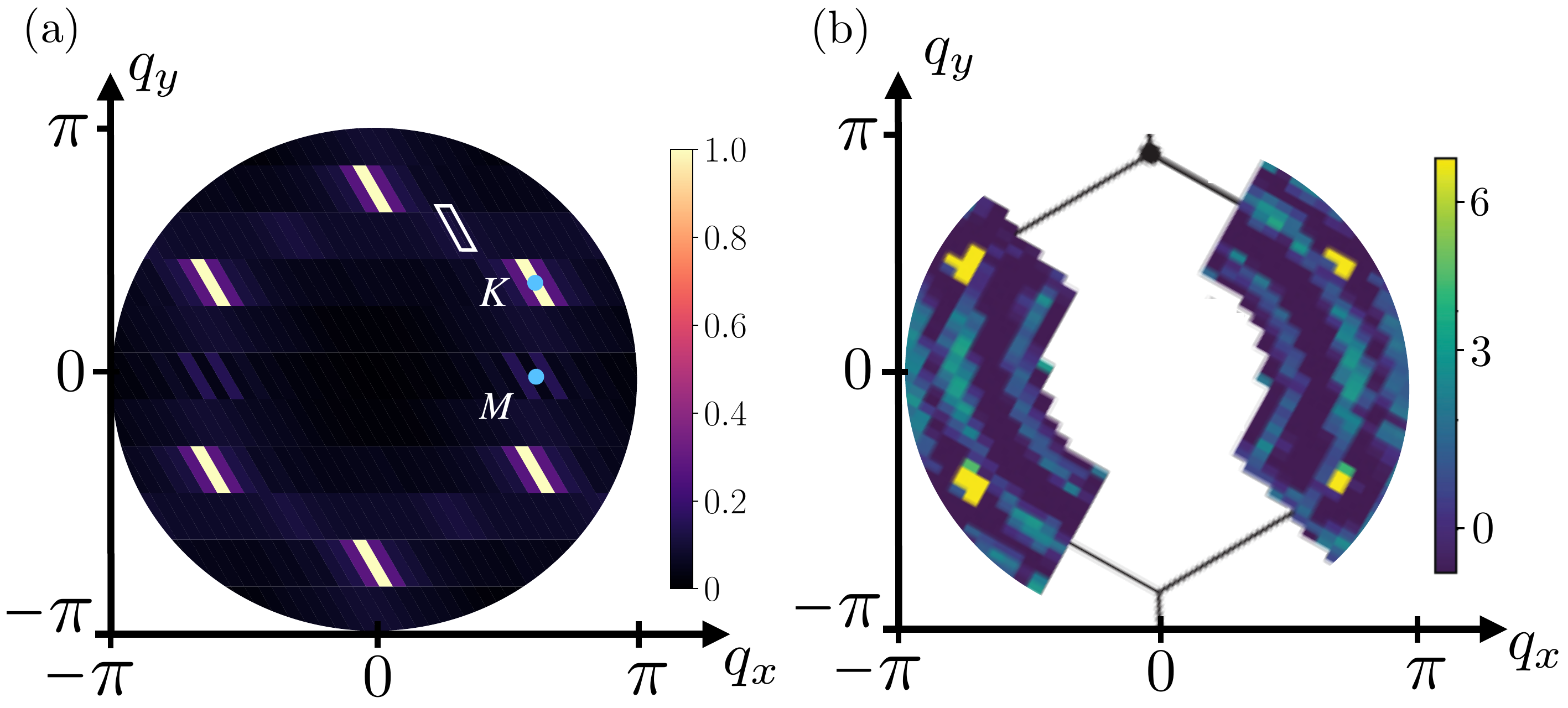}
    \caption{ Static structure factor comparison between our DMRG ground state for a $24\times 6$ cylinder geometry with circumference $L = 6$ and bond dimension up to $\chi=800$. A white rectangle was selected to indicate the DMRG resolution for $J_z = 3\  \text{meV}$,  $J_\perp = 0.2 \ \text{meV}$. In panel b)  we show the results of elastic neutron scattering of Ref. \cite{chen_phase_2024}.}
    \label{fig:ssf_exp}
\end{figure}

The static spin structure factor is defined as
\begin{align}
    \mathcal{S}(\textbf{q})=\expval{ \vec{S}(\textbf{q})\cdot \vec{S}(-\textbf{q})}.
\end{align}
We plot the results from DMRG\footnote{We note that using the finite version of DMRG, one needs to adiabatically tune the anisotropy to reach convergence, while with infinite DMRG (iDMRG), the result can be obtained without such adiabatic tuning. We confirmed that both algorithms lead to the same structure factor.} and assume a ground state in the zero magnetization sector for distinct anisotropy values in Fig.~\ref{fig:dmrg_ssf}. 
We see from the DMRG results of  Fig.~\ref{fig:dmrg_ssf} a smooth interpolation in the structure factor between the Heisenberg point and the Ising limit consistent with previous studies \cite{yamamoto_quantum_2014,sellmann_phase_2015}. The weight in between the high-intensity points gets redistributed to peak more strongly at the corners of the Brillouin Zone. Since the structure factor is calculated from the ground state wave function, we find that the ground state is ordered at the wave-vector $K$, when translated into real space, this indicates a three-sublattice structure. We furthermore confirm that such a result is consistent with the neutron scattering experiment of Ref.~\cite{chen_phase_2024} as shown in Fig.~\ref{fig:ssf_exp}. Both the experiment and the DMRG structure factor agree on the three-sublattice ordering in the ground state. We now focus on the anisotropy value of the experiment, which is $J_\perp=0.2, J_z=3$. {In appendix \ref{appendix:DMRG} we show the perpendicular and parallel structure factors and use their values at the $K,K'$ point to find consistent agreement with the order parameter values of the supersolid phase found in Ref. \cite{paramekanti_2005}}. Moreover, we confirmed that allowing for a non-zero magnetization, the DMRG simulations capture the peak at the $\Gamma$ point and a transition to the $1/3$ magnetization plateau, which is also seen in experiments for low magnetic fields. 

\subsection{Dynamical spin structure factor}
\begin{figure}[t]
    \centering
    \includegraphics[scale=0.17]{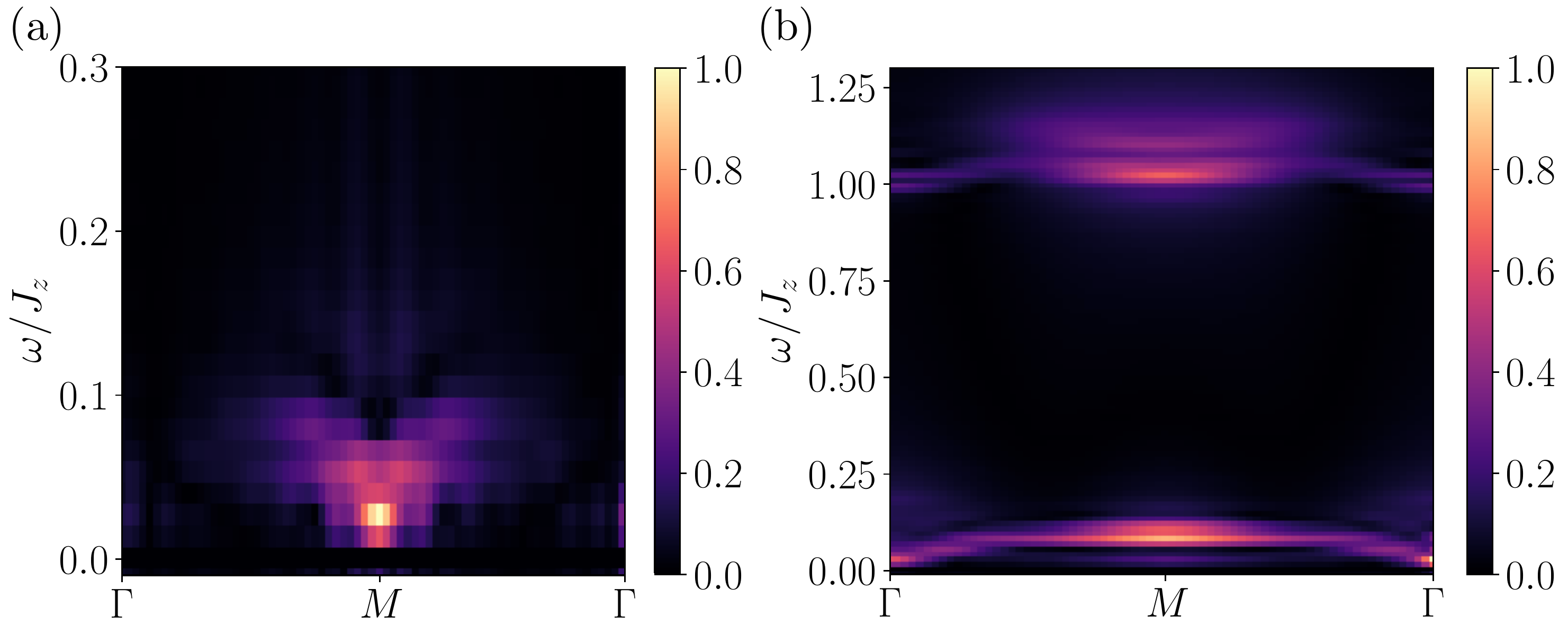}
    \caption{Dynamical spin structure factor of the spin-1/2 XXZ model calculated from the DMRG simulations with parameters as in Fig. \ref{fig:DSF_exp} with $J_z = 3\  \text{meV}$,  $J_\perp = 0.2 \ \text{meV}$  separated into the a) $S^z_iS^z_j$ and b) $S^x_iS^x_j+S^y_iS^y_j$ contributions.}
    \label{fig:Szz_Sperp_DSF}
\end{figure}

After establishing that the ground state has the three-sublattice structure consistent with other Refs.~\cite{chen_phase_2024,zhu_continuum_2024}, we now focus on the excitation spectrum via the computation of the dynamical structure factor (DSF) defined as
\begin{align}
    \mathcal{S}(\mathbf{q}, \omega)=\int \mathrm{d} t \sum_j e^{\mathrm{i} \omega t-\mathrm{i} \mathbf{q} \cdot\left(\mathbf{r}_j-\mathbf{r}_{j_c}\right)}\left\langle \vec{S}_j(t) \cdot\vec{S}_{j_c}(0)\right\rangle.
\end{align}
The algorithm we use for the time evolution is the TDVP method \cite{haegeman2011,haegeman2016,vanderstraeten2019} with $\delta_t = 0.1/J_z$ and evolves for $N_T=400$ steps for $N_x=24,N_y=6$ cylinder with $J_\perp=0.2,J_z=3$ and bond dimension $\chi=800$. One of the main results of this paper, the dynamical spin structure factor along the $\Gamma-M-\Gamma$ is shown in Fig.~\ref{fig:DSF_exp}. We indeed confirm the presence of an Ising excitation branch at $\omega \approx J_z$, see Fig.\ \ref{fig:DSF_exp}, for our model. Such an excitation branch has its origin in the Ising energy of flipping a spin with a finite mean field in the disordered triangular lattice ground state. Only the spins in the ground state with zero mean field have zero cost of flipping, while the others experience a field $h\propto J_z$. Since such states are eigenstates of the pure Ising Hamiltonian, we do not have any dispersion of the energy if $J_\perp=0$. This reasoning can also be tested by looking at the different contributions to the DSF. If we separate the parallel $S^z_iS^z_j$ contribution from the perpendicular $S^x_iS^x_j+S^y_iS^y_j$ and plot each one separately, we obtain Fig. \ref{fig:Szz_Sperp_DSF}. We observe that the spin-flipped high-energy Ising branch has a strong perpendicular spectral weight. Meanwhile, the parallel contribution carries a strong weight at the $M$ point and showcases it as the minimum of the dispersion. We see almost vanishing dispersion close to $\Gamma$ and a broader spectral weight above $M$, which gives good agreement with the Schwinger boson calculations of Sec. \ref{sec:schwinger}.

\begin{figure}[t]
    \centering
    \includegraphics[scale=0.15]{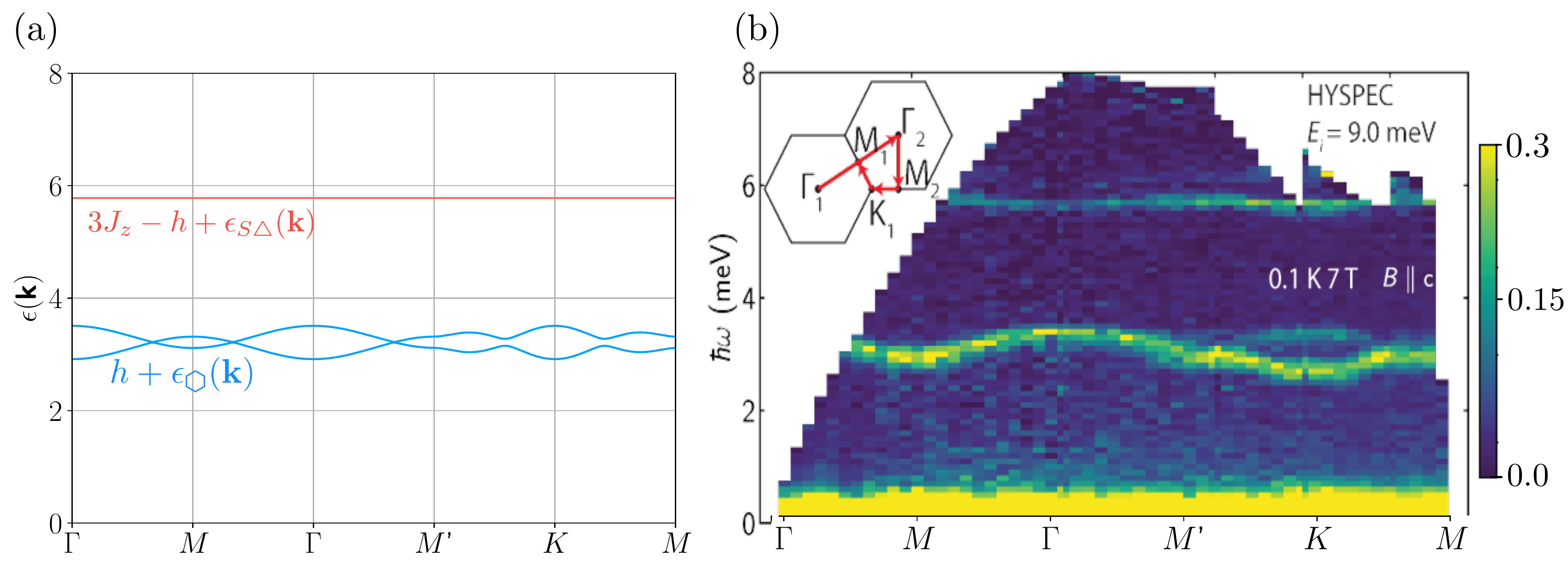}
    \caption{ Dynamical structure factor at finite magnetic field $0.1T$ from a) the excitation spectrum using perturbation theory about the UUD state and b) the experiment of Ref. \cite{chen_phase_2024}.}
    \label{fig:finite_field_exp}
\end{figure}

For finite $J_\perp$ values but close to the Ising point, we expect the defect spins to hop and disperse within an energy width $\Delta E \propto J_\perp$, which we observe in  {Fig.~\ref{fig:DSF_exp}}. A more detailed explanation of such excitations and the emergent lattice on which they hop within the $1/3$ magnetization plateau is given in section \ref{UUDexcit}. The key message from this analysis is that introducing a finite $J_\perp$ perturbatively leads to the emergence of two distinct excitation branches. { As we will show, two types of spins can be flipped, an up spin and a down spin. The up-spin flip operation corresponds to applying to the ground state a $S_i^-\propto b_i$ operator, which one can think of as creating a hole. Here, the up or down states are represented as occupied or unoccupied bosons. In contrast, flipping a down spin corresponds to applying $S_i^+\propto b_i^\dagger$, which we identify as creating a particle. } The particle and hole branches are shown in Fig.~\ref{fig:finite_field_exp}, where we rescaled our results to the energies from the experimental plot of ref \cite{chen_phase_2024}. A remarkably good agreement is seen for the particle (red) line, which has a width of order $J_\perp^2/J_z$. The hole excitations are effectively described by hopping in an emergent honeycomb lattice and have a reduced weight for certain branches due to the neutron scattering cross section \cite{zhu_wannier_2024}.  We further take inspiration from this model and from a hard-core boson mean-field to propose an effective staggered boson model (ESBM), section \ref{sec:ESBM}, which gives qualitative agreement with the DMRG simulations as shown in the turquoise dot-line of \ref{fig:DSF_exp}a.

Focusing on the low energy part of the spectrum, we observe a strong signal at the $M$ point as shown in the heat plot of Fig.~\ref{fig:DSF_exp}b. { The spectral weight is maximum at the $M$ point in momentum space and at a frequency of approximately $\omega_M\approx 0.3 J_\perp$. Away from this momentum point, the maximum in the spectral weight occurs for higher frequencies than $\omega_M$. The intensity of the spectral weight is also reduced away from the $M$ point and continues decreasing down to the $\Gamma$ point.} Such a feature will be seen again within the Schwinger boson approach (white dashed line in Fig.~\ref{fig:DSF_exp}b) in section \ref{sec:schwinger}, it is reminiscent of the vanishing weight for the photon in quantum spin ice models \cite{seeing_light}. Indeed, as we will see in section \ref{sec:qdm}, by using the single-mode approximation on the quantum dimer model, there are transverse photon-like excitations with a very similar dispersion  (blue line in Fig.~\ref{fig:DSF_exp}b). The transverse nature of such excitations at finite but small momentum can be extracted from the dimer structure factor, and indeed, quantitative agreement between the DMRG ground state and the QDM variational supersolid is seen in Fig.~\ref{fig:DMRG_QDM_dimer}.  We conclude from such an analysis that the DMRG numerical results together with the QDM supersolid variational wave function and the Schwinger boson mean field theory agree on the minimum at the $M$ point, the qualitative low-energy excitation dispersion, and its spectrum weight.

Not only is such an agreement reached between our numerical and analytical approaches, but it is also consistent with the neutron scattering experiments of Ref. \cite{zhu_continuum_2024} and \cite{chen_phase_2024} as shown in Fig.~\ref{fig:DSF_exp}d and c). The plots we present here are taken from the inelastic neutron cross-section of the compound $\text{K}_2\text{Co}(\text{SeO}_3)_2$ across the $\Gamma-M-\Gamma$ line. We observe that a pronounced minimum is seen at the $M$ point, with a strong spectral weight, while close to the $\Gamma$ point, we have a decreasing spectral weight. The broad features above the $M$ point seen in both experimental plots are consistent with the picture of higher-energy excitations decomposing into the single photon-like modes obtained from our supersolid variational wave function or from the SBT perspective, they come from fractionalized bosonic spinons. 

\begin{figure}[t]
    \centering
    \includegraphics[scale=0.16]{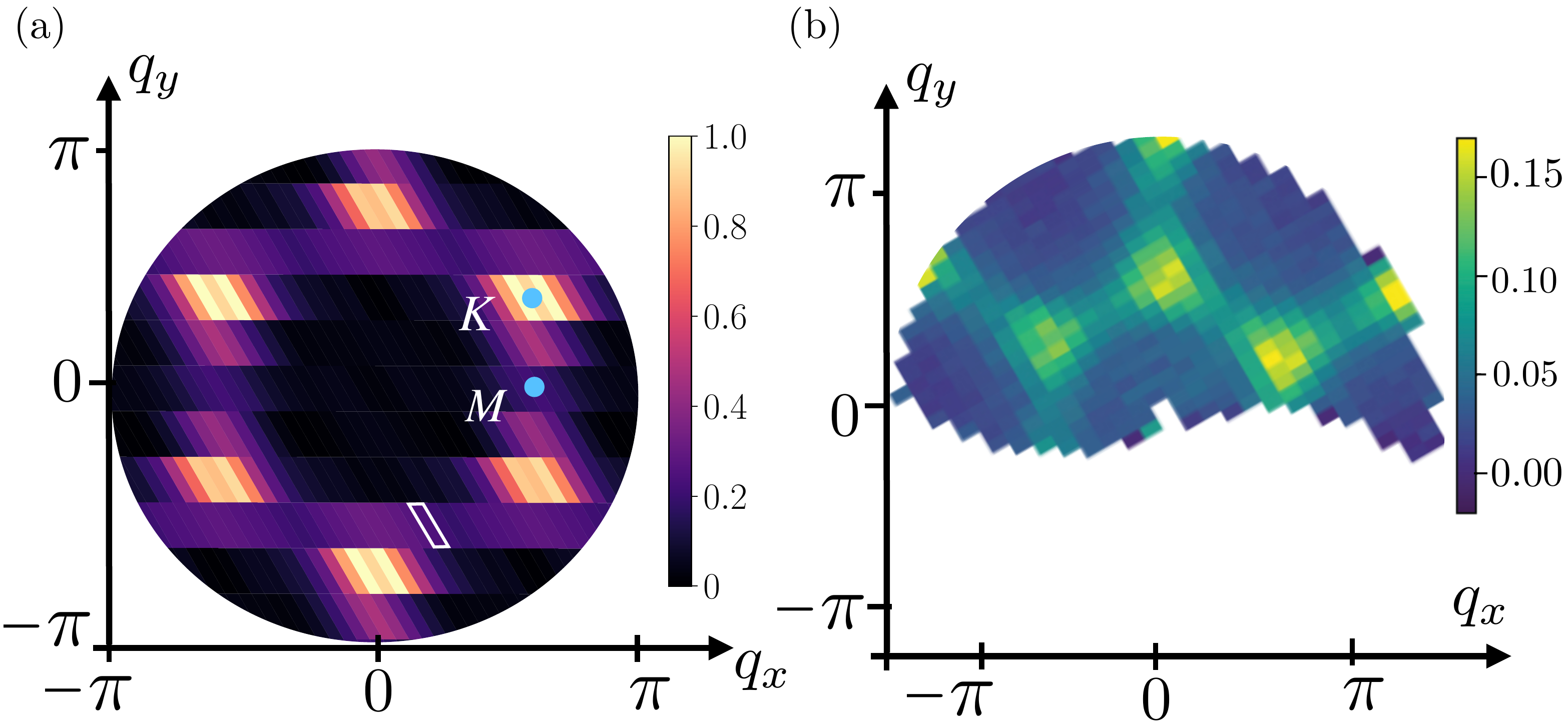}
    \caption{ Dynamical structure factor at the $XX$ exchange frequency $\omega=J_\perp = 0.2 \ \text{meV}$, $J_z = 3\  \text{meV}$ for a) our DMRG simulations of a $24\times 6$ cylinder geometry with circumference $L = 6$ and bond dimension up to $\chi=800$, a white rectangle was selected to indicate the DMRG resolution. b) shows the elastic neutron scattering of Ref. \cite{chen_phase_2024}.}
    \label{fig:DSF_J_exp}
\end{figure}

In both cases, we see that the relevant scattering of the neutrons occurs with emergent quasiparticles, leading to such a broad continuum. Indeed, we also observe such a broad persistent feature at higher energies from the DMRG simulations as seen in the persistent spectral weight extending from $\omega\propto J_\perp$ to $\omega\propto 2J_\perp$. We further analyze the dynamical structure factor at a fixed energy $\omega= J_\perp$. We take the $XX$ exchange energy since that is the observed energy scale for the low-energy branch of excitations. The resulting DMRG simulation and experimental comparison is shown in Fig.~\ref{fig:DSF_J_exp}. The spectral weight now has a pinch point structure near the $M$ point and a strong weight at the $K$ point. Both DMRG and experimental data confirm that such pinch points are broadened but follow the same kagome-like lattice with triangular cells at the $K$ point. 

It is worth mentioning that simple linear spin wave theory fails to capture the previous dynamical structure factor features. Indeed, LSWT around the 120-degree order, which is the proposed ordering at the Heisenberg point $J_\perp=J_z$ is unstable for $0<J_\perp<J_z$ as we show in appendix \ref{App:120instable}. { The instability arises from regions in momentum space where the linear spin wave spectrum has imaginary energies. These unphysical solutions imply the 120-degree state is not a stable ground state of the model}. Even the LSWT around the classical $XXZ$ ground states fails to capture the minimum of the spectrum at $M$. Such an analysis was discussed already in \cite{kleine_spin-wave_1992}, we summarize the relevant results in appendix \ref{app:LSWTXXZ} and plot the spectrum in Fig.~\ref{fig:DSF_exp}b as a dashed green line. {Since linear spin-wave theory does not capture the features of the DSF  we can conclude that there are strong interaction effects in the excitation bands. In the following sections, we present our distinct analytical approaches and their implications for the excitation spectrum near the Ising limit of the triangular-lattice XXZ model in Eq.~\eqref{XXZmodel}.}

\section{Mapping to Hard-core and Schwinger bosons}\label{sec:hardcore}

A fundamental mapping we exploit here is that of hard-core bosons \cite{streib_hard-core_2015}, we will further make contact with the dimer representation and previous literature on the XXZ  \cite{sen_variational_2008,wang_extended_2009}. We substitute our spins for bosons via
\begin{align}
    S_i^{+}=b_i, \quad S_i^{-}=b_i^{\dagger}, \quad S_i^z=1 / 2-b_i^{\dagger} b_i, 
\end{align}
we further restrict to the occupation number $n_i=0,1$, and have $\quad [b_i,b_j^\dagger]=\delta_{ij}(1-2b_i^\dagger b_i)$, this can be implemented for canonical bosons by adding {the term $H_U=U\sum_i (b_i^\dagger)^2(b_i^2)/2$ and taking $U\rightarrow \infty$} which will project the canonical bosons to new bosons in the hard-core subspace. The rest of the Hamiltonian can be transformed to bosons by defining $t=-J_\perp/2$ and $ V=J_z$ we then have
\begin{align}
    H=\sum_{\langle i j\rangle}\left[-t\left(b_i^{\dagger} b_j+b_j^{\dagger} b_i\right)+V n_i n_j \right]-\sum_i (3V+h)n_i, \label{hard-core}
\end{align}
where we also added a magnetic field $h$. Instead of analyzing the system directly at zero field, let us think first about how to approach it from the finite field limit.

\subsection{Excitation spectrum within the 1/3 magnetization plateau}\label{UUDexcit} 

\begin{figure}[t]
    \centering
    \includegraphics[scale=0.25]{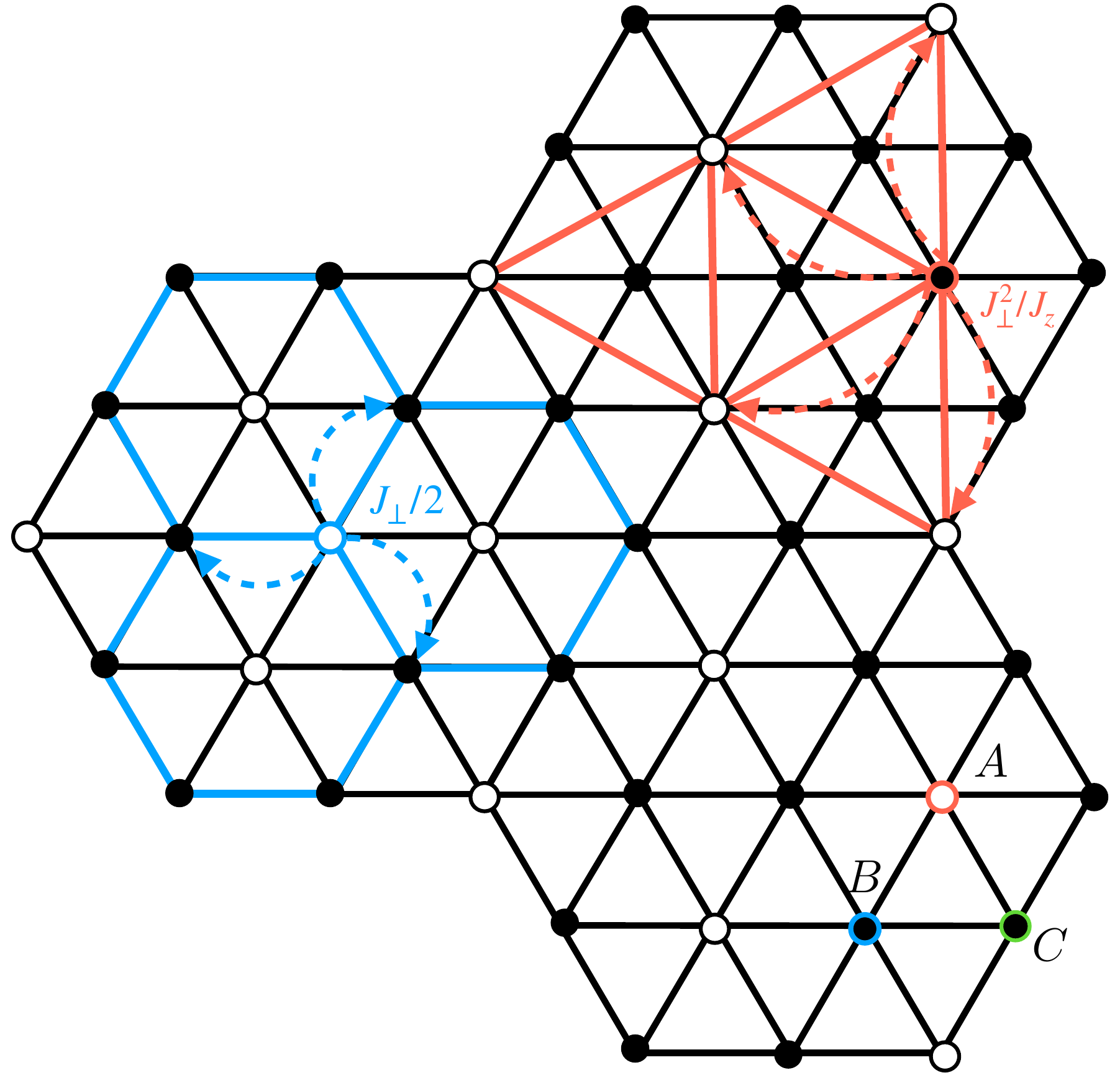}
    \caption{Schematic of one UUD state for the $1/3$ magnetization plateau and it's excitations for $J_\perp\ll J_z$. Black (white) circles denote spin up (down) states of the original triangular lattice in black. Hole excitations (blue) hop in the emergent blue honeycomb lattice while hopping of particle excitations is suppressed by $J_\perp^2/J_z$ and takes place in the supertriangular lattice (red). {The three-sublattice structure ${A,B,C}$ is shown on the bottom right.} }
    \label{fig:uudexcit}
\end{figure}

{For a strong longitudinal magnetic field, we expect to have a fully spin-polarized ground state with all spins pointing up. As the field is lowered, another phase with partially polarized spins is encountered. We can understand this from the classical Ising Hamiltonian in a field defined as } 
\begin{align}
    H_z&= J_z \sum_{\langle i j\rangle} S_i^z S_j^z-h\sum_i S_i^z \label{HIsing}\\
    &= \dfrac{J_z}{4} \sum_{\triangle} \left(S^z_{\triangle}-\dfrac{h}{3J_z}\right)^2+E_0^z
\end{align}
where we rearranged the terms in the last line and defined $S^z_{\triangle}=\displaystyle \sum_{i \in \triangle}S^z_i$ together with $E_0^z=-N_\triangle\left(3\dfrac{J_z}{16}+\dfrac{h^2}{36}\right)$ and $N_\triangle$ as the number of elementary triangles. Now it is clear to see that the ground state of such a Hamiltonian for small but finite $h$ has  $S^z_\triangle=1/2$ for each elementary triangle. This means we need two spins pointing up and one down; there are three such up-up-down(UUD) states ($\ket{\Omega_i},\ i=1,2,3$), reflecting the spontaneously broken translational symmetry to a unit cell with now three sublattices. In Fig.~\ref{fig:uudexcit}, we depict in black and white one such ground state. {Let us denote the $z$-magnetization of the ground state per unit site as $m_z$. For a fully polarized state, we have $m_z=1$ while for the UUD state we have $m_z=1/3$. Since the state is a stable phase, we expect this value to persist for a finite interval of the magnetic field strength. Indeed, this is the $1/3$ magnetization plateau which is observed experimentally as discussed in refs.~\cite{zhu_continuum_2024,chen_phase_2024}. The relevance of this phase is that experimentally the observed supersolid phases appear for magnetic fields close to the $1/3$-plateau. We deduce then that the low-energy excitations of this phase must come down in energy and condense to form the supersolid as a function of the magnetic field for a finite $J_\perp$. In the following, we study these excitations within perturbation theory on $J_\perp$. We then take intuition from the nature of these excitations to construct an effective staggered boson model for the supersolid state in the next section.  }

 Two types of spins can be flipped. First, if one flips an up spin (black), then in the boson language we have created a hole $\ket{i,-}=b_{i}\ket{\Omega_1}$. It is easy to check, since the state is again a product state, that the $H_z$ energy with respect to the ground state is $\Delta E_h = h$. However, it is also possible to flip a down spin (white), in which case we have a particle state $\ket{i,+}=b_{i}^\dagger \ket{\Omega_1}$, with excitation energy $\Delta E_p = 3J_z-h$. From this, we conclude that hole excitations have lower energy, and as the magnetic field is increased, an energy crossing occurs at $h\approx 3J_z$ for which both excitations will be degenerate in energy.

Turning now to the effect of the $XX$ term we may calculate the correction in energy to such excitations for small $J_\perp$ from $H_{XX}=\frac{J_\perp}{2}\sum_{\langle i j\rangle}\left(S_i^{\dagger} S_j+S_j^{\dagger} S_i\right)$. For the hole excitations $\ket{i,-}$, we have, at first order in degenerate perturbation theory, that a hopping process is possible. We can flip a neighboring black spin with hopping amplitude $J_\perp/2$, such that black spins form an emergent honeycomb lattice depicted in blue in Fig.~\ref{fig:uudexcit}. 

The problem of finding the energy correction becomes that of diagonalizing $\bra{i,-}H_{XX}\ket{j,-}$, which we know can be done in momentum space and results in the energy dispersion of graphene $\epsilon_{\varhexagon}(\textbf{k})$. The unit cell of the honeycomb lattice is larger, and thus the graphene-like correction to the energy will have a reduced Brillouin Zone compared to the original triangular lattice. In contrast, the hole excitations behave very differently; first-order perturbation theory doesn't connect such states as the down spins are separated by black spins. As such, one needs to flip first a black spin down and then flip it back to move the particle excitation. Mathematically, we have to go to second order in perturbation theory, such a superexchange gives us that the correction to the energy comes from hopping in the supertriangular lattice (red) in Fig.~\ref{fig:uudexcit}, suppressed by $J_\perp^2/J_z$. We show both particle and hole branches of the excitation energy spectrum in Fig.~\ref{fig:SWUUD}a, where we approximate the particle energy to be flat.
For the parameter values relevant to the experiment, we observe a very good agreement. Nevertheless, one of the honeycomb dispersion branches is absent, as has been noted before \cite{zhu_wannier_2024}. The reason for this comes from the experimental neutron measurements projecting into only one branch.

\subsection{Effective staggered boson hopping model (ESBM)} \label{sec:ESBM}

Although the previous analysis describes quite well the excitations above the UUD state, { we note that such an analysis is strictly valid for the model of eq.~\eqref{HIsing} only at finite $h$ since the zero field limit is disordered. Nevertheless, we may suggest that the zero field model of eq. \eqref{XXZmodel} may perhaps host similar features to the finite field model for small enough $J_\perp$.} In this case, let us consider that the three-sublattice structure persists and motivate an effective model from a mean field perspective on the hard-core boson model eq.~\eqref{hard-core}. We label sites of the triangular lattice by $\lambda=A,B,C$, {as shown in Fig.~\ref{fig:uudexcit}}. Small fluctuations around an unknown mean field value results in $ n_i n_j\approx \left\langle n_i\right\rangle n_j+n_i\left\langle n_j\right\rangle-\left\langle n_i\right\rangle\left\langle n_j\right\rangle$ we re-express the hard-core boson Hamiltonian terms with such values and assume $\expval{n_{\tilde{l}B}}=\expval{n_{\tilde{l}C}}=\tilde{\mu}_2,\expval{n_{\tilde{l}A}}=\tilde{\mu}_1$ for all $\tilde{l}$ sites of the three sub-lattice unit cell. The effective staggered boson model is
\begin{align}
    H_{\text{eff}}=-t\sum_{\langle i j\rangle}\left(b_i^{\dagger} b_j+b_j^{\dagger} b_i\right)-\sum_{\tilde{l}} \mu_2 (n_{\tilde{l}B}+ n_{\tilde{l}C})+\mu_1 n_{\tilde{l}A}. \label{effstagg}
\end{align}
We defined $\mu_1= 6V\tilde{\mu_2}-3V-h$ and $\mu_2=3V(\tilde{\mu}_1+\tilde{\mu}_2-3V-h)$. As an effective model, we can think of the parameters $\mu_1,\mu_2$ as tunable and wonder when the spectrum matches the previously calculated dynamical structure factor plots. The resulting values which generate the plot of Fig.~\ref{fig:DSF_exp} turn out to be $\expval{n_A}=0.525,\expval{n_B}=\expval{n_C}=0.83$. Further comparison to the previous UUD state spectrum and details of the previous calculation are given in Appendix \ref{app:hard-core}. This indicates, from a different perspective, that the system indeed chooses a three-sublattice ordering with two sublattices having approximately the same occupation and one with slightly lower occupation. From the supersolid perspective, such a sublattice has a distinct center spin. Importantly, we remark here that not only is there a good agreement with the dynamical structure factor shown in Fig.~\ref{fig:DSF_exp} along the $\Gamma-M-\Gamma$ line, but we also find gapless excitations at the $K$ point as shown in the appendix. Such low-energy excitations are consistent with the experimental results of Ref. \cite{chen_phase_2024} as seen in the inset of Fig.~\ref{fig:DSF_exp}d.

\subsection{Schwinger boson self-consistent mean field theory (SBT)} \label{sec:schwinger}
To go beyond the simple LSWT result, we must take into account the correlations of the underlying low-energy degrees of freedom. One approach that takes into account interactions already at the mean field level is using Schwinger bosons. In this representation, we can capture the interaction effects of the magnons and naturally describe fractionalization.  We follow the Schwinger boson approach of refs. \cite{chandra_quantum_1990,yukalov_gapless_2006,gazza_schwinger_1993,mezio_broken_2013,ghioldi_magnons_2015}. The quantum spin decomposes into a multiplication of spinon bosonic operators $\phi_i$ with an occupation constraining the size of the local Hilbert space, in equations
\begin{align}
    \vec{S}_{i}&=\dfrac{1}{2}\phi_{is}^\dagger \vec{\sigma}_{ss'}\phi_{is'},\\
    \ [\phi_{is}^\dagger ,\ \phi_{js'}]&=\delta_{ss'}\delta_{i,j}, \ \sum_s \phi_{is}^\dagger \phi_{is}=2S
\end{align}

\begin{figure}[t]
    \centering
    \includegraphics[scale=0.35]{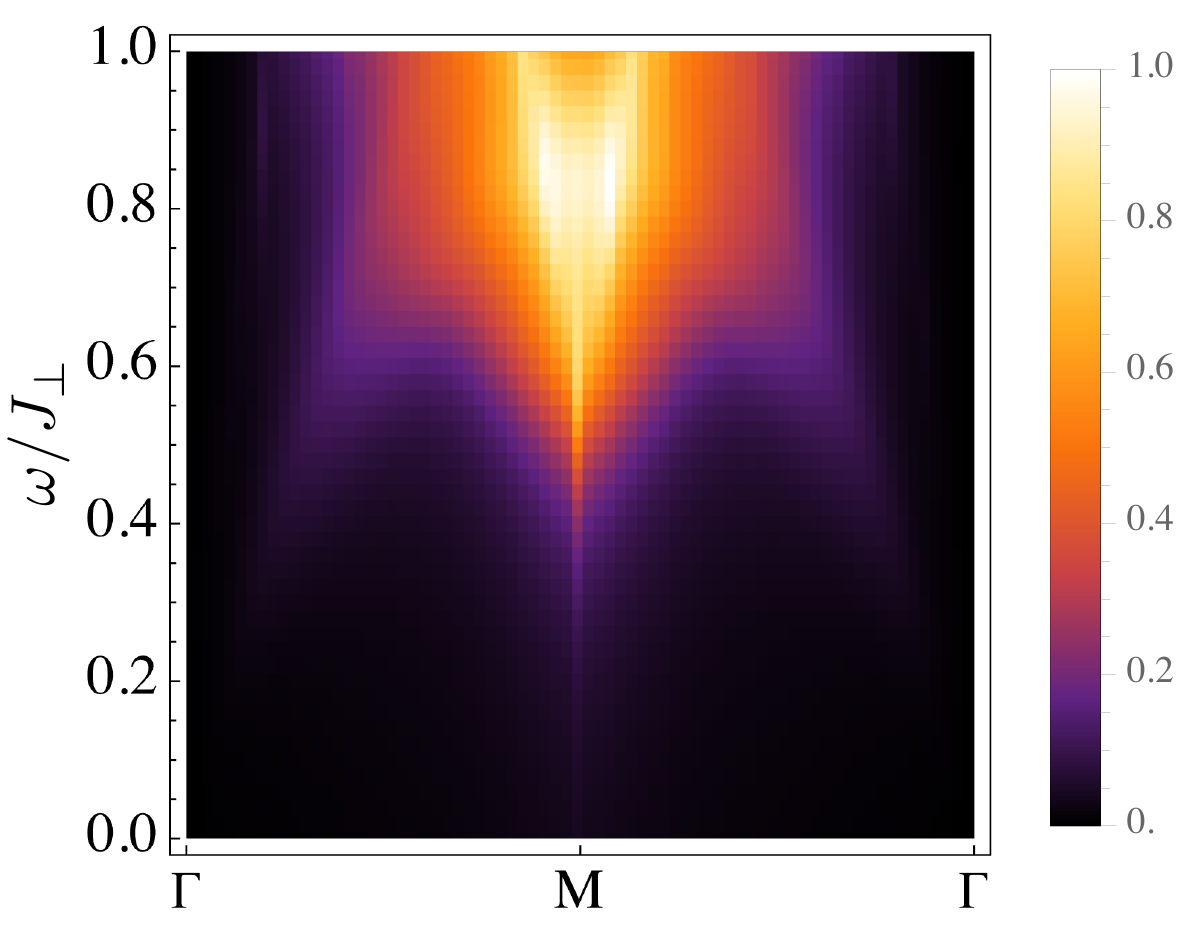}
    \caption{$S_z$ dynamical spin structure factor from Schwinger boson mean field theory for $J_z = 3\  \text{meV}$,  $J_\perp = 0.2 \ \text{meV}$.}
    \label{fig:Schwinger}
\end{figure}

The last constraint makes the Hilbert space of the right dimension, and we consider in the following $S=1/2$. We may substitute in the original spin Hamiltonian to obtain a quartic interaction. To interpret the mean field, it is convenient to define first the bond operators
\begin{align}
    &\hat{A}_{i j} =\dfrac{1}{2}\phi_i(i\sigma^y)\phi_j\\
    &\hat{B}_{i j} =\dfrac{1}{2}\phi_i \phi_j^\dagger\\
    &\hat{D}_{i j} =\dfrac{i}{2}\phi_i(\sigma^zi\sigma^y)\phi_j.
\end{align}
We interpret the first operator as an operator detecting antiferromagnetic correlations, while B detects ferromagnetic correlations, and the last operator comes from the anisotropic interactions since it disappears for the Heisenberg limit. Using the boson particle constraint and defining the normal ordering operation with double dots, we can express the Hamiltonian exactly as
\begin{align}
H= & \frac{1}{2} \sum_{\langle i j\rangle}\left[\left(J_\perp+J_z\right)\left(: \hat{B}_{i j}^{\dagger} \hat{B}_{i j}:-\hat{A}_{i j}^{\dagger} \hat{A}_{i j}\right)\right.\\
& \hspace{1cm} \left. -\left(J_\perp-J_z\right)\left(S^2-2\hat{D}_{i j}^{\dagger} \hat{D}_{i j}\right)\right] . \notag \label{ABD_Ham}
\end{align}
We may now do a mean field analysis by considering expectation values of the bond operators such that the Hamiltonian becomes quadratic in the Schwinger boson operators. Mathematically we consider $\hat{X}_{i j}^{\dagger} \hat{X}_{i j} \approx\left\langle\hat{X}_{i j}^{\dagger}\right\rangle \hat{X}_{i j}+\hat{X}_{i j}^{\dagger}\left\langle\hat{X}_{i j}\right\rangle-\left\langle\hat{X}_{i j}^{\dagger}\right\rangle\left\langle\hat{X}_{i j}\right\rangle$,
where $X=A,B,D$, we further assume translational symmetry for the three distinct bonds of the triangular lattice, meaning $X_{ij}\rightarrow X_{\bm{\delta}}$. In Appendix \ref{app:schwinger}, we calculate the dispersion from going to momentum space and applying a Bogoliubov transformation. The values of the mean field are calculated similarly to superconductors, where we numerically solve the self-consistent gap equation for each order $A,B,D$. Since each of these operators represents a different mean field ground state, solving the coupled gap equations is akin to an intertwined charge density wave and superconducting order. Because of this, there exist multiple instabilities, and we focus on the calculation which converges for $J_\perp=0.2,J_z=3$. The resulting energy dispersion along the $\Gamma-M-\Gamma$ line is plotted as a dashed line denoting Schwinger Boson Theory (SBT) in Fig.~\ref{fig:DSF_exp}. The magnitude of the energy can change dramatically depending on how the constraints imposed on the bond operators are handled. An example comes with the $A,B$ operators, which satisfy always $:B_{ij}^\dagger B_{ij}:+A_{ij}^\dagger A_{ij}=S^2$ because of the commutation relations and the constraint imposed on the $\phi_i$ spinons. If one uses the constraint in the Heisenberg model to express everything in terms of only the $A$ operators, it has been shown \cite{yukalov_gapless_2006,gazza_schwinger_1993,mezio_broken_2013,ghioldi_magnons_2015} that the energy differs from the one calculated with both the $A,B$ operators by the exchange constant in front of the Heisenberg case $J_z=J_\perp$. In our case, we use both $A,B$ operators but reduce the anisotropic complexity by using the occupation constraint to have only one $D$ operator. To compare with the other methods, we divide by the exchange constant in front of this operator $J_z-J_\perp$ and we plot this rescaled dispersion in Fig.~\ref{fig:DSF_exp}. Indeed, we find a minimum at the $M$ point and a linear dispersion near the $\Gamma$ point. Additionally, the dispersion can be evaluated directly to find gapless excitations at the $K$ point. We can go further and calculate directly the dynamical spin structure factor, at least for the $z$ component of spin, as shown in Ref. \cite{ghioldi_magnons_2015}, indeed the general expression is
\begin{align}
S^{z z}(\mathbf{k}, \omega)= & \frac{1}{4 N} \sum_{\mathbf{q}}w(\textbf{k},\textbf{q})\delta\left(\omega-\left(\omega_{\mathbf{q} \uparrow}+\omega_{\mathbf{k}-\mathbf{q} \downarrow}\right)\right).
\end{align}
Where $w(\textbf{k},\textbf{q})=\left(u_{\mathbf{q}} v_{\mathbf{k}+\mathbf{q}}-u_{\mathbf{k}+\mathbf{q}} v_{\mathbf{q}}\right)^2$ and the weights $u_{\mathbf{q}},v_{\mathbf{q}}$ come from the Bogoliubov transformation. The result is plotted in Fig.~\ref{fig:Schwinger}, where we find again a minimum at the $M$ point and a gapless $\Gamma$ point as seen also in the DMRG simulations. The comparison can be done directly since the DMRG simulations allow us to get the $z$ component directly as plotted in the appendix \ref{appendix:DMRG} in Fig.~\ref{fig:DSFpanels}c. Astonishingly, we find both the Schwinger boson mean field theory and the DMRG simulations show a bright spot at the $M$ point minimum, together with a decreased intensity as we approach the $\Gamma$ point. Such a decreased spectral weight has been seen in experiments as shown in Fig.~\ref{fig:DSF_exp}c and \ref{fig:DSF_exp}d. It also appears in the three-dimensional case of quantum spin ice, where this has been interpreted as a signal for the emergent photons. As we will see in the next section, such an interpretation can be followed for our case if we employ the quantum dimer formulation. We will find that there exist transverse photon-like excited states described by a single-mode approximation from a variational supersolid wave function.

\begin{figure}[t]
    \centering
    \includegraphics[width=0.7\linewidth]{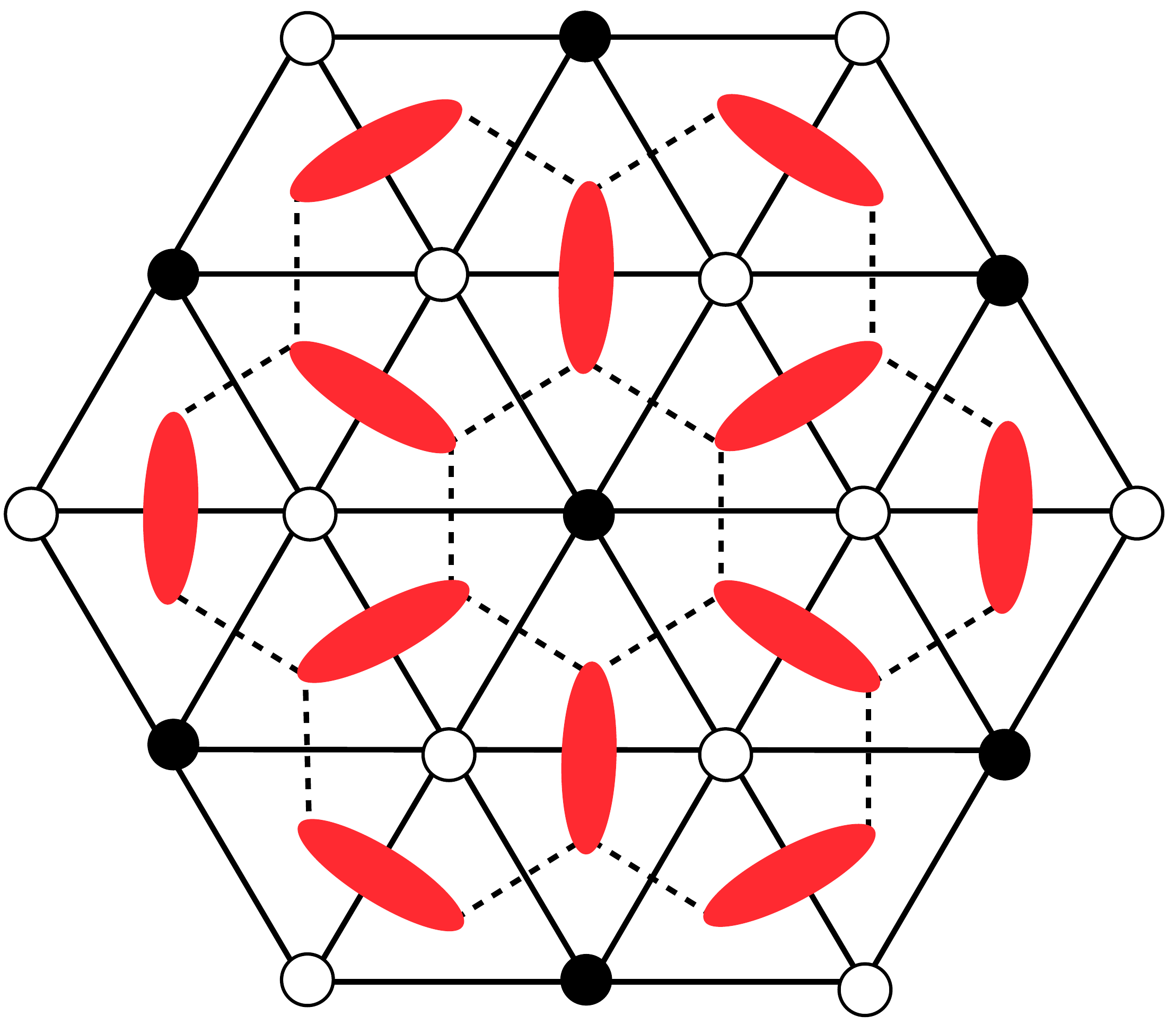}
    \caption{ Example of one possible hard-core dimer covering of the honeycomb lattice within a triangular lattice. The white (black) circles represent the down (up) Ising spins. We have denoted the honeycomb lattice bonds with a black dashed line and dimers in red. A dimer is placed for every frustrated bond. The antiferromagnetic Ising ground state manifold in the original triangular lattice maps to all such hard-core dimer coverings on the dashed honeycomb lattice. }
    \label{fig:dimers}
\end{figure}

\section{Quantum dimer model (QDM) and the variational supersolid wave function} \label{sec:qdm}

We are interested in the properties of the $XXZ$ model of Eq. \eqref{XXZmodel} close to the Ising limit where $J_z\gg J_\perp$. In the pure Ising limit, the model reduces to the antiferromagnetic Ising model on the triangular lattice. As is well known, the ground state of such a model is disordered due to the triangular lattice frustration. The correlation functions of the Ising spins have algebraic decay as shown by Wannier in 1950 \cite{wannier_antiferromagnetism_1950}. Of relevance to us is the Ising ground state manifold, since for small $J_\perp$ we expect perturbations to act within this space only. Generally, they will lift the degeneracy and select an ordered state via a quantum order by disorder mechanism. In the following, we will argue that the variational supersolid wave function discussed in previous studies \cite{sen_variational_2008,wang_extended_2009} also captures the low-energy properties as well as the nature of the correlations. {Indeed, as hinted at in the introduction, we will first map the Ising ground state manifold to hard-core dimer coverings of the honeycomb lattice, an example covering is shown in Fig. \ref{fig:dimers}. The structure of this manifold will reveal an emergent Gauss law, which, for a translation-invariant variational wavefunction composed of the equal-weight superposition of all dimer coverings, will lead to a dimer-dimer structure factor exhibiting a characteristic pinch-point singularity in momentum space as shown in Fig. \ref{fig:structfactdimerRK}b. Such a singularity, well known from the theory of classical spin liquids \cite{Castelnovo2012,Knolle2019,Moessner1998a,Moessner1998b,Isakov2004,Henley2005,Yan2024}, occurs when the intensity along the transverse (longitudinal) direction is maximal (minimal). In turn, only transverse correlations are present while longitudinal ones are suppressed. We will find that when the supersolid variational wavefunction is considered, a small spontaneous translational symmetry-breaking parameter will gap the pinch point while allowing only for transverse-like excitations. }

\subsection{Ising GS manifold}

The antiferromagnetic Ising GS manifold in the triangular lattice is characterized by triangles with only one frustrated bond being ferromagnetic. One can move from one degenerate ground state to another by flipping a spin that has no net field, since that is a zero-energy move. Since we know the ferromagnetic bonds are rare but always present, it makes sense to adopt a formalism that incorporates them directly. Indeed, one can better understand the Ising ground state manifold by mapping it to a dimer model. For every frustrated bond, we put a dimer at the midpoint perpendicular to the bond, as seen for a specific configuration in Fig.~\ref{fig:dimers}. {The endpoints of the dimers will then form a honeycomb lattice, which is depicted with dashed lines.} If we focus on the ground state, it will have the minimum number of frustrated bonds possible. As such, each dimer endpoint in the ground state will touch only one site of the honeycomb lattice. We can then identify the low-energy space of the Ising Hamiltonian as the hard-core dimer coverings of the honeycomb lattice.

The low energy theory for the extensively degenerate Ising GS manifold can be easily shown by going to alternating variables defined as $B_i(\textbf{x})= \sigma_x(n_i(\textbf{x})-1/3)$ where $\sigma_x=\pm 1$ for $A,B $ sublattices of the honeycomb lattice, here  $n_i(\textbf{x})$ with $i=1,2,3$ is the dimer occupation between sites $\textbf{x}$ and $\textbf{x}+\bm{\delta}_i$. The hard core dimer constraint manifests as $\sum_i n_i(\textbf{x})+n_i(\textbf{x}-\bm{\delta}_i)=1$  which in the new variables implies an emergent Gauss law
\begin{align}
    \div{\textbf{B}}=0.
\end{align}
\subsection{Dimer resonance from perturbation theory}

For finite $J_\perp$ we consider the effective Hamiltonian from first-order perturbation theory. This will be given by the processes that flip two neighboring spins such that the hard-core dimer constraint is respected. In dimer language, we have 
\begin{align}
    \mathcal{H}_{\text{eff}} = -t \hspace{-0.4cm} \sum_{\hspace{0.4cm} \includegraphics[width=0.5cm]{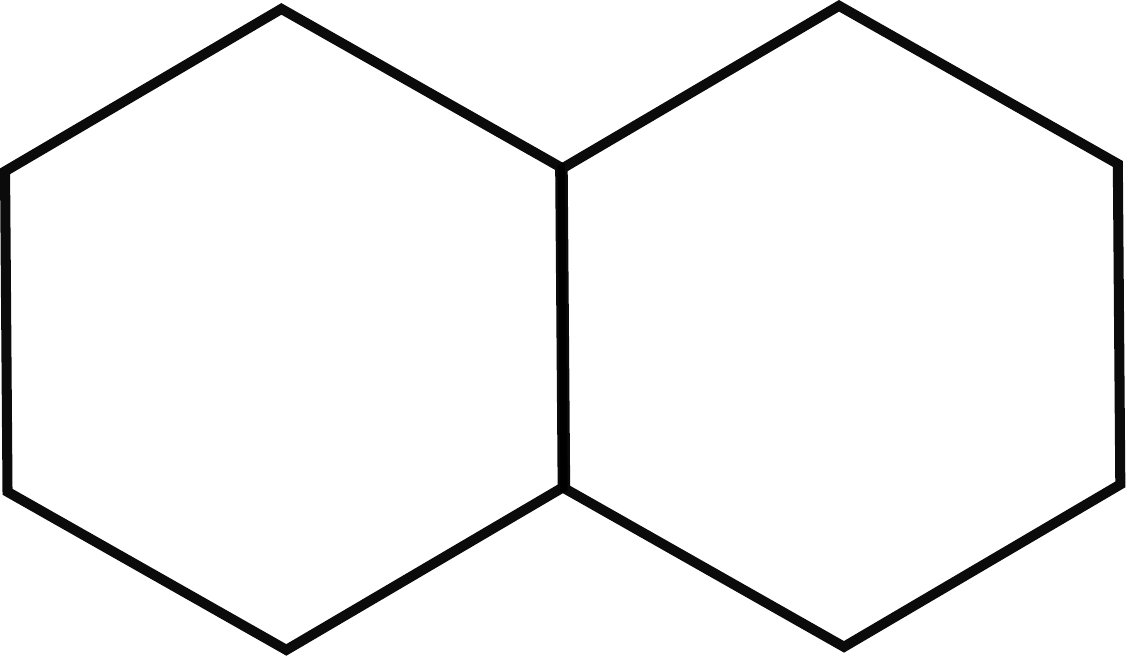}_{\langle ij \rangle}} \hspace{-0.4cm} \left( \left| \includegraphics[width=0.5cm]{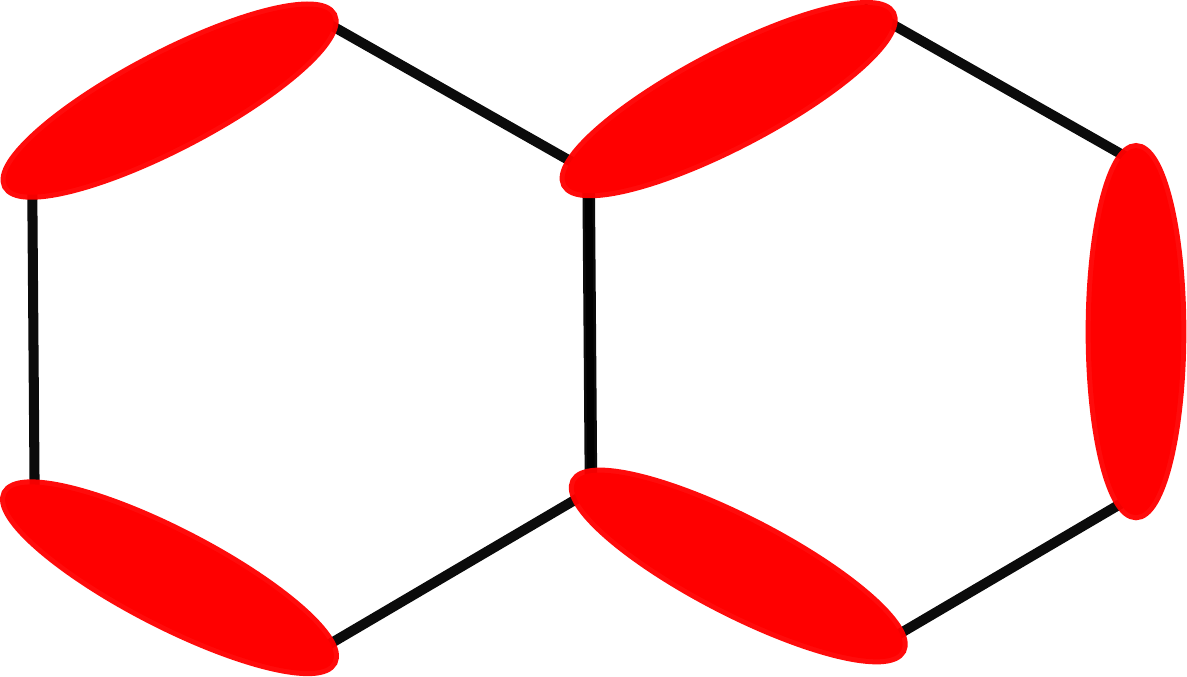}_{\langle ij \rangle} \right\rangle \left\langle \includegraphics[width=0.5cm]{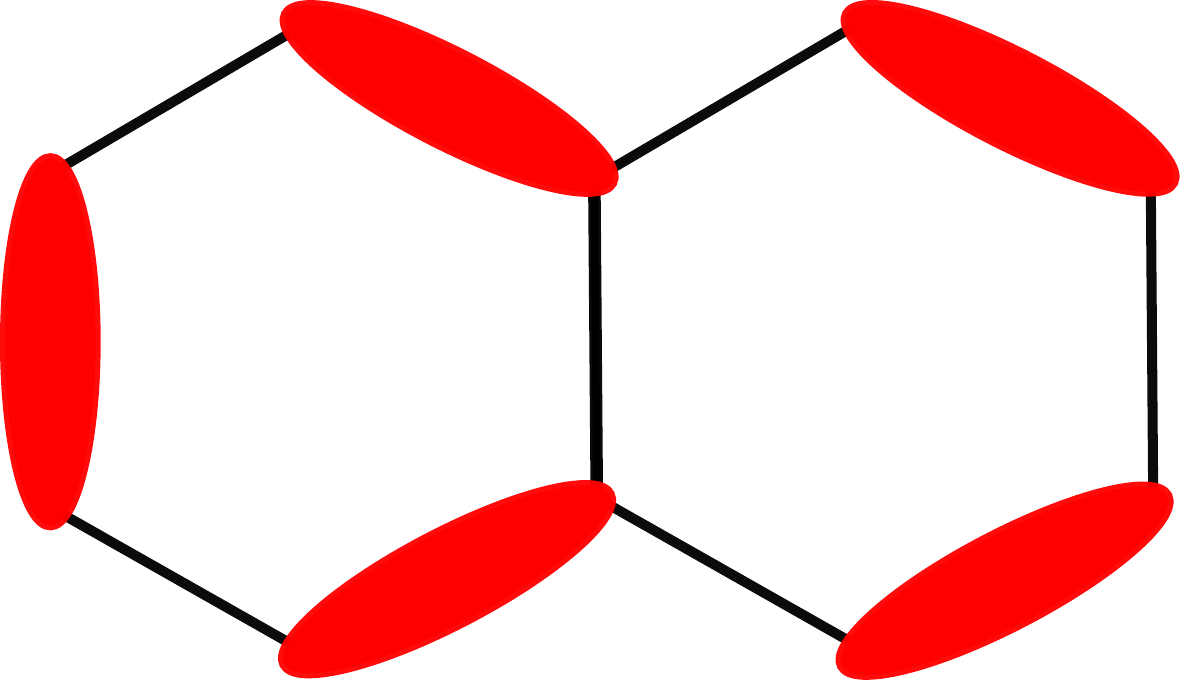}_{\langle ij \rangle} \right| + \text{h.c.} \right). \label{DoubleRes}
\end{align}
In contrast, when we consider a transverse field, we only have to lowest order single hexagon resonances. The extra structure from the nearest neighbor flips will contribute to the minima observed at the $M$ point.

\subsection{Supersolid variational wave function}

We now review the supersolid wave function that was proposed for the XXZ model of Eq. \eqref{XXZmodel} in refs.  \cite{sen_variational_2008,wang_extended_2009}. First, from the perspective of the hard-core bosons, we see that to get a superfluid wave function $\ket{\Psi}$, we need a finite expectation value $\expval{b_i}{\Psi}\neq 0$. Such a requirement in the dimer language amounts to the presence { of dimer configurations with one elementary plaquette of the lattice having an alternating dimer occupancy, which we call a flippable plaquette}. The natural wave function to postulate is an equal weight superposition of all hard-core dimer coverings of the honeycomb lattice; this is the Rohksar-Kivelson (RK) wavefunction. Such an idea goes back to Anderson’s resonating valence bond (RVB) state from 1973, which features an incompressible quantum liquid \cite{anderson_resonating_1973}, as he revisited it in the context of a mechanism for high-temperature superconductivity in 1987 \cite{anderson_resonating_1987}. The Rokhsar–Kivelson model, postulated in 1988 \cite{RKpoint}, recognizes the previously mentioned wavefunction as the $U(1)$ quantum spin liquid ground state of a quantum dimer model Hamiltonian at the RK point. We denote each hard-core covering by $\mathcal{C}$ and its corresponding state in the Hilbert space by $\ket{\mathcal{C}}$. Now, such a state has full translational symmetry, while we are interested in a supersolid with a broken translational symmetry. Since we know that the Bragg peaks are at the $K,K'$ points, it is natural to think of a wave function with a unit cell that allows the original crystal momentum vectors of $K,K'$ to be folded down to the new $\Gamma$ point. The easiest way to implement this is to change the weights such that they are positive and reflect this pattern. We then define the supersolid variational wavefunction as
\begin{align}
    &\ket{\Psi}= \sum_{\mathcal{C}}\phi(\mathcal{C})\ket{\mathcal{C}},\quad Z = \sum_{\mathcal{C}} W(\mathcal{C}),  \label{VA_SS}\\
    &\quad P(\mathcal{C}) = W(\mathcal{C})/Z, \quad \phi(\mathcal{C})=\sqrt{P(\mathcal{C})}.
\end{align}
The weights $W(\mathcal{C})$  are shown in Fig.~\ref{fig:honeycomb}. The red or blue colors represent the $w_{ab}=z$ and $w_{ab}=1$ weight for each dimer, respectively. {The weight $z$ is found by minimizing the expectation value of $\expval{\mathcal{H}_{\text{eff}}}{\Psi}/\braket{\Psi}$. Indeed, this was done by the authors of ref.  \cite{sen_variational_2008} where they found the minimum occurred for $z=0.9250$, we use this value in the following}. Since we are still summing over hard-core dimer coverings, we expect still flippable plaquettes to be present and in this way  $\expval{b_i}{\Psi}\neq 0$. Now we would like to understand the excitations above this ground state; for this, we use the single-mode approximation (SMA).

\subsection{Single Mode Approximation} \label{sec:SMA}

\begin{figure}[t]
    \centering
    \includegraphics[scale=0.23]{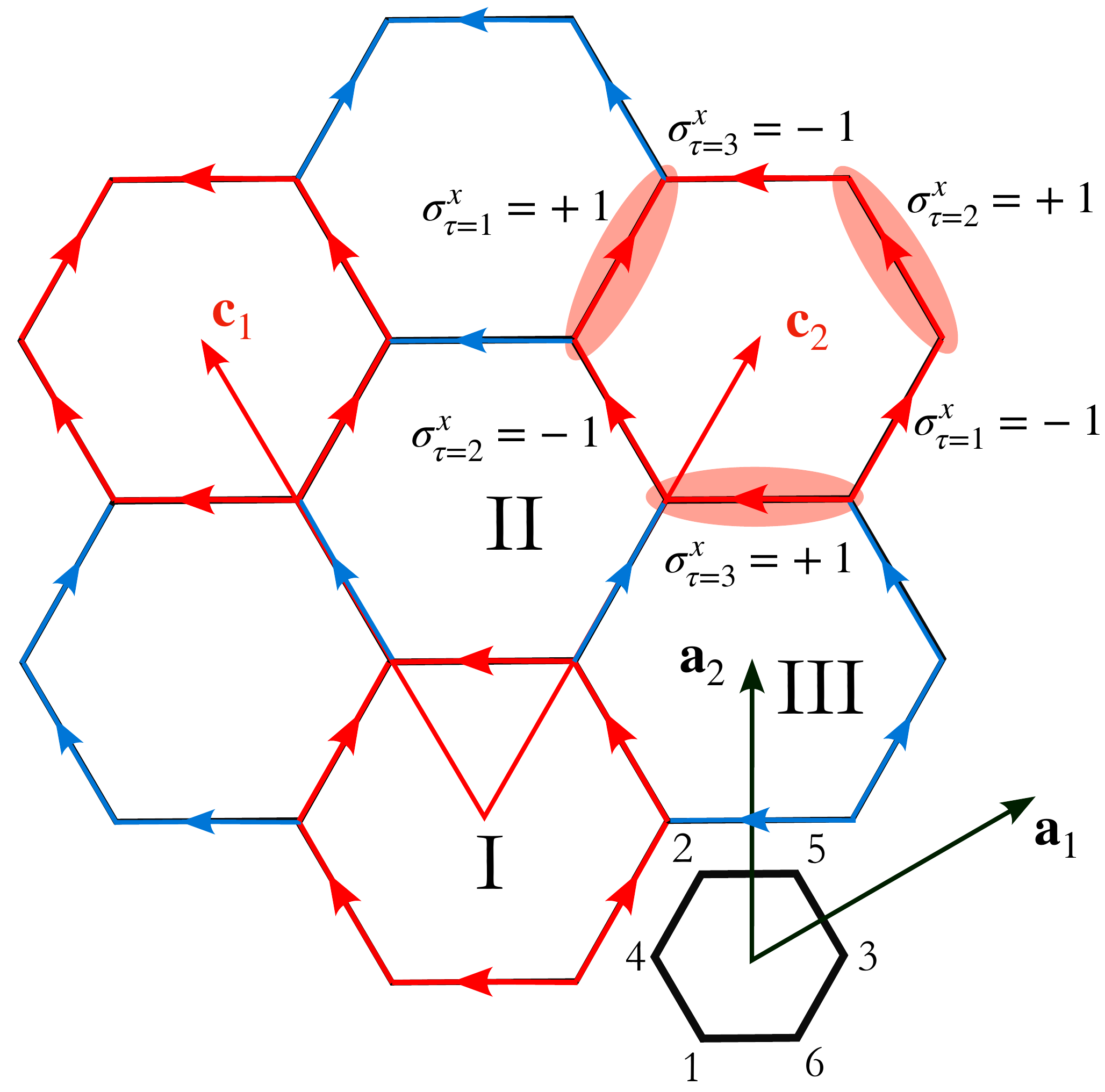}
    \caption{{Honeycomb lattice for the dimer model representation of the XXZ model in the triangular lattice, we use the convention of Ref. \cite{sen_variational_2008}. We show the triangular lattice vectors in black and the variational wavefunction wave vectors showcasing three sublattice ordering in red. The arrows correspond to a given Kasteleyn orientation. We name the six sublattices according to the small black bottom-right hexagon. The labels $I,II,III$ denote the three symmetry inequivalent hexagons under the three-sublattice ordering. The red lines have a weight of $z$ while the blue lines have a weight of $1$.}}
    \label{fig:honeycomb}
\end{figure}
The SMA was originally introduced by Feynman in 1954 to investigate the collective excitations of superfluid $\text{He}^4$ \cite{feynman1954}, its use has been widely applied to various systems including cold atom systems, BCS-BEC crossover, phonons in crystals, metals, quantum spins and fractional quantum hall systems \cite{griffin2009bose,toennies2004superfluid,haegeman2013elementary,yi2002single,bruschi2010unruh,lauchli2008dynamical,manousakis1991spin,chen2005bcs,girvin1986}. The spirit of the approximation relies on noticing that for a quantum harmonic oscillator the first excited state can be constructed by multiplying by the position operator, since $\hat{x}\propto a+a^\dagger$. This creates an orthogonal state to the ground state, as the average position of the ground state is zero. {For a superfluid, the particle density and phase operators satisfy analogous position-momentum commutation relations. Thus, it is instructive to consider the wavefunction with an applied density operator on the ground state as a possible low-energy excitation. In our case this corresponds to $n_i(\textbf{R})\ket{\Psi}$. Let us define for convenience the operator $\sigma^x_\tau(\textbf{R})=2n_{\hat{\tau}}(\textbf{R})-1=\pm 1$, minus when the dimer in the $\tau=1,2,3$ direction of the unit cell located at $\textbf{R}$ is unoccupied and plus for occupied. Where $\textbf{R}$ lives in the original triangular lattice $\textbf{R}=l \textbf{a}_1+k\textbf{a}_2$. We show an example of the values of this operator in Fig. \ref{fig:honeycomb}, where the red bonds of the Honeycomb lattice that are occupied by dimers lead to $\sigma^x_\tau = +1$}. Now we expect the Fourier modes to be the relevant decoupled modes, so we define
\begin{align}
    \tilde{\sigma}_{\hat{\tau}}^x(\mathbf{q}) = \dfrac{1}{\sqrt{N}} \sum_{\textbf{R}}\sigma_{\hat{\tau}}^x(\textbf{R})e^{i\textbf{q}\cdot \textbf{R}},
\end{align}
where $N$ is the number of lattice sites. A candidate for a low-energy excited state becomes 
\begin{align}
    \ket{\textbf{q},\tau}= \tilde{\sigma}_{\hat{\tau}}^x(\mathbf{q}) \ket{\Psi}.
\end{align}
{ The energy of this excitation will not necessarily match the true lowest excitation energy, which we call $E(\mathbf{q}, \hat{\tau})$. Nevertheless, as long as the excitation is orthogonal to the ground state, the energy of our candidate wavefunction will act as an upper bound to $E(\mathbf{q}, \hat{\tau})$. To calculate the variational energy of our state, we first define the norm of our candidate excitation as $D_\tau (\mathbf{q})=\braket{\textbf{q},\tau}$. This quantity will also turn out to be the dimer-dimer structure factor. Next, we need the expectation value of the effective Hamiltonian $\expval{\mathcal{H}_{\text{eff}}}{\textbf{q},\tau}$. For computing this expectation value, it is better to rewrite this in terms of commutators
\begin{align}
    E(\mathbf{q}, \hat{\tau}) \leqslant {\dfrac{1}{2}}\frac{\left\langle \Psi\left|\left[\tilde{\sigma}_{\hat{\tau}}^x(-\mathbf{q}),\left[\mathcal{H}_{\text{eff}}, \tilde{\sigma}_{\hat{\tau}}^x(\mathbf{q})\right]\right]\right| \Psi\right\rangle}{\left\langle \Psi\left|\tilde{\sigma}_{\hat{\tau}}^x(-\mathbf{q}) \tilde{\sigma}_{\hat{\tau}}^x(\mathbf{q})\right| \Psi\right\rangle} \equiv {\dfrac{1}{2}}\frac{f_\tau(\mathbf{q})}{D_\tau (\mathbf{q})},\label{sma}
\end{align}
where $f_\tau(\mathbf{q})$ is called the oscillator strength. In the following, we calculate both quantities separately, starting with the oscillator strength.}
 
\subsubsection{Oscillator strength}

We calculate the {numerator} of the SMA energy by using the commutator $ [\sigma^{\pm},\sigma^x]=\mp 2 \sigma^{\pm}$ and expressing all resonance operators from Eq. \eqref{DoubleRes} in terms of $\sigma^{\pm}$. As shown in the appendix \ref{app:oscstrength}, this depends on $z,$ but for the lowest order contribution where $z\approx 1,$ we can write 

\begin{align}
    &f_{3}(\mathbf{q})= \frac{4\abs{E_0}}{3 N}\left( \abs{e^{i\textbf{q}\cdot(\textbf{a}_1-\textbf{a}_2)}+e^{-i\textbf{q}\cdot \textbf{a}_2}-e^{i\textbf{q}\cdot\textbf{a}_1}-1}^2 + \notag  \right.  \\
   & \left. 4\sin^2(\textbf{q}\cdot \textbf{a}_2) +\abs{1+e^{i\textbf{q}\cdot (\textbf{a}_1-\textbf{a}_2)}-e^{i\textbf{q}\cdot (\textbf{a}_1-2\textbf{a}_2)}-e^{-i\textbf{q}\cdot \textbf{a}_2}}^2 \right),
\end{align}

 where we assumed the resonance operators along all nearest neighbor directions have all the same expectation value to first order in $1-z$ for the variational wave function, so that the oscillator strengths turn out to be proportional to the minimized energy per site {$\abs{E_0}/N=0.1377J_\perp$ which we take from ref.~\cite{sen_variational_2008}}. The form of the oscillator strength in other directions is shown in the appendix \ref{app:oscstrength}. 

Before calculating the structure factor, let us mention that the oscillator strength has a gapless zero line from $\Gamma$ to $M$ for the $\tau=3$ dimers. Such a line also appears for the square lattice dimer model with single plaquette resonances. In that case for the dimers in the $\tau=\hat{x}$ direction there is a conserved quantity $\sigma_{\tau}^x(\textbf{q}_0)$ with $\textbf{q}_0=(q_x,\pi)$  so $f(\textbf{q}_0)=0$ and the oscillator strength for $\textbf{q}_0+\textbf{k}$ has the form $f(\textbf{q}_0+\textbf{k})\propto (\textbf{k}\times \hat{\tau})^2$. 

In our case we also find that the quantum dynamics creates and destroys pairs of dimers, and hence  for $\tau=\hat{x}=\hat{3}$ we find that $\sigma_{\tau}^x(\textbf{q}_0)$ is conserved with $\textbf{q}_0=(q_x,0)$ we can expand to obtain (assuming resonances have {approximately} the same expectation value)
\begin{align}
    f_{\hat{\tau}=\hat{x}}(\textbf{q}_0+\textbf{k})\propto (\textbf{k}\times \hat{\tau})^2+O(\abs{\textbf{k}}^3)\label{osc_stren}
\end{align}

\begin{figure}[t]
    \centering
    \includegraphics[scale=0.15]{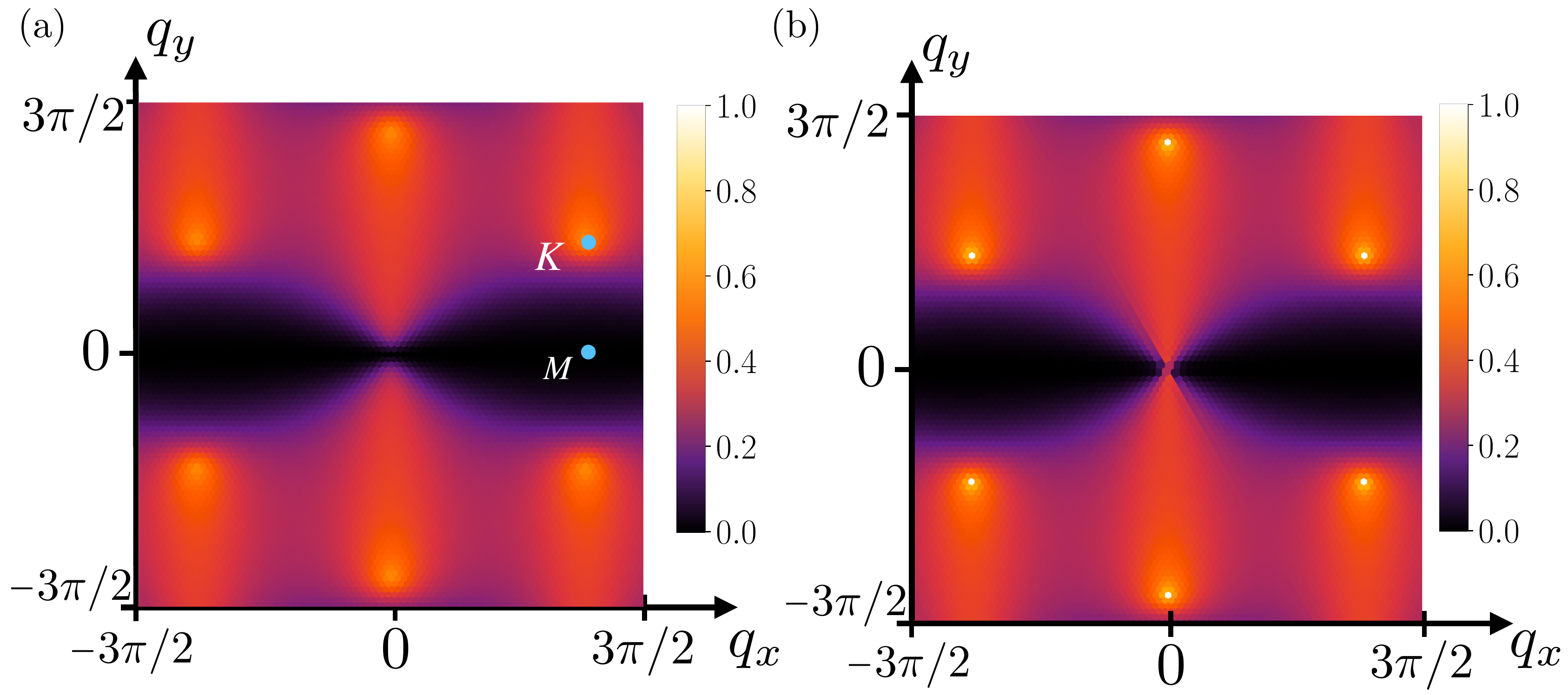}
    \caption{Connected component of the structure factor for the $\tau=3$ bonds $D_c(\textbf{q})$ for a) $z=0.9250$ and b) $z=1$ corresponding to the supersolid variational wave function and RK honeycomb wavefunction, respectively.}
    \label{fig:structfactdimerRK}
\end{figure}

\begin{figure}[t]
    \centering
    \includegraphics[scale=0.15]{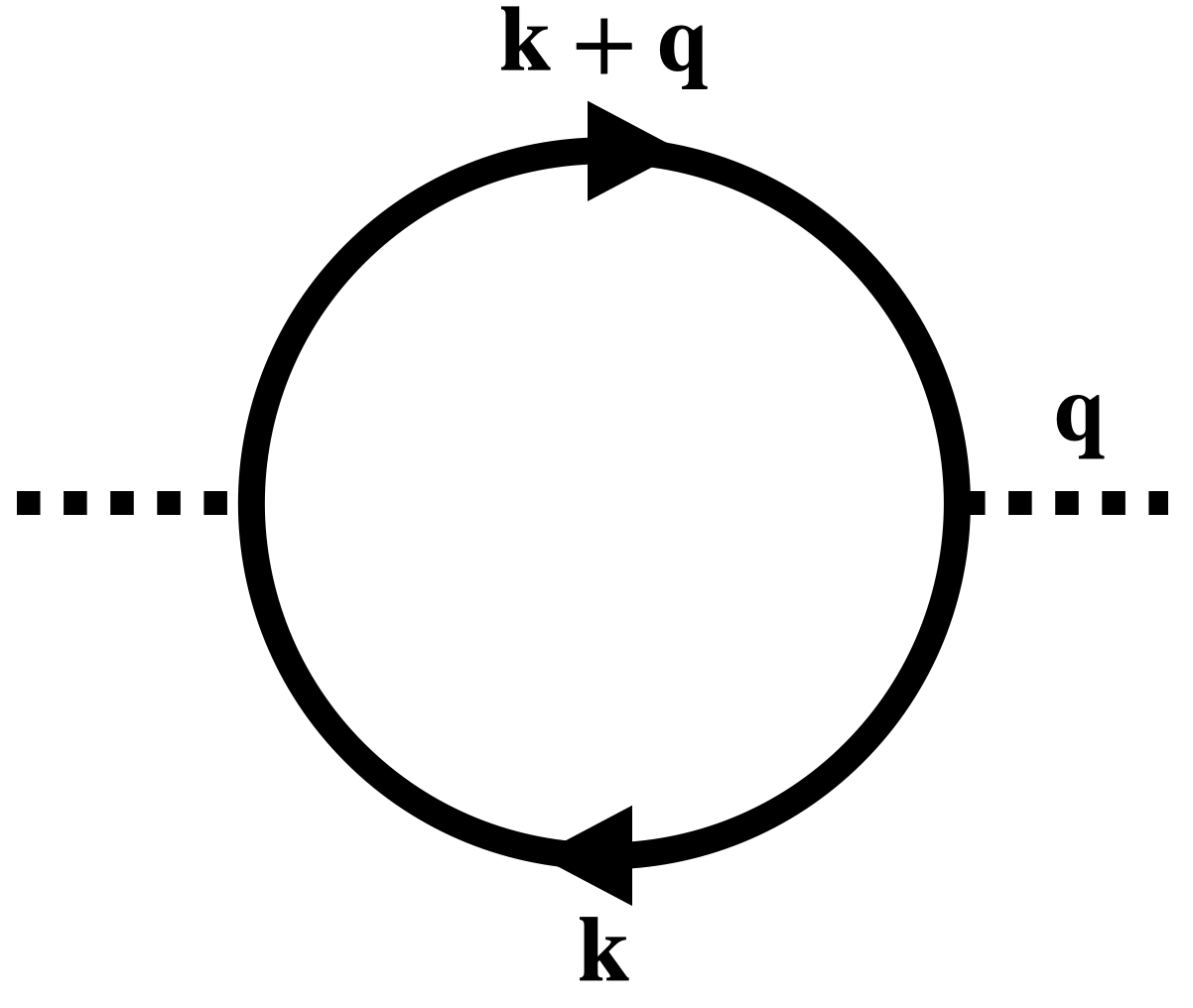}
    \caption{Feynman diagram representing the connected part of the dimer structure factor for the supersolid variational wave function. The mapping to Kasteleyn fermions leads to the form of a polarization bubble with external momentum $\textbf{q}$. We emphasize that, in contrast to usual diagrams, there is no frequency summation since we have classical fermions. }
    \label{fig:polbubble}
\end{figure}

\subsubsection{Dimer structure factor}

Next, we calculate the denominator of Eq. \eqref{sma}, which is given by the dimer structure factor as can be seen by reexpressing the $\sigma^x$ correlator in terms of the occupations in real space. We end up needing to calculate the Fourier transform of
\begin{align}
   &\left\langle \Psi\left|\tilde{\sigma}_{\hat{\tau}}^x(\textbf{R}_1) \tilde{\sigma}_{\hat{\tau}}^x(\textbf{R}_2)\right| \Psi\right\rangle=
4\left\langle \Psi\left| n_{\hat{\tau}}(\textbf{R}_1)n_{\hat{\tau}}(\textbf{R}_2)\right| \Psi\right\rangle \notag\\
&\hspace{1cm}-2 \left\langle \Psi\left| n_{\hat{\tau}}(\textbf{R}_1)\right| \Psi\right\rangle-2 \left\langle \Psi\left| n_{\hat{\tau}}(\textbf{R}_2)\right| \Psi\right\rangle+1
\end{align}
Since we are first treating the $t>0$ model (we will use the unitary mapping from Ref. \cite{wang_extended_2009} to convert to the frustrated regime), we can calculate the correlation functions from classical dimer models. Any planar graph close-packed classical dimer model can be solved in terms of Pfaffians via Kasteleyn's theorem. In more modern language, one can invoke fermionic variables that represent the hard-core dimers. In this case, we have a free fermion action:
\begin{align}
    Z=\abs{\int \mathcal{D} \psi \  e^{S}}, \quad S= \dfrac{1}{2} \sum_{ab}\psi_a K_{ab} \psi_b,
\end{align}
 where the so-called Kasteleyn fermions live in the sites of the honeycomb lattice $\psi_a$, where $a,b$ label the sites of the honeycomb lattice. $K_{ab}$ is known as the Kasteleyn matrix and is defined in terms of the dimer model weights or fugacities $w_{ab}$, where $ W(\mathcal{C})= \prod_{\text{covered}\expval{a,b}}w_{ab}$, an a chosen orientation of the bonds in the lattice also known as a Kasteleyn orientation. This orientation needs to satisfy that when traversing clockwise, each elementary plaquette of the lattice, here hexagons, has an odd number of arrows in the clockwise direction. For the variational wave function exhibiting supersolid order, we have $w_{ab}$ is $z$ or $1$, and the Kasteleyn orientation is taken as in the Fig.~\ref{fig:honeycomb}. The Kasteleyn matrix entries are then defined as $K_{ab}=w_{ab}$ if an arrow goes from $a$ to $b$ and $K_{ab}=-w_{ab}$ if it goes from $b$ to $a$ instead.

The correspondence for correlation functions says that we should replace $n_{ab}\rightarrow K_{ab}\psi_a \psi_b$. To write the positions of the fermions we rename $\textbf{r}_1=\textbf{R}_1+\bm{\tau}_A$ since then we can identify $\tau=1,2,3$ with the bonds of the honeycomb $\bm{\delta}_{1},\bm{\delta}_{2},\bm{\delta}_{3}$ pointing away from the site. We note that the mapping to fermions fails for correlations of the double occupancy $\expval{n^2_\tau(\textbf{R})}$. In this case, we need to add the occupations by hand, which will result in a constant shift of the values for the structure factor as a function of momentum.

The full structure factor from just the fermionic mapping is given by $D(\textbf{q})=  D_c(\textbf{q}) + D_d(\textbf{q})$, where $D_d(\textbf{q})$ is the disconnected structure factor. We used Wick's theorem to decompose the four-point fermion correlator in terms of two-point correlators. It has the general form $D_d(\textbf{q})\propto N\delta_{q,\textbf{G}}+cte.$ The connected part of the structure factor is given by the polarization bubble of the Kasteleyn fermions, in diagrammatic form it is shown in Fig.~\ref{fig:polbubble}, the explicit expression is shown in the appendix \ref{app:hard-core}.

\begin{figure}[t]
    \centering
    \includegraphics[scale=0.17]{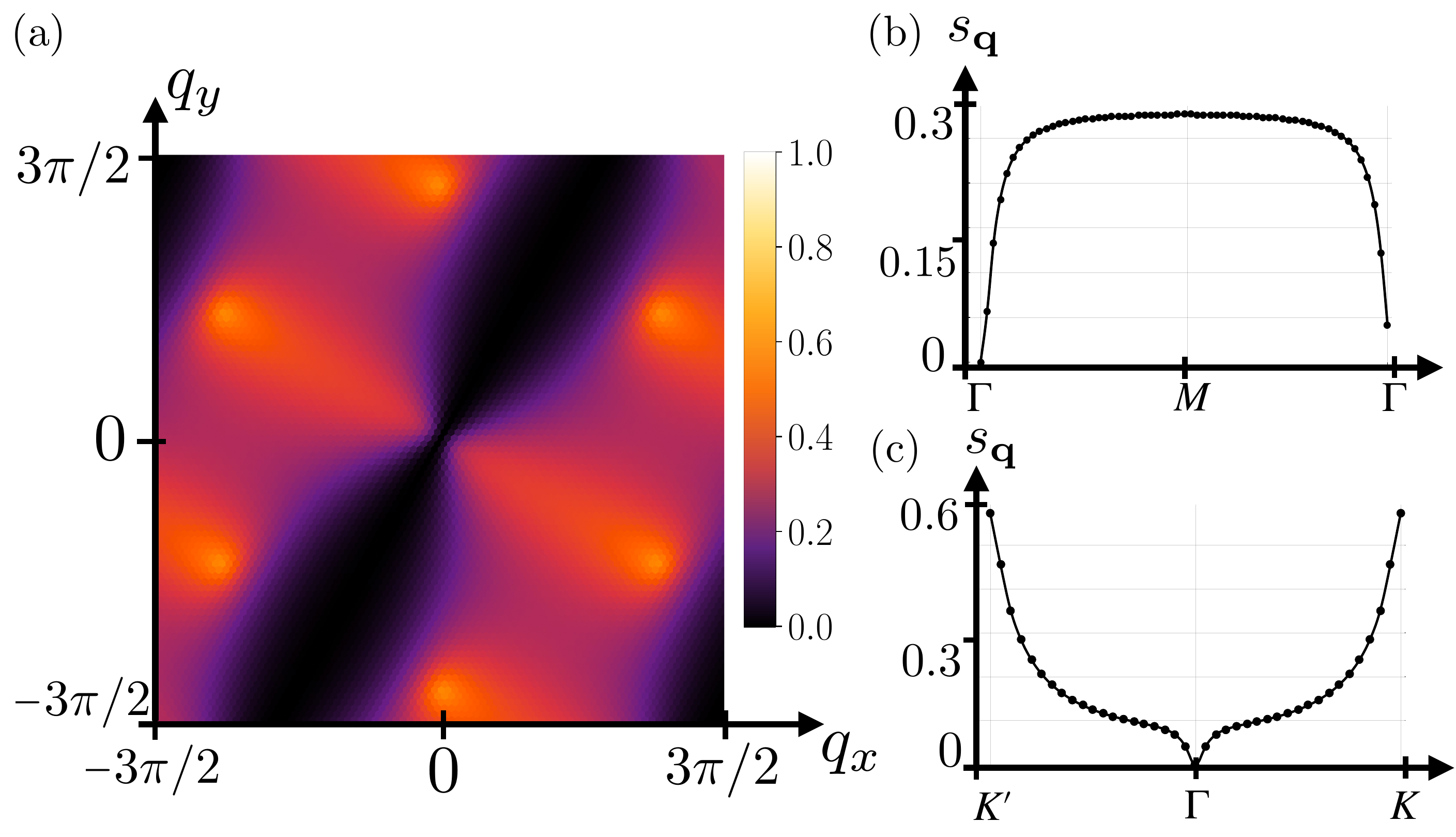}
    \caption{ a)  Connected component of the structure factor for the $\tau=1$ bonds $D_c(\textbf{q})$ for the $z=0.9250$ variational wave function b) Cuts along the original BZ.}
    \label{fig:structfacttau1}
\end{figure}
Performing the corresponding integrals in momentum space for horizontal dimers, we obtain as a main result the plot of Fig.~\ref{fig:DMRG_QDM_dimer}, where we also plot the dimer structure factor from our DMRG wavefunction. We see a remarkably good agreement between the two distinct methods. Indeed, both methods show zero weight for wavevectors propagating parallel to the dimers, meaning $(q_x,0)$. This is consistent with the vanishing found for the oscillator strength for the same momentum line, necessary so that there is no line of gapless excitations. Moreover, there is a region where the structure factor weight is very close to zero (shown in black) between the $K$ and $K'$ points with both methods. The main intensity peaks are at $\Gamma$ and the $K,K'$ points, as suspected since the variational wavefunction has the three-sub-lattice structure; nevertheless, this is not explicitly so with the DMRG, yet we also see a peak at these wavevectors. Otherwise, the intensity increase is mainly seen in the transverse {direction}, meaning for crystal momentum perpendicular to the dimers. The structure of the excitations from the SMA is therefore reminiscent of photons with only transverse propagation. Nevertheless, it is worth noting that this pinch point structure is an emergent feature that appears for a momentum not arbitrarily close to zero but only above the small parameter $1-z$. 

This can be seen more clearly by comparing the connected dimer structure factor of both the supersolid wavefunction and the Rohksar-Kivelson (RK) wavefunction, as shown in Fig.~\ref{fig:structfactdimerRK}. The RK wavefunction shown in panel b) has the pinch point singularity for arbitrarily small momentum, while this singularity is cut off for the supersolid wavefunction at a very small but finite momentum of the order $1-z$. Another notable difference is the reduced peak at $K,K',$ which can be understood since the RK wavefunction sits at the border of two distinct crystalline phases; this implies the presence of gapless {pi0n modes, a term coined in Ref. \cite{moessner_three-dimensional_2003},} indicating the incipient order. For the supersolid variational wavefunction of Eq. \eqref{VA_SS}, the symmetry is explicitly broken, and the logarithmic divergence found for the RK point becomes a full extensive divergence present in the disconnected dimer structure factor. Finally, we note that the same analysis holds for the dimer structure factor along different directions, for example, $\tau=1$ dimers as shown in Fig.~\ref{fig:structfacttau1}. Since the dimer direction is changed, we now have only transverse correlations along the perpendicular direction to the dimers as seen in panel a). Meanwhile, for a momentum path through the $\Gamma-M$ points, we get a finite structure factor going to zero at the $\Gamma$ point and a maximum at $K,K'$ as seen in b),c). This will become the finite energy excitations, and at the $M$ point, a roton-like minimum will emerge as shown in the next section.

\subsubsection{Variational energy bound and effective field theory} \label{eff_fieldtheory}

\begin{figure}[t]
    \centering
    \includegraphics[scale=0.23]{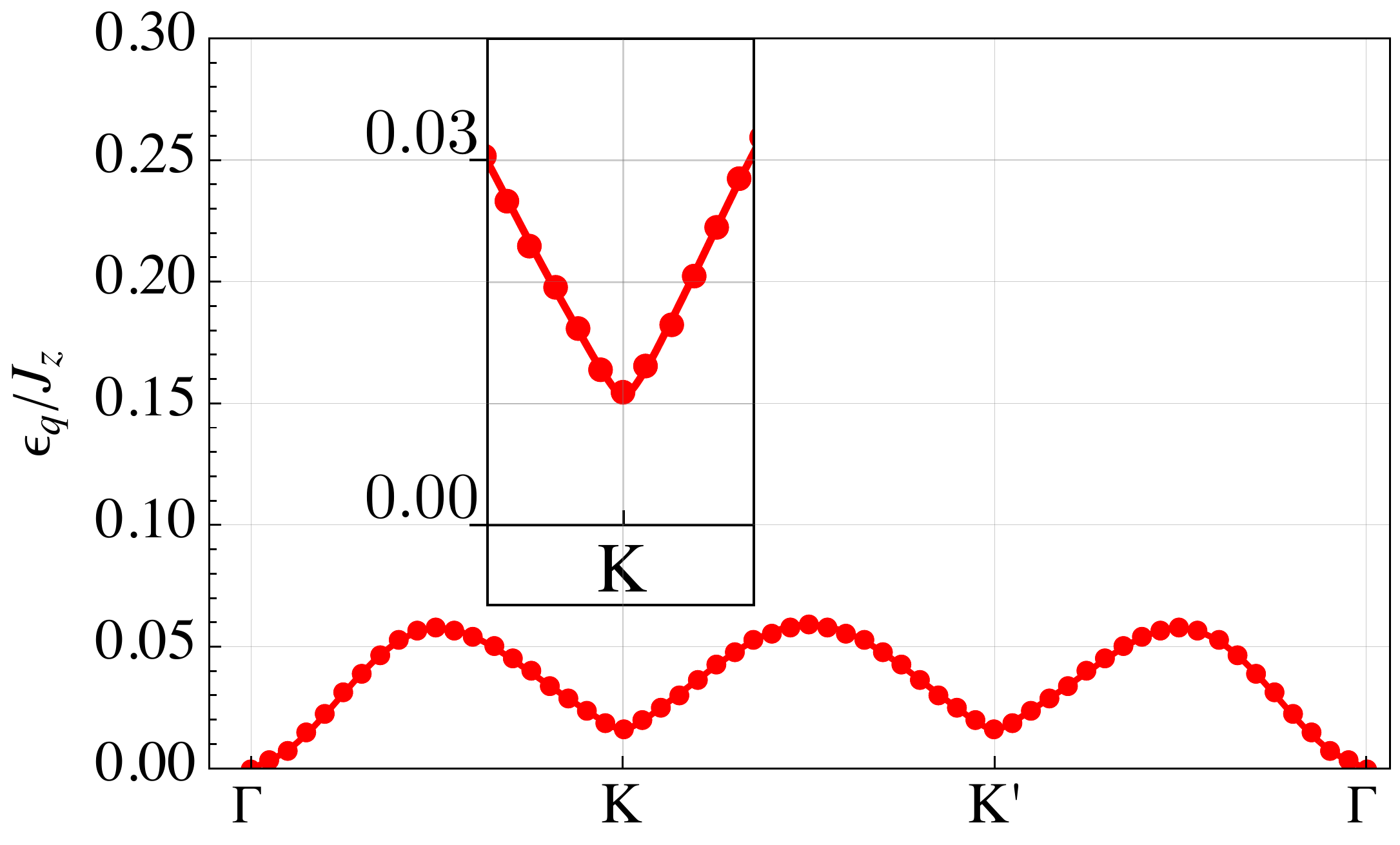}
    \caption{SMA energies for the variational wave function with $z=0.9250$ the momentum cut goes through the $K$ and $K'$ points. The disconnected component of the structure factor gives a divergent contribution to $K,K',\Gamma$ indicated as red lines.}
    \label{fig:SMA}
\end{figure}

The analysis from the previous sections can be combined now to calculate the variational energy along the $\textbf{q}_0=(q_x,0)$ line. As can be seen from Fig.~\ref{fig:DMRG_QDM_dimer} or \ref{fig:structfactdimerRK}, this line has zero structure factor for the horizontal $\tau=3$ dimers. This implies that the wave function $\sigma_{\tau=3}(\textbf{q})\ket{\Psi}$ is not a normalizable state. We can conclude, then, that there are no collective dimer density excitations that propagate parallel to the horizontal dimer direction. In other words, because of the proximity to the RK wavefunction, we have that there are only transverse density fluctuations near $\textbf{q}=0$. Nevertheless, for $\tau=1,2$ we can consider the same line, which will now have a finite structure factor. Because of this, there will be finite energy excitations. Since the oscillator strength has a minimum at the $M$ point, we also have such a feature as shown in Fig.~\ref{fig:DSF_exp} in blue. This minimum corresponds to a finite momentum dimer density excitation which resembles the roton minimum of superfluid $\text{He}^4$. We also present here the result along the $\Gamma-K-K'$ line in Fig.~\ref{fig:SMA}. The vertical red line indicates the jump to the divergent contribution coming from the disconnected part of the dimer structure factor. Such a gap is actually also seen in experiments, as can be noted from the inset of Fig.~\ref{fig:DSF_exp}d.

{For the RK wavefunction with $z=1$ and dimers in the $\tau=1$  direction, $D(q_x,0)$ takes a constant value with a discontinuous behaviour at $\Gamma$ where it goes to another constant. Since the oscillator strength depends quadratically on momentum as shown in Eq. \eqref{osc_stren}, the dispersion of the excitation is of the form $\omega_{RK} = ck^2 $. This is indeed what has been found in the original RK paper \cite{rokhsar_superconductivity_1988} for the square lattice and later generalizations \cite{moessner_three-dimensional_2003}. At the level of an effective action, this corresponds to \cite{moessner_three-dimensional_2003}}

\begin{align}
    \mathcal{S}=\int d^2 x d t\left[\left(\partial_t \mathbf{A}\right)^2-\rho_2(\boldsymbol{\nabla} \times \mathbf{A})^2-\rho_4(\boldsymbol{\nabla} \times \boldsymbol{\nabla} \times \mathbf{A})^2\right] 
\end{align}

{where we defined the emergent electromagnetic potential $\textbf{A}=\varphi\hat{z}$ determining $\textbf{B}=\curl{\textbf{A}}$ and $\varphi$ the height field. The height field is introduced in 2D dimer models as an integer-valued field defined on the dual lattice. The changes of height around any given site are zero, and the height change between plaquettes depends on the presence or absence of a dimer. $\textbf{B}$ is the emergent magnetic field introduced at the beginning of this section, satisfying a Gauss law, characteristic of most classical spin liquids \cite{Castelnovo2012,Knolle2019,Moessner1998a,
Moessner1998b,Isakov2004,Henley2005,Yan2024,fragile_SL,irr_mom_prl}. The RK point has the peculiarity that $\rho_2=0$, resulting in the quadratic dispersion mentioned before.}

In our case, for $z\neq1$ we instead have the connected component vanishing at $\Gamma$ and a crossover between a quadratic dependence and a linear behavior as a function of momentum along the $\textbf{q}_0=(q_x,0)$ line. Indeed, the change in behaviour is linked to the center black region of Fig.~\ref{fig:DMRG_QDM_dimer}, which has a radius proportional to $1-z\approx 0.08$, we know that for $z=1$ there is a pinch point singularity. It is clear then that for a region of momentum that is farther than $1-z$ but close to $\Gamma$ the dispersion can be calculated from the oscillator strength which goes like $(\textbf{k}\times \bm{\tau})^2$ and the structure factor with a linear dependence $\propto \abs{k}$ so that there are only transverse excitations with SMA energy $\omega_{\perp} \leq c \ \abs{\textbf{k}}$. { The effective action related to this transverse mode is  $S_\varphi=\int d^2 x d t \ (\partial_t \varphi)^2+\rho_2(\grad{\varphi})^2-g \cos(2\pi \varphi)+\rho_4 (\laplacian \varphi)^2$, where a pinning field term with coupling $g$ was introduced as it is relevant in the renormalization group (RG) sense for $d=(2+1)$ dimensions. The microscopic origin of the pinning term comes from the integer-valuedness of the height field.}

{For a complete effective field theory, we may also include the three-sublattice ordering, which, following refs. \cite{da_liao_phase_2021,biswas_singular_2018}, is characterized by an order parameter $\psi(\textbf{r},t)=e^{i\theta(\textbf{r},t)}\abs{\psi(\textbf{r},t)}= \sum_i m_i^z e^{i\textbf{K}\cdot\textbf{R}_i}$ where $m_i^z$ is the local magnetization in the unit cell $\textbf{r}$ for the three sublattice sites $\textbf{R}_i,\ i=1,2,3$. Here we have summed over positions for a fixed wavevector $\textbf{K}$ since that is the relevant three-sublattice wavevector where we expect ordering. Moreover, for an ordered phase, there will be a finite magnetization; therefore, fluctuations in the magnitude $\abs{\psi(\textbf{r},t)}$ will be small and can be neglected when compared to fluctuations in the phase, coming from the $\theta(\textbf{r},t)$ field. From a symmetry approach, similar to refs. \cite{da_liao_phase_2021,biswas_singular_2018}, we expect an action of the form $S_\theta=\int d^2 x d t \ g_1 (\partial_t\theta)^2+g_2(\grad{\theta})^2-\lambda \cos(6\theta)$. The first two terms come from space-time fluctuations of $\theta$, while the last term allows for pinning of the phase into six minima, which is the lowest order symmetry allowed coupling between the magnitude and the phase, here $\lambda\propto \abs{\psi}^6$. Additionally, the supersolid variational wave function, just like the RK wave function, is composed of a superposition of all hard-core dimer coverings; thus, it will have a non-zero expectation value of flipping a plaquette. In the language of the hard-core bosons in the original triangular lattice, this means $\ev{b_i}{\Psi}=cte.$ which implies XY ordering of the spins. The goldstone boson $\phi$ associated to the spontaneous $U(1)$ symmetry breaking will contribute to lowest order in the field with a free action, so that the total effective action is now}
\begin{align}
    &S= \int d^2 x d t \left[\ g_1 (\partial_t\theta)^2+g_2(\grad{\theta})-\lambda \cos(6\theta)\right.\notag \\ 
    &+(\partial_t \varphi)^2+\rho_2(\grad{\varphi})^2-g \cos(2\pi \varphi)+\rho_4 (\laplacian \varphi)^2 \notag \\ 
    & \left. +(\partial_t \phi)^2+\rho_0(\grad{\phi})^2+\partial_t \theta \partial_t \varphi+\grad{\theta}\cdot \grad{\varphi}+\dots)\right].
\end{align}
In terms of the RK field theory, this shows a non-zero $\rho_2$. We additionally added some symmetry allowed couplings between the $\theta$ and $\varphi$ modes, although more couplings mixing the fields are in principle possible. The nature of the transitions as a function of magnetic field or further neighbor interactions is the subject of future work.

\section{Conclusions and Outlook} \label{sec:conclusions}

In this work, we have thoroughly analyzed the excitation spectrum of the triangular lattice XXZ model close to the Ising point. Our study provides a comprehensive and unified description of the supersolid phase by combining numerical simulations with multiple complementary analytical techniques. Our matrix product state (DMRG) simulations robustly capture the emergence of a three-sublattice ordered ground state, as evidenced by pronounced peaks in the static spin structure factor, and reveal a rich dynamical spectrum marked by a distinctive roton-like energy minimum at the $M$ point and photon-like transverse excitations near the $\Gamma$ point. These findings are reinforced by complementary analytical treatments: perturbative expansions around the Ising limit expose emergent graphene-like dispersions for hole excitations, while our effective staggered boson hopping model quantitatively reproduces the DMRG-calculated dynamical features. In addition, the Schwinger boson self-consistent mean-field approach and a variational quantum dimer model analyzed via the single-mode approximation offer independent confirmation of the low-energy spectral properties. Consistently, the QDM and DMRG simulations yield nearly identical dimer structure factors, both displaying pronounced transverse photon-like excitations.

The excellent quantitative and qualitative agreement among these varied methodologies not only substantiates the microscopic mechanisms driving the coexistence of crystalline and superfluid orders but also reconciles our theoretical predictions with recent inelastic neutron scattering experiments. By demonstrating that our effective models capture both the subtle interplay of anisotropic interactions and quantum fluctuations, as well as the emergent low-energy excitations inherent to frustrated magnetic systems, our work solidifies the understanding of supersolidity in quantum magnets.

The observed agreement in both the ground state and low-lying excitations may well persist to higher energies. In particular, the QDM model may be suitable for understanding the excitation branch at the Ising energy for zero magnetic field, a subject of future work. Furthermore, another interesting direction to continue is to study the transition out of the supersolid phase into a $U(1)$ symmetric or a translational invariant state. In the continuum, this has been linked to a dual critical vector-tensor gauge theory \cite{pretko-supersolid}, which may be linked to the experimentally observed second-order phase transition at finite temperature and finite magnetic field. Finally, our methods could also be used to investigate the microscopic origin of the re-entrant high-field supersolid phase observed in ref. \cite{chen_phase_2024}.

\begin{acknowledgments}
We thank Sylvain Capponi for his helpful comments. This work was in part supported by the Deutsche Forschungsgemeinschaft under grants SFB 1143 (project-id 247310070) and the cluster of excellence ct.qmat (EXC 2147, project-id 390858490). FP acknowledges support by the Deutsche Forschungsgemeinschaft (DFG, German Research Foundation) under Germany’s Excellence Strategy EXC-2111-390814868, TRR360 (project-id492547816), and the Munich Quantum Valley, which is supported by the Bavarian state government with funds from the Hightech Agenda Bayern Plus. {Moreover, we would like to thank the Referee for the considerable time and effort invested in providing such a detailed and thoughtful report.}
\end{acknowledgments}

\appendix

\section{Details on the DMRG time evolution and DMRG}\label{appendix:DMRG}

\begin{figure}[b]
    \centering
    \includegraphics[scale=0.17]{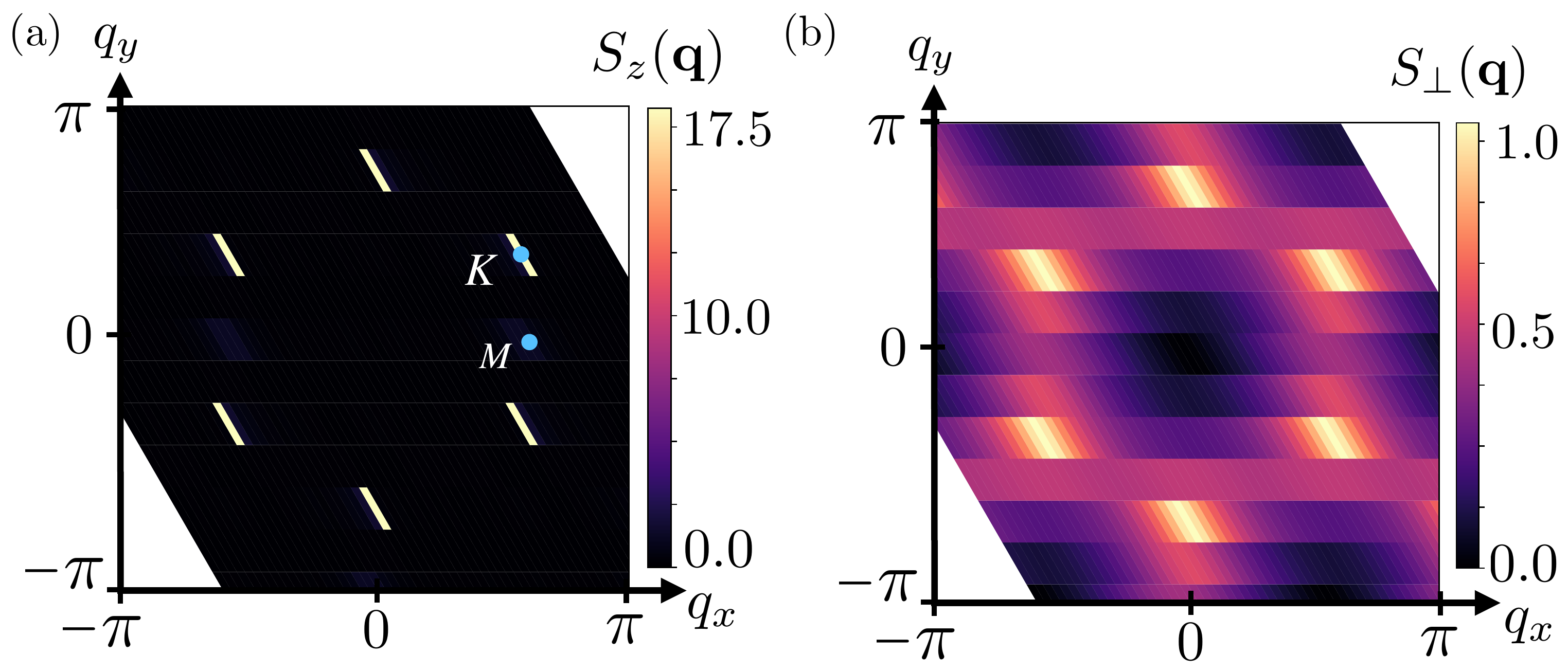}
    \caption{ Static structure factor from the matrix product state wavefunction of $N=36\times 6$ with $\chi=800$ calculated with DMRG for $J_z=3,J_\perp=0.2$. a) $z$ component of the structure factor defined as the equal time Fourier transform of the correlation function $\expval{S^z_iS^z_j}$ b) $\perp$ component of the structure factor defined as the equal time Fourier transform of the correlation function $\expval{S^x_iS^x_j+S^y_iS^y_j}$.  }
\label{fig:struct_decomp}
\end{figure}

\begin{figure}[t]
    \centering
    \includegraphics[scale=0.17]{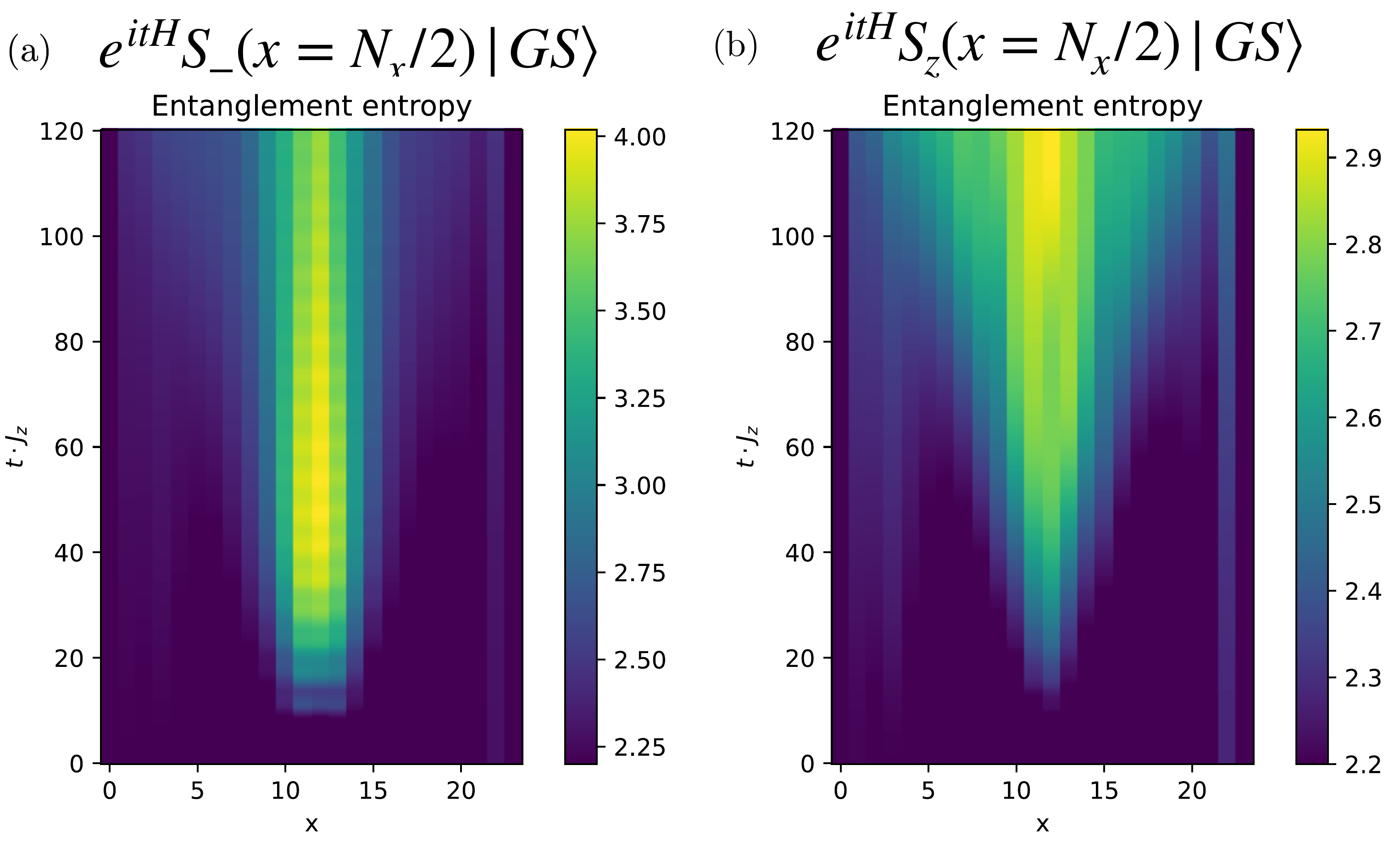}
    \caption{Entanglement entropy production as a function of time for the time-evolved ground state after a local spin operator is applied for a $24\times 6$ cylinder geometry with circumference $L = 6$ and bond dimensions up to $\chi = 700$. For our TDVP simulations, we used a maximum time $T_{max}=400$ and a time interval of $\Delta t=0.1$. }
    \label{fig:ent}
\end{figure}

The ground state wavefunction was found using both iDMRG and finite DMRG, for cylinders with periodic $y$ direction. {  We observed a Bragg peak of the $z$ component of the structure factor at the $K,K'$ vectors as shown in Fig.~\ref{fig:struct_decomp}a) confirming the three-sublattice order of the ground state. Furthermore, we also observed a peak for the perpendicular component of the structure factor, as shown in Fig.~\ref{fig:struct_decomp}b). Our values of the superfluid and three-sublattice order parameters, $\expval{m_\perp}^2 = S_\perp(K)/N \approx 0.0044$ and $\expval{m_z}^2 = S_z(K)/N \approx 0.08$, are within 10\% and 3\% of the values $\expval{m_\perp}^2 = 0.002$ and $\expval{m_z}^2 = 0.06$, found  in ref. \cite{paramekanti_2005} in the large $J_z$ limit. Together, these signatures indicate a supersolid ground state with a spontaneously broken $U(1)$ and translational symmetry. This result is consistent with the extensive numerical studies indicating supersolidity of references \cite{sellmann_phase_2015,HeidarianParamekanti2010,wang_extended_2009,paramekanti_2005}.} The time evolution was done using the TDVP algorithm, we use a Gaussian kernel for the final Fourier transform as in \cite{Sherman_2023} with the form $e^{-\eta^2 t^2}$ and we use $\eta=T_{\text{max}}^{-1}$. A check was done to ensure the entropy built from perturbing on the ground state and time evolving does not hit the boundary of the system, as shown in Fig.~\ref{fig:ent}. For the main plots of Fig.~\ref{fig:DSF_exp} we further process the DMRG time-evolved data by first doing a linear prediction on the time series for each momentum point, then interpolating the momentum with a cubic spline, the time data is then multiplied by the window function and Fourier transformed in time. Finally, a cubic spline interpolation is performed on the frequency space, and for the final plots, the zero frequency component is subtracted for better comparison with the experiment.

\begin{figure}[ht]
    \centering
    \includegraphics[scale=0.35]{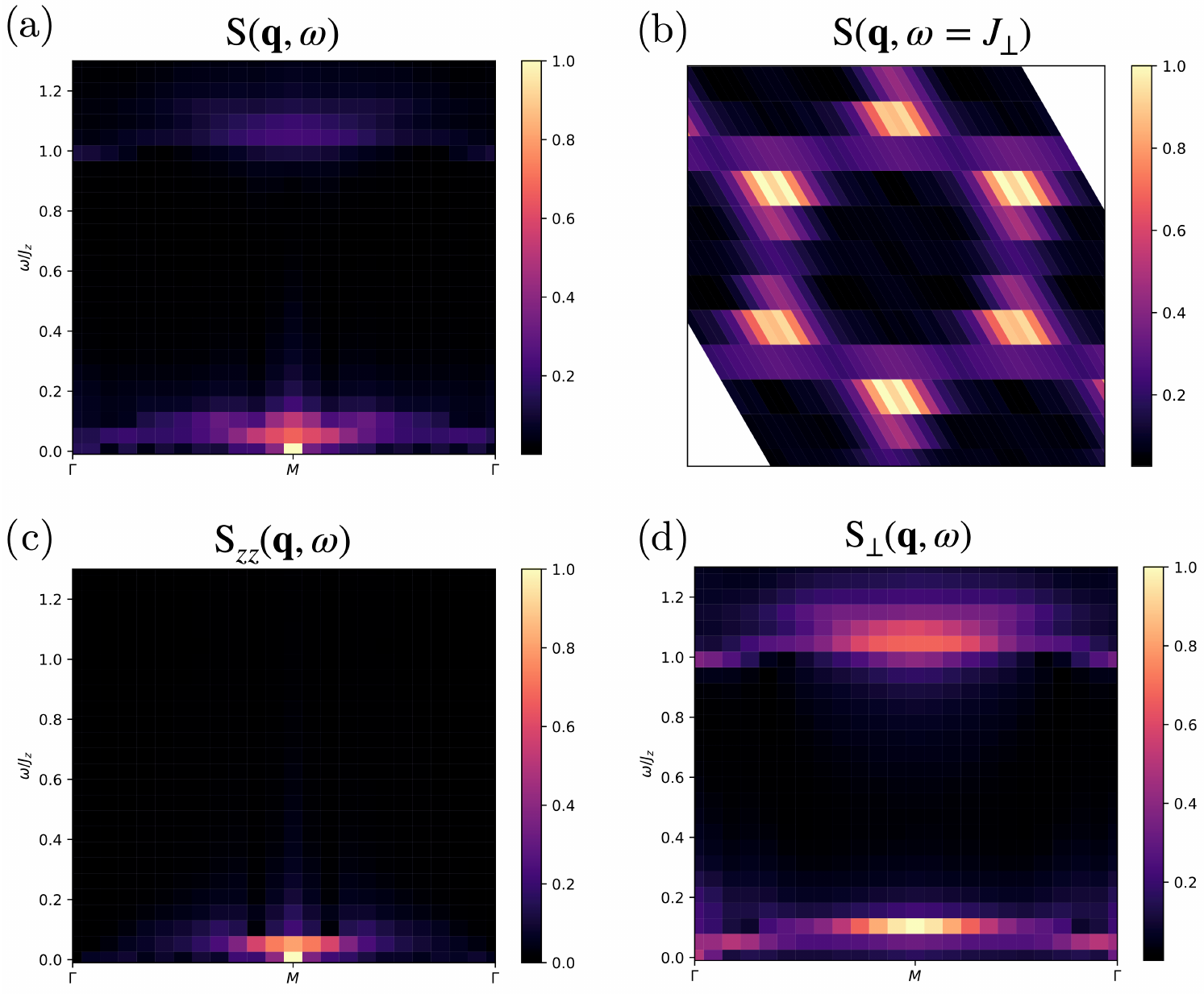}
    \caption{ Dynamical spin structure factor obtained from our DMRG simulations with no further data processing. We simulate a $24\times 6$ cylinder geometry with circumference $L = 6$ and bond dimensions up to $\chi = 800$ for $J_\perp=0.0666J_z$. a) Shows the full DSF, b) Shows a frequency cut at the $XY$ exchange. Bottom panels show the DSF for the c) $z$ and d)$xy$ components of spin.}
    \label{fig:DSFpanels}
\end{figure}

We also present in Fig.~\ref{fig:DSFpanels} the dynamical spin structure factor, a),b), separated into the c) $S^zS^z$ and d) the perpendicular parts, the first panel shows the raw uninterpolated data of Fig.~\ref{fig:DSF_exp}, while b) is evaluated at the XX exchange value. We can see from the separated DSF that the Ising fluctuations happen very close to the $M$ point, as expected from our quantum dimer results and the Schwinger boson theory (the weight at $\Gamma$ decreasing is also predicted by this approach). Intuitively, the $S_\perp$ excitations correspond to an $S^+S^-$ term that flips spins and generates a dispersion for the original Ising excitations near $J_z$. Meanwhile, the $S_{z}S_z$ term has a soft branch near the $M$ point, we see a minimum at this momentum point.  The perpendicular component instead has a bigger intensity near the $\Gamma$ point, as expected for a superfluid order. 


\section{Linear Spin Wave theory (LSWT)}
In this section, we review the known linear spin wave theory results (LSWT), which, as we will see, indicate that the 120-degree order of the Heisenberg point is unstable as soon as anisotropy is added. We then focus on LSWT starting from the classical XXZ ground states and conclude that the simple picture fails to capture the $M$ point excitation minimum shown in our DMRG results and experiments. This indicates that we must approach the problem from a strongly correlated perspective.
\subsection{ Instability of the 120 degree order} \label{App:120instable}
For the simplest theory, we note that from previous studies, the Heisenberg point seems to have 120-degree order. Let us assume the order persists up to the Ising point and investigate if it is stable. We expand around the classical state. We label the spin at the sublattice $\lambda=1,2,3$ at position $i$ as $\bm{S}_{\lambda}^i$, each sublattice has the same ordering vector. We set the $\lambda=1$ ordering {vector} in the $y$ direction of a new coordinate system, and the other directions are separated by 120 degrees. We want the $S_z$ direction of our system to point in the ordering directions. We do this by rotating the reference frame of each sublattice with the transformation
\begin{align}
    \begin{pmatrix}
        S^\lambda_x\\S^\lambda_y\\S^\lambda_z
    \end{pmatrix}\rightarrow
    \begin{pmatrix}
        0 && -\sin\theta_\lambda && \cos\theta_\lambda\\
        0 && \cos\theta_\lambda && \sin\theta_\lambda\\
        1 && 0 && 0
    \end{pmatrix}
     \begin{pmatrix}
        S^\lambda_x\\S^\lambda_y\\S^\lambda_z
    \end{pmatrix}
\end{align}
 Where we have the angles $\theta_1=\pi/2,\theta_2=\pi/2-2\pi/3,\theta_3=\pi/2+2\pi/3$, note $\cos(\theta_\lambda-\theta_{\lambda'})=-1/2$. 
 We will use the lowest-order Holstein-Primakoff expansion to convert to hard-core bosons by
\begin{align}
    &S_{\lambda,z}^{\textbf{R}}=S-a^{\dagger}_{\textbf{R}\lambda}a_{\textbf{R}\lambda},   \ S_{\lambda,+}^{\textbf{R}}=a_{\textbf{R}\lambda}(2S)^{1/2}, \\ & \qquad \qquad S_{\lambda,-}^{\textbf{R}}=a_{\textbf{R}\lambda}^\dagger (2S)^{1/2}
\end{align}
Doing now the lowest expansion in $S$ corresponding to LSWT has no sub-lattice dependent factor, and so one can regroup the terms into the original unit cell to give:
\begin{align}
    &H=H_0+H_2, \qquad H_0=-\dfrac{3}{2}JS^2 N\\
    &H_2 = \dfrac{S}{2} \sum_{\textbf{r}}\sum_{\bm{\delta}} \left\{\left(J_z-\dfrac{J_\perp}{2}\right)\left[a^\dagger_{\textbf{r}} a_{\textbf{r}+\bm{\delta} }+ a^\dagger_{\textbf{r}+\bm{\delta} }a_{\textbf{r}}\right]\right. \notag\\
      &\hspace{1.5cm} -\left(J_z+\dfrac{J_\perp}{2}\right)\left[a_{\textbf{r}} a_{\textbf{r}+\bm{\delta} }+ a^\dagger_{\textbf{r}+\bm{\delta} }a^\dagger_{\textbf{r}}\right]\\
    & \hspace{1.5cm} \left. +J_\perp\left[a^\dagger_{\textbf{r}} a_{\textbf{r} }
    + a^\dagger_{\textbf{r}+\bm{\delta} }a_{\textbf{r}+\bm{\delta} }\right] \right\}\notag
\end{align}
We can diagonalize the Hamiltonian by going to momentum space with a Bogoliubov transformation to obtain:
\begin{align}
& H= \sum_\textbf{k} \varepsilon_{\textbf{k}} \left(b_{\textbf{k}}^\dagger b_{\textbf{k}}+\dfrac{1}{2}\right)\\
& \varepsilon_{\textbf{k}}=3 J_\perp S \sqrt{\left(1-\gamma_{\mathbf{k}}\right)\left(1+2 \alpha \gamma_{\mathbf{k}}\right)} , \quad \gamma_{\textbf{k}} = \dfrac{1}{6} \sum_{\bm{\delta}}e^{i\textbf{k}\cdot \bm{\delta}}
\end{align}
This Hamiltonian actually has regions of imaginary energy, unstable solutions for $J_z>J_\perp>0,\alpha=J_z/J_\perp$, indicating that 120-degree order is not the classical ground state. 

\begin{figure}[t]
    \centering
    \includegraphics[scale=0.3]{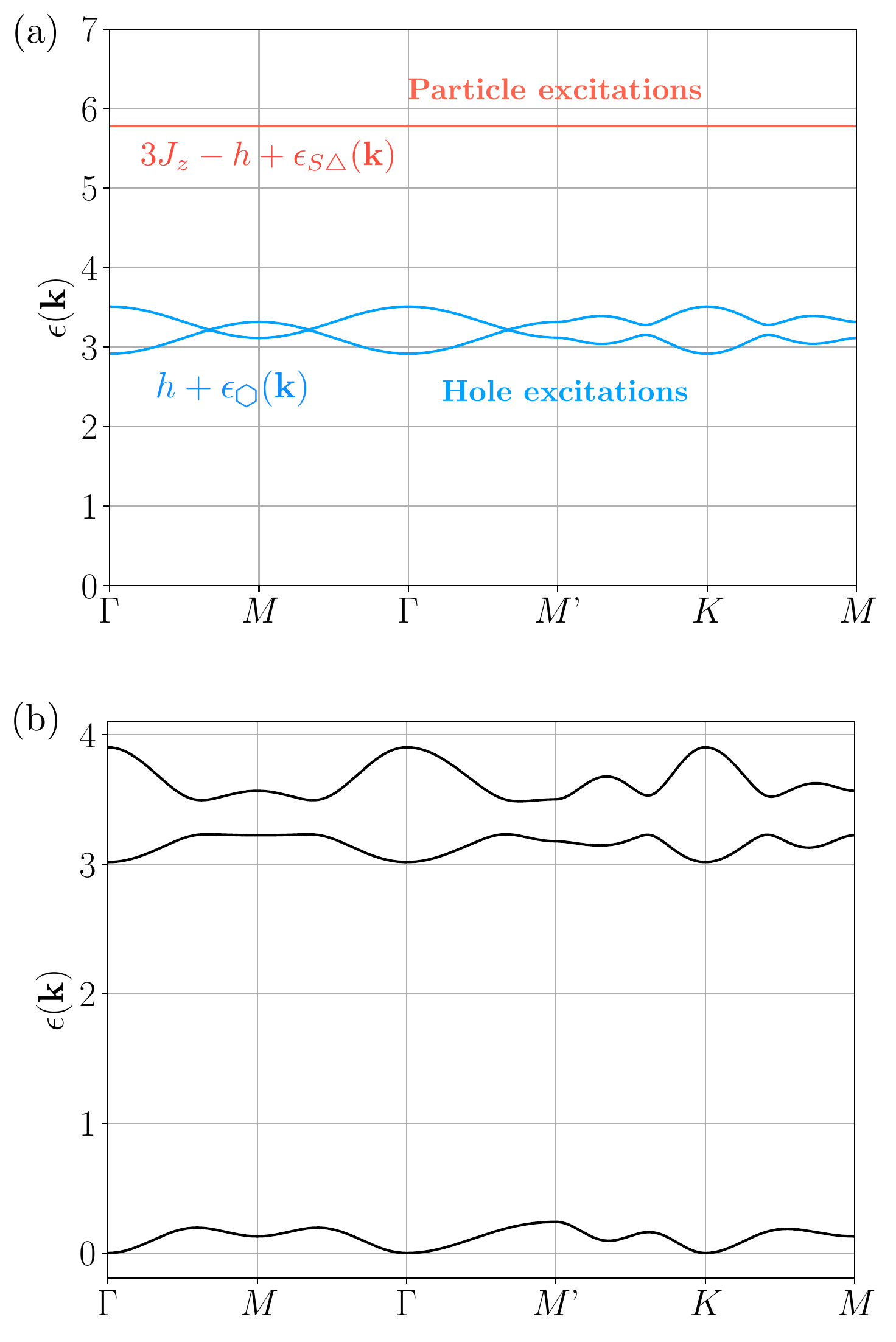}
    \caption{a) Excitation spectrum for the particle and hole branches of the UUD state within the $1/3$ magnetization plateau (here $h=3.22$). Hole excitations move within an emergent honeycomb lattice with the graphene dispersion $\epsilon_{\varhexagon}(\textbf{k})$ and hopping $t=-J_\perp/2$, while Particle excitations are higher in energy for $h<3J_z$ and are supressed by hopping $t=J_\perp^2/J_z$ in the super triangular lattice with energy $\epsilon_{S\triangle}(\textbf{k})$. b) Effective staggered boson hopping model spectrum with a three sublattice structure at zero magnetic field, $J_z=3,J_\perp=0.2$ and two chemical potentials $\mu_1= 3J_z+h-6\expval{n_B},\mu_2=3J_z+h-3\expval{n_B}+3\expval{n_A}$, values are chosen as in Fig.~\ref{fig:DSF_exp}, $\expval{n_A}=0.525,\expval{n_B}=0.83 $.}
    \label{fig:SWUUD}
\end{figure}

\subsection{LSWT about the classical XXZ ground states }\label{app:LSWTXXZ}
Instead of a 120-degree order, we can take $\theta_\lambda$ to be arbitrary and vary as a function of $J_\perp/J_z$. We choose the angle that gives the classical ground state energy. The resulting spin-wave spectrum was studied in Ref. \cite{kleine_spin-wave_1992}, although not plotted explicitly, we present the result in Fig.~\ref{fig:DSF_exp} for convenience. The model gets simplified if one goes close to the Ising limit. An effective Hamiltonian in terms of a magnetic field term $V=-h\sum_i S_i^z$ is then obtained in the honeycomb lattice (site $b,c$), with one spin in the original triangular lattice frozen, one gets:
\begin{align}
    h_{\mathrm{eff}}=&  \sum_{b, c}\left[S_b^z S_c^z+\frac{J_\perp}{J_z}\left(S_b^x S_c^x+S_b^y S_c^y\right)\right] \\
    &+3 S\left(1-\frac{h}{6}\right)\left(\sum_b S_b^z+\sum_c S_c^z\right)
\end{align}

The spin wave spectrum for zero field is plotted in Fig.~\ref{fig:DSF_exp}. Analytically, one has the two branches:
\begin{align}
    \omega_1^2(\mathbf{k})&=\frac{J_\perp^2}{4J_z^2}\left(9-f^* f\right)+\dfrac{J_\perp h}{J_z} \left(\frac{1}{4}+\frac{|f|}{12}\right),\\  \omega_2^2(\mathbf{k})&=\frac{J_\perp^2}{4J_z^2}\left(9-f^* f\right)+ \dfrac{J_\perp h}{J_z}\left(-\frac{1}{4}+\frac{|f|}{12}\right),
\end{align}
with the function $f(\mathbf{k})=e^{i \mathbf{k} \boldsymbol{\delta}_1}+e^{i \mathbf{k} \delta_2}+e^{i \mathbf{k} \boldsymbol{\delta}_3}$ for the $\delta_i$ neirest neighbours in the triangular lattice. Near the Ising limit one obtains a doubly degenerate LSW branch that has a maximum at the $M$ point, not a minimum.

\section{Effective staggered boson hopping model}\label{app:hard-core}

Here, we present a mean-field analysis assuming three-sublattice ordering and expanding the interaction term in the following way

\begin{align}
   & H_{V}=V \sum_{\langle i j\rangle}  n_i n_j -\sum_i (3V+h)n_i\\
    &\approx V \sum_{\langle i j\rangle}  \expval{n_i}n_j+\expval{n_j}n_i-\expval{n_i}\expval{n_j}-\sum_i (3V+h)n_i\\
    &\approx V \sum_{\textbf{T}}  \expval{n_{\textbf{T}A}}(n_{\textbf{T}-\textbf{c}_1B}+n_{\textbf{T}-\textbf{c}_1C}+n_{\textbf{T}B}\\
    &+n_{\textbf{T}C}+n_{\textbf{T}-\textbf{c}_2C}+n_{\textbf{T}-\textbf{c}_1+\textbf{c}_2B})\\
    &+n_{\textbf{T}A}(\expval{n_{\textbf{T}-\textbf{c}_1B}}+\expval{n_{\textbf{T}-\textbf{c}_1C}}+\expval{n_{\textbf{T}B}}\\
    &+\expval{n_{\textbf{T}C}}+\expval{n_{\textbf{T}-\textbf{c}_2C}}+\expval{n_{\textbf{T}-\textbf{c}_1+\textbf{c}_2B}})\\
    &+\expval{n_{\textbf{T}B}}n_{\textbf{T}C}+\expval{n_{\textbf{T}C}}n_{\textbf{T}B}+\expval{n_{\textbf{T}B}}n_{\textbf{T}-\textbf{c}_2 C}\\
    &+\expval{n_{\textbf{T}-\textbf{c}_2 C}}n_{\textbf{T} B}+\expval{n_{\textbf{T}B}}n_{\textbf{T}-\textbf{c}_2+\textbf{c}_1C}+\expval{n_{\textbf{T}-\textbf{c}_2+\textbf{c}_1C}}n_{\textbf{T}B}\\
    &-\sum_{\textbf{T}\lambda} (3V+h)n_{\textbf{T}\lambda}
\end{align}
where we distinguished between the sites connected from the occupation at a site $A$ to $B$ and $C$ to the ones between $B$ and $C$ alone. Because of the three sublattice translation symmetry we assume $\expval{n_{\textbf{T}B}}=\expval{n_{\textbf{T}C}}=\tilde{\mu}_2$ for all $\textbf{T}$, and $\expval{n_{\textbf{T}A}}=\tilde{\mu}_1$

\begin{align}
   & H_{V}\approx V \sum_{\textbf{T}}  3 \tilde{\mu}_1 n_{\textbf{T}B}+3 \tilde{\mu}_1 n_{\textbf{T}C}+6 \tilde{\mu}_2 n_{\textbf{T}A}+3 \tilde{\mu}_2 n_{\textbf{T}C}\notag\\
    &+3 \tilde{\mu}_2 n_{\textbf{T}B}-\sum_{\textbf{T}\lambda} (3V+h)n_{\textbf{T}\lambda}\\
    &=  V \sum_{\textbf{T}}  3 (\tilde{\mu}_1+\tilde{\mu}_2) (n_{\textbf{T}B}+ n_{\textbf{T}C})+6 \tilde{\mu}_2 n_{\textbf{T}A}\\
    &-\sum_{\textbf{T}\lambda} (3V+h)n_{\textbf{T}\lambda}\\
    &= \sum_{\textbf{T}}   (3V\tilde{\mu}_1+3V\tilde{\mu}_2-3V-h) (n_{\textbf{T}B}+ n_{\textbf{T}C})\\
    &+(6 V \tilde{\mu}_2-3V-h) n_{\textbf{T}A}\\
    &=\sum_{\textbf{T}} -\mu_2 (n_{\textbf{T}B}+ n_{\textbf{T}C})-\mu_1 n_{\textbf{T}A}
\end{align}
As described in the main text, we tune the occupations to reach a good agreement with the dynamical structure factor obtained from the DMRG simulations. The resulting plot and comparison with the UUD spectrum is shown in Fig.~\ref{fig:SWUUD}.

\begin{figure}[t]
    \includegraphics[scale=0.2]{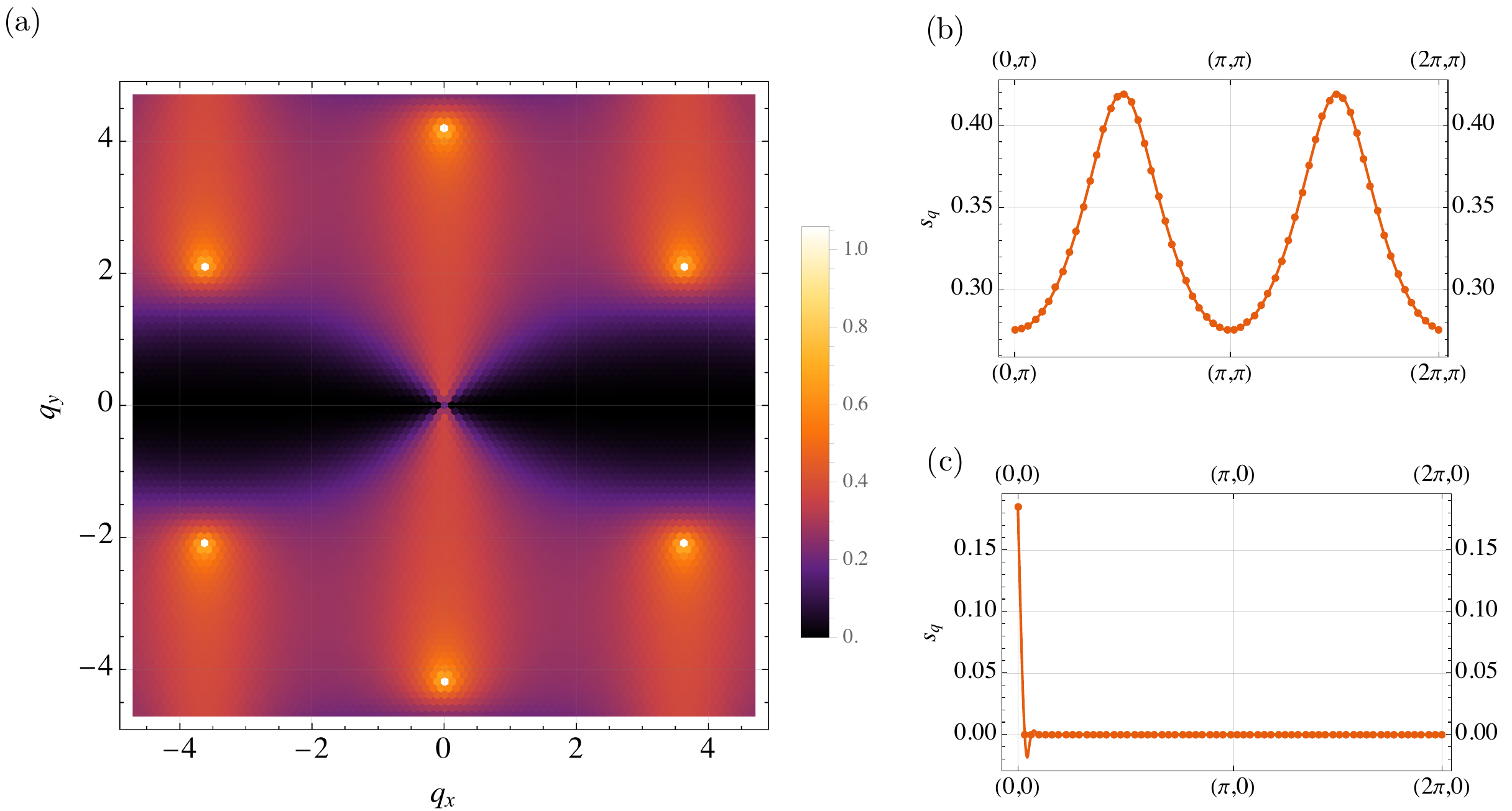}
    \caption{Connected component of the structure factor for the $\tau=3$   bonds a) $s_c(\textbf{q})$ and b) $\textbf{q}=q_x\textbf{b}_1+\pi \textbf{b}_2$ c) $\textbf{q}=q_x\textbf{b}_1$ }
    \label{fig:structfactHoney}
\end{figure}
\begin{figure}[t]
    \includegraphics[scale=0.3]{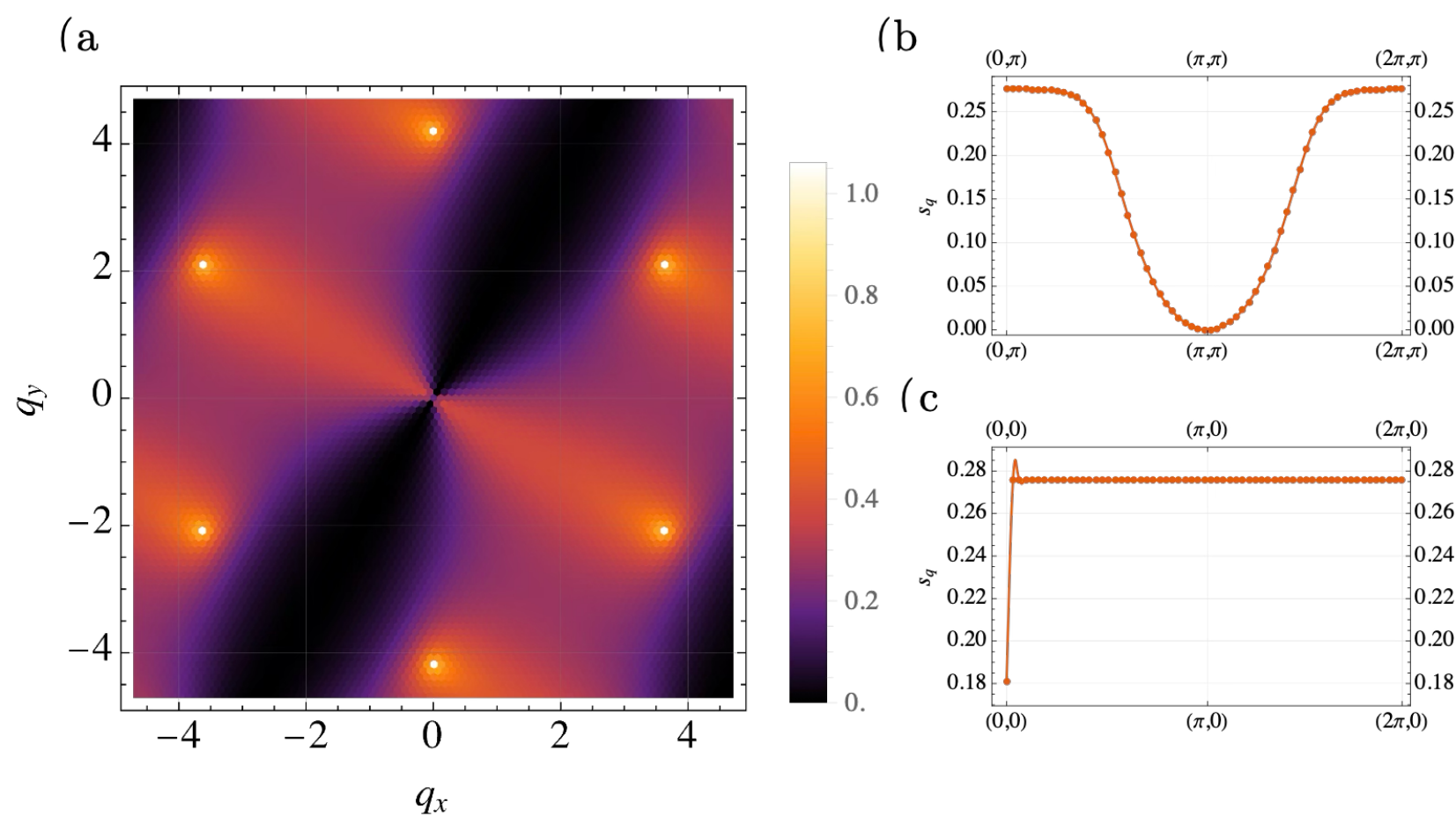}
    \caption{Connected component of the structure factor for the $\tau=1$   bonds a) $s_c(\textbf{q})$ and b) $\textbf{q}=q_x\textbf{b}_1+\pi \textbf{b}_2$ c) $\textbf{q}=q_x\textbf{b}_1$ }
    \label{fig:structfactHoneytau1}
\end{figure}
\section{Details on the Schwinger boson mean field theory}\label{app:schwinger}

After the mean field decoupling and rewriting in momentum space, we obtain a Hamiltonian Eq. \eqref{ABD_Ham} with terms of the form $\phi^\dagger \phi^\dagger$ so that a Bogoliubov transformation is necessary. The final expression in terms of the Bogolons $\alpha_{\textbf{k}s}$ leads to the Mean field Hamiltonian
\begin{align}
    H_{MF}= \sum_{\textbf{k}} \omega_{\textbf{k},\uparrow} \alpha_{\textbf{k},\uparrow}^\dagger \alpha_{\textbf{k},\uparrow}+\omega_{-\textbf{k},\downarrow} \alpha_{-\textbf{k},\downarrow}^\dagger \alpha_{-\textbf{k},\downarrow}+E_{MF}
\end{align}
where we defined the quasiparticle energies as
\begin{align}
&\omega_{\mathbf{k} \uparrow}=\omega_{-\mathbf{k} \downarrow}=\omega_k=\sqrt{\left[\Gamma_{\mathbf{k}}^{\mathrm{B}}+\lambda\right]^2-\left[\Gamma_{\mathbf{k}}^{\mathrm{AD}}\right]^2} \\
&\Gamma_{\mathbf{k}}^{\mathrm{B}}=\frac{1}{2}\left(1+\frac{J_z}{J_\perp}\right) \gamma_{\mathbf{k}}^B \\
&\Gamma_{\mathbf{k}}^{\mathrm{AD}}=\frac{1}{2}\left(1+\frac{J_z}{J_\perp}\right) \gamma_{\mathbf{k}}^A-\left(1-\frac{J_z}{J_\perp}\right) \gamma_{\mathbf{k}}^D
\end{align}
With the following functions
\begin{align}
\gamma_{\mathbf{k}}^A=\sum_{\delta>0} &J_\perp A_\delta \sin (\mathbf{k} \cdot \boldsymbol{\delta}), \quad   \gamma_{\mathbf{k}}^B=\sum_{\delta>0} J_\perp B_\delta \cos (\mathbf{k} \cdot \boldsymbol{\delta}),\\
& \gamma_{\mathbf{k}}^D=\sum_{\delta>0} J_\perp D_\delta \cos (\mathbf{k} \cdot \boldsymbol{\delta}).
\end{align}
The self-consistent mean field equations become
\begin{align}
    \begin{gathered}
S+\frac{1}{2}=\frac{1}{2 N} \sum_{\mathbf{k}} \frac{\Gamma_{\mathbf{k}}^{\mathrm{BC}}+\lambda}{\omega_{\mathbf{k}}}, \\
A_\delta=\frac{1}{2 N} \sum_{\mathbf{k}} \frac{\Gamma_{\mathbf{k}}^{\mathrm{AD}}}{\omega_{\mathbf{k}}} \sin (\mathbf{k} \cdot \boldsymbol{\delta}), \\
B_\delta=\frac{1}{2 N} \sum_{\mathbf{k}} \frac{\Gamma_{\mathbf{k}}^{\mathrm{BC}}+\lambda}{\omega_{\mathbf{k}}} \cos (\mathbf{k} \cdot \boldsymbol{\delta}), \\
C_\delta=\frac{1}{2 N} \sum_{\mathbf{k}} \frac{\Gamma_{\mathbf{k}}^{\mathrm{BC}}+\lambda}{\omega_{\mathbf{k}}} \sin (\mathbf{k} \cdot \boldsymbol{\delta}), \\
D_\delta=\frac{1}{2 N} \sum_{\mathbf{k}} \frac{\Gamma_{\mathbf{k}}^{\mathrm{AD}}}{\omega_{\mathbf{k}}} \cos (\mathbf{k} \cdot \boldsymbol{\delta}).
\end{gathered}
\end{align}

The dispersion can be calculated and has a minimum at the $M$ point. Moreover, the $S_z$ contribution to the dynamical spin structure factor can also be calculated as \cite{ghioldi_magnons_2015}:
\begin{align}
S^{z z}(\mathbf{k}, \omega)= & \frac{1}{4 N} \sum_{\mathbf{q}}w(\textbf{q},\textbf{p})^2\delta\left(\omega-\left(\omega_{\mathbf{q} \uparrow}+\omega_{\mathbf{k}-\mathbf{q} \downarrow}\right)\right).
\end{align}
The weights $u_{\mathbf{q}},v_{\mathbf{q}}$ come from the Bogoliubov transformation as explained in the main text.
\section{RK dimer structure factor on the honeycomb lattice} \label{app:honey}
Here benchmark the dimer-dimer correlation functions of the RK wave function on the Honeycomb lattice, which for the Honeycomb has been calculated in real space along a certain direction. Here we generalize and calculate the full momentum space correlation function. We begin as in the main text and focus on the Honeycomb: 
\begin{align}
    s(\mathbf{q})&=\left\langle \Psi\left|\tilde{\sigma}_{\hat{\tau}}^x(-\mathbf{q}) \tilde{\sigma}_{\hat{\tau}}^x(\mathbf{q})\right| \Psi\right\rangle\\ 
    &= \dfrac{1}{N}\sum_{\textbf{R}_1,\textbf{R}_2} \left\langle \Psi\left|\sigma_{\hat{\tau}}^x(\textbf{R}_1) \sigma_{\hat{\tau}}^x(\textbf{R}_2)\right| \Psi\right\rangle e^{i\textbf{q}\cdot \left(\textbf{R}_2-\textbf{R}_1\right)}\notag
\end{align}
 where $\textbf{R}$ lives in the original triangular lattice $\textbf{R}=l \textbf{a}_1+k\textbf{a}_2$,  we denote $\tau=1,2,3$ the three distinct bonds on the hexagonal lattice and $\tilde{\sigma}_{\hat{\tau}}^x(\textbf{R})$ is $1$ for a dimer at $\textbf{R}$ with link $\tau$ or it has the value $-1$ when no dimer is present. Let us write the dimer density $n_{\hat{\tau}}(\textbf{R})$ so that it has values $1$ or $0$, we then have the relation $\tilde{\sigma}_{\hat{\tau}}^x(\textbf{R})=2n_{\hat{\tau}}(\textbf{R})-1$. We continue to rewrite
 \begin{align}
     &\left\langle \Psi\left|\tilde{\sigma}_{\hat{\tau}}^x(\textbf{R}_1) \tilde{\sigma}_{\hat{\tau}}^x(\textbf{R}_2)\right| \Psi\right\rangle
\\
&=\left\langle \Psi\left| (2n_{\hat{\tau}}(\textbf{R}_1)-1)(2n_{\hat{\tau}}(\textbf{R}_2)-1)\right| \Psi\right\rangle\\
 &=
4\left\langle \Psi\left| n_{\hat{\tau}}(\textbf{R}_1)n_{\hat{\tau}}(\textbf{R}_2)\right| \Psi\right\rangle-2 \left\langle \Psi\left| n_{\hat{\tau}}(\textbf{R}_1)\right| \Psi\right\rangle\\
& \qquad -2 \left\langle \Psi\left| n_{\hat{\tau}}(\textbf{R}_2)\right| \Psi\right\rangle+1 \notag
 \end{align}
The correspondence for correlation functions says that we should replace $n_{ab}\rightarrow K_{ab}\psi_a \psi_b$. To write the positions of the fermions we rename $\textbf{r}_1=\textbf{R}_1+\bm{\tau}_A$ since then we can identify $\tau=1,2$ with the bonds of the square lattice $\bm{\delta}_{1}=-x,\bm{\delta}_{2}=y$ pointing away from the site. 
A very subtle issue arises because the fermionic mapping fails to give the correct result, since for the dimers we have 
\begin{align}
    \expval{n^2_{\tau}(\textbf{R})}=   \expval{n_{\tau}(\textbf{R})}=\dfrac{1}{3}
\end{align}
While for the fermions:
\begin{align}
    &\expval{n^2_{\tau}(\textbf{R})}_F= K_{\textbf{r},\textbf{r}_1+\bm{\tau}}K_{\textbf{r},\textbf{r}+\bm{\tau}}\expval{\psi_{\textbf{r}}\psi_{\textbf{r}+\bm{\tau}}\psi_{\textbf{r}}\psi_{\textbf{r}+\bm{\tau}}}\\
    &=\expval{\psi_{\textbf{r}}\psi_{\textbf{r}+\bm{\tau}}}\expval{\psi_{\textbf{r}}\psi_{\textbf{r}+\bm{\tau}}}-\expval{\psi_{\textbf{r}}\psi_{\textbf{r}}}\expval{\psi_{\textbf{r}+\bm{\tau}}\psi_{\textbf{r}+\bm{\tau}}}\\
    &+\expval{\psi_{\textbf{r}}\psi_{\textbf{r}+\bm{\tau}}}\expval{\psi_{\textbf{r}+\bm{\tau}}\psi_{\textbf{r}}}\\
    & =\expval{\psi_{\textbf{r}}\psi_{\textbf{r}+\bm{\tau}}}\expval{\psi_{\textbf{r}}\psi_{\textbf{r}+\bm{\tau}}}\\
    &-0 \expval{\psi_{\textbf{r}+\bm{\tau}}\psi_{\textbf{r}+\bm{\tau}}}+\expval{\psi_{\textbf{r}}\psi_{\textbf{r}+\bm{\tau}}}(-\expval{\psi_{\textbf{r}}\psi_{\textbf{r}+\bm{\tau}}})=0
\end{align}
Therefore, we need to sum this term independently to calculate the true $s_c(\textbf{q})$.
We can write the correlation function as
 \begin{align}
     &\left\langle \Psi\left|\tilde{\sigma}_{\hat{\tau}}^x(\textbf{R}_1) \tilde{\sigma}_{\hat{\tau}}^x(\textbf{R}_2)\right| \Psi\right\rangle\\
     &=
     4\expval{K_{\textbf{r}_1,\textbf{r}_1+\bm{\tau}} \psi_{\textbf{r}_1} \psi_{\textbf{r}_1+\bm{\tau}}K_{\textbf{r}_2,\textbf{r}_2+\bm{\tau}} \psi_{\textbf{r}_2} \psi_{\textbf{r}_2+\bm{\tau}}  }\\
     &\qquad +4\delta_{\textbf{R}_1,\textbf{R}_2}\expval{n_{\tau}(\textbf{R}_1)}-2\expval{K_{\textbf{r}_1,\textbf{r}_1+\bm{\tau}} \psi_{\textbf{r}_1} \psi_{\textbf{r}_1+\bm{\tau}} } \notag\\
     &\qquad -2\expval{K_{\textbf{r}_2,\textbf{r}_2+\bm{\tau}} \psi_{\textbf{r}_2} \psi_{\textbf{r}_2+\bm{\tau}} }+1 \notag
 \end{align}
 
 We now use Wick's theorem for the free fermionic four-point correlation function (without the correct $n^2$):
 
 \begin{align}
     &\left\langle \Psi\left|\tilde{\sigma}_{\hat{\tau}}^x(\textbf{R}_1) \tilde{\sigma}_{\hat{\tau}}^x(\textbf{R}_2)\right| \Psi\right\rangle_F\\
     &=
     4K_{\textbf{r}_1,\textbf{r}_1+\bm{\tau}} K_{\textbf{r}_2,\textbf{r}_2+\bm{\tau}} \left( \expval{\psi_{\textbf{r}_1} \psi_{\textbf{r}_1+\bm{\tau}} } \expval{\psi_{\textbf{r}_2} \psi_{\textbf{r}_2+\bm{\tau}}}\right. \\
     &\left.-\expval{\psi_{\textbf{r}_1} \psi_{\textbf{r}_2}}\expval{\psi_{\textbf{r}_1+\bm{\tau}} \psi_{\textbf{r}_2+\bm{\tau}}  }+\expval{\psi_{\textbf{r}_1}\psi_{\textbf{r}_2+\bm{\tau}} }\expval{\psi_{\textbf{r}_1+\bm{\tau}}  \psi_{\textbf{r}_2} } \right)\notag\\
     &-2K_{\textbf{r}_1,\textbf{r}_1+\bm{\tau}}\expval{ \psi_{\textbf{r}_1} \psi_{\textbf{r}_1+\bm{\tau}} }-2K_{\textbf{r}_2,\textbf{r}_2+\bm{\tau}} \expval{\psi_{\textbf{r}_2} \psi_{\textbf{r}_2+\bm{\tau}} }+1
 \end{align}
 Now since the $\textbf{r}_1$ and $\textbf{r}_2$ are located in the same sublattice and the correlation function for the same sublattice is zero, we have $\expval{\psi_{\textbf{r}_1} \psi_{\textbf{r}_2} }=0$, as well as the ones with the same $\tau $ displacement we have then a simplified expression
 \begin{align}
     &\left\langle \Psi\left|\tilde{\sigma}_{\hat{\tau}}^x(\textbf{R}_1) \tilde{\sigma}_{\hat{\tau}}^x(\textbf{R}_2)\right| \Psi\right\rangle_F\\
     &=
     4K_{\textbf{r}_1,\textbf{r}_1+\bm{\tau}} K_{\textbf{r}_2,\textbf{r}_2+\bm{\tau}} \left( \expval{\psi_{\textbf{r}_1} \psi_{\textbf{r}_1+\bm{\tau}} } \expval{\psi_{\textbf{r}_2} \psi_{\textbf{r}_2+\bm{\tau}}}\right. \\
     &\left.+\expval{\psi_{\textbf{r}_1}\psi_{\textbf{r}_2+\bm{\tau}} }\expval{\psi_{\textbf{r}_1+\bm{\tau}}  \psi_{\textbf{r}_2} } \right)\notag\\
     &-2K_{\textbf{r}_1,\textbf{r}_1+\bm{\tau}}\expval{ \psi_{\textbf{r}_1} \psi_{\textbf{r}_1+\bm{\tau}} }-2K_{\textbf{r}_2,\textbf{r}_2+\bm{\tau}} \expval{\psi_{\textbf{r}_2} \psi_{\textbf{r}_2+\bm{\tau}} }+1 \notag \label{commonEQ}
 \end{align}
The Fourier transform of the disconnected components will give contributions of $\delta_{\textbf{q},\textbf{G}}$. To see this, we use the translational invariance:
\begin{align}
    \expval{\psi^\alpha_{ \textbf{x}} \psi^\beta_{\textbf{y}}}= \dfrac{1}{\tilde{N}}\sum_{\textbf{k}} e^{-i \textbf{k} \cdot(\textbf{x}-\textbf{y})}\left[h^{-1}(\textbf{k})\right]_{\beta,\alpha}
\end{align}
Then, for fixed $\tau$ say horizontal direction, and in the convention where they all have the same sign $+1$ we have
\begin{align}
   & \sum_{\textbf{R}_1,\textbf{R}_2} K_{\textbf{r}_1,\textbf{r}_1+\bm{\tau}} K_{\textbf{r}_2,\textbf{r}_2+\bm{\tau}}  \expval{\psi_{\textbf{r}_1} \psi_{\textbf{r}_1+\bm{\tau}} } \expval{\psi_{\textbf{r}_2} \psi_{\textbf{r}_2+\bm{\tau}}} e^{i\textbf{q}\cdot \left(\textbf{R}_2-\textbf{R}_1\right)}\\
   &=  \sum_{\textbf{R}_1}   \expval{\psi_{\textbf{r}_1} \psi_{\textbf{r}_1+\bm{\tau}} }e^{-i\textbf{q}\cdot \textbf{R}_1}\sum_{\textbf{R}_2}  \expval{\psi_{\textbf{r}_2} \psi_{\textbf{r}_2+\bm{\tau}}} e^{i\textbf{q}\cdot \textbf{R}_2}\\
    =& \sum_{\textbf{R}_1}   \dfrac{1}{\tilde{N}}\sum_{\textbf{k}} e^{-i \textbf{k} \cdot(\textbf{r}_1-\textbf{r}_1)}h^{-1}_{\beta,\alpha}e^{-i\textbf{q}\cdot \textbf{R}_1}\\
    & \times \sum_{\textbf{R}_2} \dfrac{1}{\tilde{N}}\sum_{\textbf{k}} e^{-i \textbf{k} \cdot(\textbf{r}_2-\textbf{r}_2)} h^{-1}_{\beta',\alpha'}(\textbf{k}) e^{i\textbf{q}\cdot \textbf{R}_2}\\
    &= \dfrac{1}{\tilde{N}^2}\sum_{\textbf{R}_1} e^{-i\textbf{q}\cdot \textbf{R}_1} \sum_{\textbf{R}_2} e^{i\textbf{q}\cdot \textbf{R}_2} C_{\alpha\beta}^{\alpha'\beta'}\\
    &=\delta_{\textbf{q},\textbf{G}}\delta_{\textbf{q},-\tilde{\textbf{G}}}C_{\alpha\beta}^{\alpha'\beta'}= \delta_{\textbf{q},\textbf{G}}C_{\alpha\beta}^{\alpha'\beta'}.
\end{align}
The same logic applies to the $\expval{n_x(\textbf{R})}$ terms, giving a Fourier transform $\delta_{\textbf{q},\textbf{G}}$, so that for non-zero momentum we can restrict to the connected contribution
 \begin{align}
     &\tilde{s}_c(\textbf{q})=\dfrac{1}{N}\sum_{\textbf{R}_1,\textbf{R}_2} e^{i\textbf{q}\cdot \left(\textbf{R}_2-\textbf{R}_1\right)}
     4K_{\textbf{r}_1,\textbf{r}_1+\bm{\tau}} K_{\textbf{r}_2,\textbf{r}_2+\bm{\tau}} \notag \\
     &\qquad \times\expval{\psi_{\textbf{r}_1}\psi_{\textbf{r}_2+\bm{\tau}} }\expval{\psi_{\textbf{r}_1+\bm{\tau}}  \psi_{\textbf{r}_2} } \\
     & s_c(\textbf{q})=\dfrac{1}{N}\sum_{\textbf{R}_1,\textbf{R}_2} e^{i\textbf{q}\cdot \left(\textbf{R}_2-\textbf{R}_1\right)}
     4K_{\textbf{r}_1,\textbf{r}_1+\bm{\tau}} K_{\textbf{r}_2,\textbf{r}_2+\bm{\tau}} \notag\\
     & \qquad \times \expval{\psi_{\textbf{r}_1}\psi_{\textbf{r}_2+\bm{\tau}} }\expval{\psi_{\textbf{r}_1+\bm{\tau}}  \psi_{\textbf{r}_2} }\notag\\
     &\qquad +\sum_{\textbf{R}_1,\textbf{R}_2} e^{i\textbf{q}\cdot \left(\textbf{R}_2-\textbf{R}_1\right)} \expval{n_{\tau}(\textbf{R}_1)} \delta_{\textbf{R}_1,\textbf{R}_2}\\
     & s_c(\textbf{q})=\dfrac{1}{N}\sum_{\textbf{R}_1,\textbf{R}_2} e^{i\textbf{q}\cdot \left(\textbf{R}_2-\textbf{R}_1\right)}
     4K_{\textbf{r}_1,\textbf{r}_1+\bm{\tau}} K_{\textbf{r}_2,\textbf{r}_2+\bm{\tau}} \\
     &\expval{\psi_{\textbf{r}_1}\psi_{\textbf{r}_2+\bm{\tau}} }\expval{\psi_{\textbf{r}_1+\bm{\tau}}  \psi_{\textbf{r}_2} }+\dfrac{1}{N}\sum_{\textbf{R}} \expval{n_{\tau}(\textbf{R})} 
 \end{align}
 The  Kasteleyn matrix for a case where we have all Fourier conventions with the physical positions of the fermions and the Kasteleyn convention has the A sites with only outward flowing arrows and $f(\textbf{k})=e^{i\textbf{k}\cdot \bm{\delta}_1}+e^{i\textbf{k}\cdot \bm{\delta}_2}+e^{i\textbf{k}\cdot \bm{\delta}_3}$
 \begin{align}
 h(\textbf{k}) = \begin{pmatrix}
     0 && f(\textbf{k})\\
    -f(\textbf{k})&& 0 
 \end{pmatrix}
 \end{align}
 Now the Kasteleyn orientation is the same as the unit cell with two sublattices, so for the orientation with positive Kasteleyn elements in the horizontal direction, we write (with 1/N normalization for $\sigma^x(\textbf{q})$):
 
 \begin{align}
     &\tilde{s}^3_c(\textbf{q})=\frac{1}{N}\sum_{\textbf{r}_1,\textbf{r}_2} e^{i\textbf{q}\cdot \left(\textbf{r}_2-\textbf{r}_1\right)}
     4K_{\textbf{r}_1,\textbf{r}_1+\bm{\tau}} K_{\textbf{r}_2,\textbf{r}_2+\bm{\tau}} \\
     & \times \expval{\psi_{\textbf{r}_1}\psi_{\textbf{r}_2+\bm{\tau}} }\expval{\psi_{\textbf{r}_1+\bm{\tau}}  \psi_{\textbf{r}_2} } \\
     &=4\frac{1}{N} \sum_{\textbf{r}_1,\textbf{r}_2} e^{i\textbf{q}\cdot \left(\textbf{r}_2-\textbf{r}_1\right)}
   \expval{\psi^1_{\textbf{r}_1}\psi^2_{\textbf{r}_2+\bm{\delta}_3} }\expval{\psi^2_{\textbf{r}_1+\bm{\delta}_3}  \psi^1_{\textbf{r}_2} } \\
   &=\frac{4}{N} \sum_{\textbf{r}_1,\textbf{r}_2}e^{i\textbf{q}\cdot \left(\textbf{r}_2-\textbf{r}_1\right)}\dfrac{1}{N}\sum_{\textbf{k}} e^{i \textbf{k} \cdot(\textbf{r}_2+\bm{\delta}_3-\textbf{r}_1)}\\
   & \times h^{-1}_{2,1}(\textbf{k}) \dfrac{1}{N}\sum_{\textbf{k}'} e^{i \textbf{k}' \cdot(\textbf{r}_2-\textbf{r}_1-\bm{\delta}_3)}h^{-1}_{1,2}(\textbf{k}') \\
   &=\frac{4}{N} \sum_{\textbf{k},\textbf{k}'} \sum_{\textbf{r}_1}\sum_{\textbf{r}_2 } e^{-i \textbf{r}_1\cdot \left(\textbf{q}+\textbf{k}+\textbf{k}'\right)} e^{i \textbf{r}_2\cdot \left(\textbf{q}+\textbf{k}+\textbf{k}'\right)}\\
    & \times \dfrac{1}{ N^2 }h^{-1}_{2,1}(\textbf{k})  e^{i \bm{\delta}_3 \cdot (\textbf{k}-\textbf{k}')}h^{-1}_{1,2}(\textbf{k}')\\
   &= \frac{4}{N} \sum_{\textbf{k},\textbf{k}'} \delta_{\textbf{q}+\textbf{k}+\textbf{k}',\textbf{G}}\delta_{\textbf{q}+\textbf{k}+\textbf{k}',-\tilde{\textbf{G}}}h^{-1}_{2,1}(\textbf{k})  h^{-1}_{1,2}(\textbf{k}')e^{i \bm{\delta}_3 \cdot (\textbf{k}-\textbf{k}')} \\
   &= \frac{4}{N} \sum_{\textbf{k}} e^{i \bm{\delta}_3 \cdot (2\textbf{k}+\textbf{q})} h^{-1}_{2,1}(\textbf{k})  h^{-1}_{1,2}(-\textbf{q}-\textbf{k})\\
   &= \frac{4}{4\pi^2}\iint \dd^2 \textbf{k}\  e^{i \bm{\delta}_3 \cdot (2\textbf{k}+\textbf{q})} h^{-1}_{2,1}(\textbf{k})  h^{-1}_{1,2}(-\textbf{q}-\textbf{k})
 \end{align}
 For this Hamiltonian, we have the function:
 \begin{align}
     h^{-1}(\textbf{k}) =  \begin{pmatrix}
     0 && - \dfrac{e^{i (k_x+k_y)/3}}{1+e^{i k_x}+e^{i k_y}}\\
   \dfrac{e^{-i (k_x+k_y)/3}}{1+e^{-i k_x}+e^{-i k_y}} && 0 
 \end{pmatrix}
 \end{align}
 Then for  $\textbf{q}=q_x \textbf{b}_1$  we have 

 \begin{widetext}
 \begin{align}
     &\tilde{s}^3_c(q_x \textbf{b}_1+q_y \textbf{b}_2) = \dfrac{4}{4\pi^2} \int_{-\pi}^{\pi} d k_x \int_{-\pi}^{\pi} d k_y \ \dfrac{-e^{i (2 k_y+q_y)}}{(e^{i k_x}+e^{i k_y}+e^{i(k_x+k_y)})(e^{i (k_x+q_x)}+e^{i (k_y+q_y)}+e^{i(k_x+k_y+q_x+q_y)})}
 \end{align}
 \end{widetext}
 
 Doing this integral numerically, we find, as expected a zero contribution for parallel excitations $\textbf{q}=(q,0)$ for $\tau=3$ see Fig.~\ref{fig:structfactHoney} and non zero for $\tau=1$, see Fig.~\ref{fig:structfactHoneytau1}.

\section{SMA oscillator strength } \label{app:oscstrength}

For single hexagon resonances, it is easy to calculate and indeed find as in Ref. \cite{moessner_three-dimensional_2003}
\begin{align}
     &{\left[\tilde{\sigma}_{\hat{\tau}}^x(-\mathbf{q}),\left[-t \hat{T}_{\circ}, \tilde{\sigma}_{\hat{\tau}}^x(\mathbf{q})\right]\right]} =8 t \hat{T}_{\circ} [1-\cos ((\mathbf{Q}+\mathbf{k}) \cdot \mathbf{r})],
\end{align}
In this case $f(\bm{0})=0$ with $  T=-t \sum_\circ T_\circ$. We may generically express the resonance term as a sum $T=T_2+T_2^\dagger$. In this case it will be enough to calculate $f$ for just one of this terms since $f(\mathbf{q})= \expval{f_2(\mathbf{q})}+\expval{f_2^\dagger(-\mathbf{q})}$ as can be seen from taking the Hermitian conjugate of the expression defining $f$ and $f_2$ is defined just as $f$ but with $T$ replaced by $T_2$. Now, for the double hexagon resonance of equation \eqref{DoubleRes}, we have
\begin{widetext}
\begin{align}
    \begin{split}
        &T_2 = -t\sum_{\textbf{R},\delta} T_2(\textbf{R},\delta)= -t\sum_{\textbf{R}} \sigma_2^+(\textbf{R}-\textbf{a}_1)\sigma_1^-(\textbf{R}-\textbf{a}_1) \sigma_3^+(\textbf{R})\sigma_1^-(\textbf{R})\sigma_3^+(\textbf{R}+\textbf{a}_1)\\
        &\times \sigma_2^-(\textbf{R}+\textbf{a}_1)\sigma_1^+(\textbf{R}-\textbf{a}_2+\textbf{a}_1)\sigma_3^-(\textbf{R}-\textbf{a}_2+\textbf{a}_1)\sigma_1^+(\textbf{R}-\textbf{a}_2)\sigma_3^-(\textbf{R}-\textbf{a}_2)\\[2ex]
        +&\sigma_2^+(\textbf{R}-\textbf{a}_1)\sigma_1^-(\textbf{R}-\textbf{a}_1)\sigma_2^+(\textbf{R}-\textbf{a}_1+\textbf{a}_2)\sigma_1^-(\textbf{R}-\textbf{a}_1+\textbf{a}_2) \sigma_3^+(\textbf{R}+\textbf{a}_2)\\
        & \times \sigma_2^-(\textbf{R}+\textbf{a}_2)\sigma_1^+(\textbf{R}) \sigma_2^-(\textbf{R}) \sigma_1^+(\textbf{R}-\textbf{a}_2)\sigma_3^-(\textbf{R}-\textbf{a}_2)\\
       + & \sigma_{2}^{+}(\textbf{R}-\textbf{a}_1)\sigma_{1}^{-}(\textbf{R}-\textbf{a}_1) \sigma_3^{+}(\textbf{R})\sigma_2^-(\textbf{R}) \sigma_{3}^+(\textbf{R}+\textbf{a}_1-\textbf{a}_2)\sigma_{2}^-(\textbf{R}+\textbf{a}_1-\textbf{a}_2)\\
       &\times \sigma_1^+(\textbf{R}+\textbf{a}_1-2\textbf{a}_2) \sigma_3^-(\textbf{R}+\textbf{a}_1-2\textbf{a}_2 )\sigma_2^+(\textbf{R}-\textbf{a}_2)\sigma_3^-(\textbf{R}-\textbf{a}_2).
    \end{split}
\end{align}
\end{widetext}
Where $\delta=1,2,3$ are the three nearest neighbor resonance processes represented by the three summands. Let us define $\tilde{T}_2$ such that $T_2=-t\ \tilde{T}_2$.
Next, we calculate the commutator with the variational field. Let us look first at the $\tau=3$ bond

\begin{align}
    &[\tilde{T}_2,\tilde{\sigma}_{3}^x(\mathbf{q}) ]=\sum_{\textbf{R}'}   [\tilde{T}_2,\sigma_{3}^x(\textbf{R}')]e^{i\textbf{q}\cdot \textbf{R}'}\\
    &= \sum_{\textbf{R}'}   [\sum_{\textbf{R},\delta} T_2(\textbf{R},\delta),\sigma_{3}^x(\textbf{R}')]e^{i\textbf{q}\cdot \textbf{R}'} \\
    &= \sum_{\textbf{R},\delta} \sum_{\textbf{R}'}   [T_2(\textbf{R},\delta),\sigma_{3}^x(\textbf{R}')]e^{i\textbf{q}\cdot \textbf{R}'} 
\end{align}
We use now the identity
\begin{align}
    [\sigma^{\pm},\sigma^x]=\mp 2 \sigma^{\pm}
\end{align}
For the calculation it is useful to think of the $\textbf{R}$ as fixed and identifying the $\sigma_3^\alpha(\textbf{R}')$ that appear in $\tilde{T}_2$ since the other operators $\sigma^x_\tau$ commute with $\sigma_3^x(\textbf{R}')$. We have then for the first commutator with $\tilde{T}_2(\textbf{R},1)$
\begin{widetext}
\begin{align}
   &[\tilde{T}_2(\textbf{R},1),\tilde{\sigma}_{3}^x(\mathbf{q}) ]= \sum_{\textbf{R}'} [\tilde{T}_2(\textbf{R},1),\tilde{\sigma}_{3}^x(\mathbf{R}') ]e^{i\textbf{q}\cdot \textbf{R}'}= \sum_{\textbf{R}'} [\sigma_2^+(\textbf{R}-\textbf{a}_1)\sigma_1^-(\textbf{R}-\textbf{a}_1) \sigma_3^+(\textbf{R})\sigma_1^-(\textbf{R})\sigma_3^+(\textbf{R}+\textbf{a}_1)\\
   & \times \sigma_2^-(\textbf{R}+\textbf{a}_1)\sigma_1^+(\textbf{R}-\textbf{a}_2+\textbf{a}_1)\sigma_3^-(\textbf{R}-\textbf{a}_2+\textbf{a}_1)\sigma_1^+(\textbf{R}-\textbf{a}_2)\sigma_3^-(\textbf{R}-\textbf{a}_2),\sigma_3^x(\textbf{R}')]e^{i\textbf{q}\cdot \textbf{R}'}\\
   &=\sigma_2^+(\textbf{R}-\textbf{a}_1)\sigma_1^-(\textbf{R}-\textbf{a}_1) \sigma_3^+(\textbf{R})[\sigma_1^-(\textbf{R})\sigma_3^+(\textbf{R}+\textbf{a}_1) \sigma_2^-(\textbf{R}+\textbf{a}_1)\sigma_1^+(\textbf{R}-\textbf{a}_2+\textbf{a}_1)\sigma_3^-(\textbf{R}-\textbf{a}_2+\textbf{a}_1)\\
  & \times  \sigma_1^+(\textbf{R}-\textbf{a}_2)\sigma_3^-(\textbf{R}-\textbf{a}_2),\sigma_3^x(\textbf{R})]e^{i\textbf{q}\cdot \textbf{R}}+ [\sigma_2^+(\textbf{R}-\textbf{a}_1)\sigma_1^-(\textbf{R}-\textbf{a}_1) \sigma_3^+(\textbf{R}),\sigma_3^x(\textbf{R})]\sigma_1^-(\textbf{R})\sigma_3^+(\textbf{R}+\textbf{a}_1)\\
  &\times \sigma_2^-(\textbf{R}+\textbf{a}_1)\sigma_1^+(\textbf{R}-\textbf{a}_2+\textbf{a}_1)\sigma_3^-(\textbf{R}-\textbf{a}_2+\textbf{a}_1) \sigma_1^+(\textbf{R}-\textbf{a}_2)\sigma_3^-(\textbf{R}-\textbf{a}_2)e^{i\textbf{q}\cdot \textbf{R}}+\sum_{\textbf{R}'\neq \textbf{R}}(\dots) \\
  &= [\sigma_2^+(\textbf{R}-\textbf{a}_1)\sigma_1^-(\textbf{R}-\textbf{a}_1) \sigma_3^+(\textbf{R}),\sigma_3^x(\textbf{R})]\sigma_1^-(\textbf{R})\sigma_3^+(\textbf{R}+\textbf{a}_1)\sigma_2^-(\textbf{R}+\textbf{a}_1)\sigma_1^+(\textbf{R}-\textbf{a}_2+\textbf{a}_1)\\
  &\times \sigma_3^-(\textbf{R}-\textbf{a}_2+\textbf{a}_1) \sigma_1^+(\textbf{R}-\textbf{a}_2)\sigma_3^-(\textbf{R}-\textbf{a}_2)e^{i\textbf{q}\cdot \textbf{R}}+\sum_{\textbf{R}'\neq \textbf{R}}(\dots)\\
  &=-2 \sigma_2^+(\textbf{R}-\textbf{a}_1)\sigma_1^-(\textbf{R}-\textbf{a}_1) \sigma_3^+(\textbf{R})\sigma_1^-(\textbf{R})\sigma_3^+(\textbf{R}+\textbf{a}_1)\sigma_2^-(\textbf{R}+\textbf{a}_1)\sigma_1^+(\textbf{R}-\textbf{a}_2+\textbf{a}_1)\\
  &\times \sigma_3^-(\textbf{R}-\textbf{a}_2+\textbf{a}_1) \sigma_1^+(\textbf{R}-\textbf{a}_2)\sigma_3^-(\textbf{R}-\textbf{a}_2)e^{i\textbf{q}\cdot \textbf{R}}+\sum_{\textbf{R}'\neq \textbf{R}}(\dots)\\
  &=-2\tilde{T}_2(\textbf{R},1)e^{i\textbf{q}\cdot\textbf{R}}+\sum_{\textbf{R}'\neq \textbf{R}}(\dots)
\end{align}
\end{widetext}
We used the commutator identity for two operators and included the rest of terms in the dots. It is clear that since the commutation of $\sigma^x$ and $\sigma^{\pm}$ gives the same $\sigma^\pm$ that we will always obtain in the end the whole resonance operator $\tilde{T}_2(\textbf{R},1)$ ( indeed even if the ordering changes all operators in $\tilde{T}_2(\textbf{R},1)$ commute with each other). We also get an exponential factor picking up the position in space and a $-2\mp$. So that the whole sum will be equal to
\begin{align}
     &[\tilde{T}_2(\textbf{R},1),\tilde{\sigma}_{3}^x(\mathbf{q}) ]= \sum_{\textbf{R}'} [\tilde{T}_2(\textbf{R},1),\tilde{\sigma}_{3}^x(\mathbf{R}') ]e^{i\textbf{q}\cdot \textbf{R}'}\\
     &=2\tilde{T}_2(\textbf{R},1)(-e^{i\textbf{q}\cdot\textbf{R}} -e^{i\textbf{q}\cdot(\textbf{R}+\textbf{a}_1)}\\
     & \qquad +e^{i\textbf{q}\cdot(\textbf{R}+\textbf{a}_1-\textbf{a}_2)}+e^{i\textbf{q}\cdot(\textbf{R}-\textbf{a}_2)})\\
     &=2\tilde{T}_2(\textbf{R},1) e^{i\textbf{q}\cdot \textbf{R}}(e^{i\textbf{q}\cdot(\textbf{a}_1-\textbf{a}_2)}+e^{-i\textbf{q}\cdot \textbf{a}_2}-e^{i\textbf{q}\cdot\textbf{a}_1}-1)
\end{align}
Similarly we can calculate the commutator for the other resonance operators
\begin{align}
    &[\tilde{T}_2(\textbf{R},2),\tilde{\sigma}_{3}^x(\mathbf{q}) ]= \sum_{\textbf{R}'} [\tilde{T}_2(\textbf{R},2),\tilde{\sigma}_{3}^x(\mathbf{R}') ]e^{i\textbf{q}\cdot \textbf{R}'}\\
    &=2\tilde{T}_2(\textbf{R},2) (-e^{i\textbf{q}\cdot (\textbf{R}+\textbf{a}_2)}+e^{i\textbf{q}\cdot (\textbf{R}-\textbf{a}_2)})\\
    &=2\tilde{T}_2(\textbf{R},2)e^{i\textbf{q}\cdot \textbf{R}}(e^{-i\textbf{q}\cdot \textbf{a}_2}-e^{i\textbf{q}\cdot \textbf{a}_2})
\end{align}
And for the last resonance
\begin{align}
     &[\tilde{T}_2(\textbf{R},3),\tilde{\sigma}_{3}^x(\mathbf{q}) ]= \sum_{\textbf{R}'} [\tilde{T}_2(\textbf{R},3),\tilde{\sigma}_{3}^x(\mathbf{R}') ]e^{i\textbf{q}\cdot \textbf{R}'}\\
     &=2\tilde{T}_2(\textbf{R},3) e^{i\textbf{q}\cdot \textbf{R}} (-1-e^{i\textbf{q}\cdot (\textbf{a}_1-\textbf{a}_2)}+e^{i\textbf{q}\cdot (\textbf{a}_1-2\textbf{a}_2)}+e^{-i\textbf{q}\cdot \textbf{a}_2})
\end{align}
For the actual commutator in the SMA we need another time to commute this with the dimer density 
\begin{align}
     &[\tilde{\sigma}_{3}^x(-\mathbf{q}),[\tilde{T}_2(\textbf{R},1),\tilde{\sigma}_{3}^x(\mathbf{q}) ]]\\
     &=[\tilde{\sigma}_{3}^x(-\mathbf{q}),2\tilde{T}_2(\textbf{R},1) e^{i\textbf{q}\cdot \textbf{R}}\\
     &\quad (e^{i\textbf{q}\cdot(\textbf{a}_1-\textbf{a}_2)}+e^{-i\textbf{q}\cdot \textbf{a}_2}-e^{i\textbf{q}\cdot\textbf{a}_1}-1)]\\
     &=-2e^{i\textbf{q}\cdot \textbf{R}}(e^{i\textbf{q}\cdot(\textbf{a}_1-\textbf{a}_2)}+e^{-i\textbf{q}\cdot \textbf{a}_2}-e^{i\textbf{q}\cdot\textbf{a}_1}-1)\\
     & \quad [\tilde{T}_2(\textbf{R},1),\tilde{\sigma}_{3}^x(-\mathbf{q})]\\
     &=-2e^{i\textbf{q}\cdot \textbf{R}}(e^{i\textbf{q}\cdot(\textbf{a}_1-\textbf{a}_2)}+e^{-i\textbf{q}\cdot \textbf{a}_2}-e^{i\textbf{q}\cdot\textbf{a}_1}-1)\\
     &\{2\tilde{T}_2(\textbf{R},1) e^{-i\textbf{q}\cdot \textbf{R}}(e^{-i\textbf{q}\cdot(\textbf{a}_1-\textbf{a}_2)}+e^{i\textbf{q}\cdot \textbf{a}_2}-e^{-i\textbf{q}\cdot\textbf{a}_1}-1)\}\\
     &=-4 \tilde{T}_2(\textbf{R},1) \abs{e^{i\textbf{q}\cdot(\textbf{a}_1-\textbf{a}_2)}+e^{-i\textbf{q}\cdot \textbf{a}_2}-e^{i\textbf{q}\cdot\textbf{a}_1}-1}^2
\end{align}
We repeat for the other resonances and summarize
\begin{align}
    &[\tilde{\sigma}_{3}^x(-\mathbf{q}),[\tilde{T}_2(\textbf{R},1),\tilde{\sigma}_{3}^x(\mathbf{q}) ]]\\
    &=-4 \tilde{T}_2(\textbf{R},1) \abs{e^{i\textbf{q}\cdot(\textbf{a}_1-\textbf{a}_2)}+e^{-i\textbf{q}\cdot \textbf{a}_2}-e^{i\textbf{q}\cdot\textbf{a}_1}-1}^2\\
    &[\tilde{\sigma}_{3}^x(-\mathbf{q}),[\tilde{T}_2(\textbf{R},2),\tilde{\sigma}_{3}^x(\mathbf{q}) ]]=-4 \tilde{T}_2(\textbf{R},1) \abs{e^{-i\textbf{q}\cdot \textbf{a}_2}-e^{i\textbf{q}\cdot \textbf{a}_2}}^2\\
    &[\tilde{\sigma}_{3}^x(-\mathbf{q}),[\tilde{T}_2(\textbf{R},3),\tilde{\sigma}_{3}^x(\mathbf{q}) ]]=\\
    &-4 \tilde{T}_2(\textbf{R},3) \abs{1+e^{i\textbf{q}\cdot (\textbf{a}_1-\textbf{a}_2)}-e^{i\textbf{q}\cdot (\textbf{a}_1-2\textbf{a}_2)}-e^{-i\textbf{q}\cdot \textbf{a}_2}}^2
\end{align}
We sum over all positions to obtain

\begin{align}
    &f_2(\textbf{q})=[\tilde{\sigma}_{3}^x(-\mathbf{q}),[T_2,\tilde{\sigma}_{3}^x(\mathbf{q}) ]]\\
    &= -4 \sum_{\textbf{R}} \tilde{T}_2(\textbf{R},1) \abs{e^{i\textbf{q}\cdot(\textbf{a}_1-\textbf{a}_2)}+e^{-i\textbf{q}\cdot \textbf{a}_2}-e^{i\textbf{q}\cdot\textbf{a}_1}-1}^2\\
    &+ \tilde{T}_2(\textbf{R},1) \abs{e^{-i\textbf{q}\cdot \textbf{a}_2}-e^{i\textbf{q}\cdot \textbf{a}_2}}^2\notag \\
    &+\tilde{T}_2(\textbf{R},3) \abs{1+e^{i\textbf{q}\cdot (\textbf{a}_1-\textbf{a}_2)}-e^{i\textbf{q}\cdot (\textbf{a}_1-2\textbf{a}_2)}-e^{-i\textbf{q}\cdot \textbf{a}_2}}^2
\end{align}

Using now that $f(\mathbf{q})= \expval{f_2(\mathbf{q})}+\expval{f_2^\dagger(-\mathbf{q})}$ we have the final result

\begin{widetext}
\begin{align}
    f_{\tau=3}(\mathbf{q})=&\left\langle \Psi\left|\left[\tilde{\sigma}_{3}^x(-\mathbf{q}),\left[\mathcal{H}_{\text{eff}}, \tilde{\sigma}_{3}^x(\mathbf{q})\right]\right]\right| \Psi\right\rangle = -4 \sum_{\textbf{R}} \expval{T(\textbf{R},1)} \abs{e^{i\textbf{q}\cdot(\textbf{a}_1-\textbf{a}_2)}+e^{-i\textbf{q}\cdot \textbf{a}_2}-e^{i\textbf{q}\cdot\textbf{a}_1}-1}^2\\
    &+ \expval{T(\textbf{R},2)}\abs{e^{-i\textbf{q}\cdot \textbf{a}_2}-e^{i\textbf{q}\cdot \textbf{a}_2}}^2 +\expval{T(\textbf{R},3) }\abs{1+e^{i\textbf{q}\cdot (\textbf{a}_1-\textbf{a}_2)}-e^{i\textbf{q}\cdot (\textbf{a}_1-2\textbf{a}_2)}-e^{-i\textbf{q}\cdot \textbf{a}_2}}^2 \notag
\end{align}
\end{widetext}

\begin{figure}[t]
    \centering
    \includegraphics[scale=0.4]{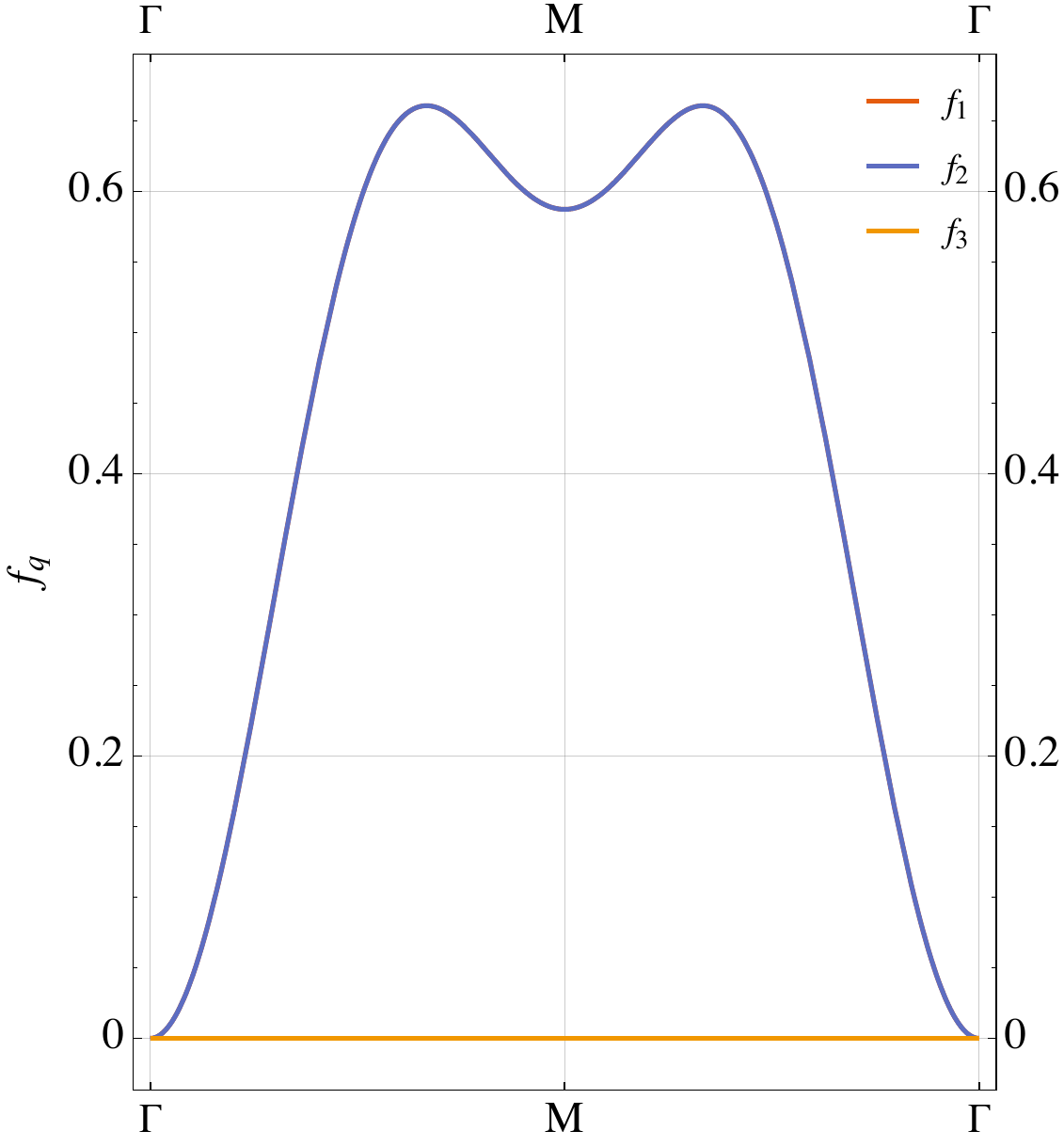}
    \caption{Oscillator strength $f(\textbf{q})$ along the $\Gamma$, $M$ line for all resonance operators having the same expectacion value $\expval{T(\delta)}=\expval{T}$. The prefactor is calculated from the variational energy found in previous studies and $t=J_\perp/2$ with $J_\perp=0.0666 J_z$. We note $f_1$ and $f_2$ lie on top of each other.}
    \label{fig:OscStrength}
\end{figure}

In contrast to other cases with single plaquette resonances, we cannot naively group all resonance operators since they can have different expectation values in the $\ket{\Psi}$ state, because of it's reduced symmetry. Nevertheless, it follows that $f(\bm{0})=0$ and at $\textbf{q}_0=\bm{b}_1/2$ the $M$ point we have $f(\textbf{q}_0)= 0$, even for different values of $\expval{T(\textbf{R},\delta)}$, more generally it is zero for all $\textbf{q}=(q_x,0)$ just as in the square lattice case.\\
For the other two types of bonds, we have a very similar reasoning leading to 
\begin{widetext}
\begin{align}
    f_{\tau=2}(\mathbf{q})=&\left\langle \Psi\left|\left[\tilde{\sigma}_{2}^x(-\mathbf{q}),\left[\mathcal{H}_{\text{eff}}, \tilde{\sigma}_{2}^x(\mathbf{q})\right]\right]\right| \Psi\right\rangle = -4 \sum_{\textbf{R}} \expval{T(\textbf{R},1)} \abs{e^{-i\textbf{q}\cdot\textbf{a}_1}-e^{i\textbf{q}\cdot\textbf{a}_1}}^2\\
    &+ \expval{T(\textbf{R},2)}\abs{e^{-i\textbf{q}\cdot\textbf{a}_1}+e^{i\textbf{q}\cdot(\textbf{a}_2-\textbf{a}_1)}
    -e^{i\textbf{q}\cdot\textbf{a}_2}-1
    }^2+\expval{T(\textbf{R},3) }\abs{e^{-i\textbf{q}\cdot\textbf{a}_1}-e^{i\textbf{q}\cdot(\textbf{a}_1-\textbf{a}_2)}
    +e^{-i\textbf{q}\cdot\textbf{a}_2}-1
     }^2 \notag
\end{align}
\end{widetext}
and 
\begin{widetext}
\begin{align}
    f_{\tau=1}(\mathbf{q})=&\left\langle \Psi\left|\left[\tilde{\sigma}_{1}^x(-\mathbf{q}),\left[\mathcal{H}_{\text{eff}}, \tilde{\sigma}_{1}^x(\mathbf{q})\right]\right]\right| \Psi\right\rangle = -4 \sum_{\textbf{R}} \expval{T(\textbf{R},1)} \abs{e^{-i\textbf{q}\cdot\textbf{a}_1}+1-e^{i\textbf{q}\cdot(\textbf{a}_1-\textbf{a}_2)}-e^{-i\textbf{q}\cdot\textbf{a}_2}}^2\\
    &+ \expval{T(\textbf{R},2)}\abs{e^{-i\textbf{q}\cdot\textbf{a}_1}+e^{i\textbf{q}\cdot(\textbf{a}_2-\textbf{a}_1)}
    -1-e^{-i\textbf{q}\cdot\textbf{a}_2}
    }^2+\expval{T(\textbf{R},3) }\abs{e^{-i\textbf{q}\cdot\textbf{a}_1}-e^{i\textbf{q}\cdot(\textbf{a}_1-2\textbf{a}_2)}
     }^2 \notag
\end{align}
\end{widetext}
We plot the oscillator strength in Fig.~\ref{fig:OscStrength} for the approximation where all resonances have on average the same value $\expval{T(\delta)}=\expval{T}$. We take that value to be the energy found in the variational study found before and plug in $t=J_\perp/2$ with $J_\perp=0.0666 J_z$.
Before calculating the structure factor, let us mention that the gapless zero line also appears for the square lattice dimer model with single plaquette resonances. In that case forthe dimers in the $\tau=\hat{x}$ direction there is a conserved quantity $\sigma_{\tau}^x(\textbf{q}_0)$ with $\textbf{q}_0=(q_x,\pi)$  so $f(\textbf{q}_0)=0$ and the oscillator strength for $\textbf{q}_0+\textbf{k}$ has the form $f(\textbf{q}_0+\textbf{k})\propto (\textbf{k}\times \hat{\tau})^2$. 
In our case we have  also that the quantum dynamics creates and destroys pairs of dimers and hence  for $\tau=\hat{x}=\hat{3}$ we have that $\sigma_{\tau}^x(\textbf{q}_0)$ is conserved with $\textbf{q}_0=(q_x,0)$ we can expand to obtain (assuming resonances have approximately the same expectation value)
\begin{align}
    f_{\hat{\tau}=\hat{x}}(\textbf{q}_0+\textbf{k})\propto (\textbf{k}\times \hat{\tau})^2+O(\abs{\textbf{k}}^3)
\end{align}\\

\section{Structure factor for the variational supersolid wave function}
Let us tackle now the more complicated problem of the supersolid variational wave function. We can go back to the calculation for the Honeycomb at an early stage \eqref{commonEQ} which we repeat here
 \begin{align}
     &\left\langle \Psi\left|\tilde{\sigma}_{\hat{\tau}}^x(\textbf{R}_1) \tilde{\sigma}_{\hat{\tau}}^x(\textbf{R}_2)\right| \Psi\right\rangle_F=\\
     &
     4K_{\textbf{r}_1,\textbf{r}_1+\bm{\tau}} K_{\textbf{r}_2,\textbf{r}_2+\bm{\tau}} \left( \expval{\psi_{\textbf{r}_1} \psi_{\textbf{r}_1+\bm{\tau}} } \expval{\psi_{\textbf{r}_2} \psi_{\textbf{r}_2+\bm{\tau}}}\right.\\
     &\left. +\expval{\psi_{\textbf{r}_1}\psi_{\textbf{r}_2+\bm{\tau}} }\expval{\psi_{\textbf{r}_1+\bm{\tau}}  \psi_{\textbf{r}_2} } \right)\notag\\
     &-2K_{\textbf{r}_1,\textbf{r}_1+\bm{\tau}}\expval{ \psi_{\textbf{r}_1} \psi_{\textbf{r}_1+\bm{\tau}} }-2K_{\textbf{r}_2,\textbf{r}_2+\bm{\tau}} \expval{\psi_{\textbf{r}_2} \psi_{\textbf{r}_2+\bm{\tau}} }+1
 \end{align}
 We need to know two correlators:
 \begin{align}
     \expval{ \psi_{\textbf{r}_1} \psi_{\textbf{r}_1+\bm{\tau}} }, \quad \expval{\psi_{\textbf{r}_1+\bm{\tau}}  \psi_{\textbf{r}_2} }
 \end{align}
 Also since the last line of the previous equation only depends on one or the other coordinate and we have a sum over both for the structure factor it is clear then that they will only contribute for $\textbf{q}=0, \ \text{mod}(\textbf{G})$, with $\textbf{G}$ reciprocal lattice vector of the extended unit cell. Let us focus then on the second-order contributions:
 \begin{align}
     & \tilde{s}_2(\textbf{q})=\dfrac{1}{N}\sum_{\textbf{R}_1,\textbf{R}_2} e^{i\textbf{q}\cdot \left(\textbf{R}_2-\textbf{R}_1\right)}
     4K_{\textbf{r}_1,\textbf{r}_1+\bm{\tau}} K_{\textbf{r}_2,\textbf{r}_2+\bm{\tau}} \\
     & \left( \expval{\psi_{\textbf{r}_1} \psi_{\textbf{r}_1+\bm{\tau}} } \expval{\psi_{\textbf{r}_2} \psi_{\textbf{r}_2+\bm{\tau}}}+\expval{\psi_{\textbf{r}_1}\psi_{\textbf{r}_2+\bm{\tau}} }\expval{\psi_{\textbf{r}_1+\bm{\tau}}  \psi_{\textbf{r}_2} } \right)
 \end{align}
 Let us focus on $\tau=3$ since is the one that has $f(\textbf{q})=0$ at the $M$ point. Now care must be taken for using the variational wave function with the symmetry breaking pattern since it has different translational symmetry and thus we must translate the original points of the Kasteleyn fermions to this basis. First we reexpress the sum as a sum over the new lattice vectors, which needs to consider also the blue plaquets which are not part of the red lattice:
 \begin{align}
     \sum_{\textbf{R}_1}\quad  = \quad  \sum^I_{\textbf{R}_1=\textbf{T}_1} \quad +\sum^{II}_{\textbf{R}_1=\textbf{T}_1-\textbf{a}_1} +\sum^{III}_{\textbf{R}_1=\textbf{T}_1-\textbf{a}_2}, 
 \end{align}
 where the Roman numeral label the three sublattices of the plaquettes. We defined the new translation vectors as $\textbf{c}_1,\textbf{c}_2$ and the lattice vectors as $\textbf{T}_1= i \textbf{c}_1+j\textbf{c}_2$, $i,j\in \mathbb{Z}$ as in  Fig.~\ref{fig:honeycomb}. It is clear that $\textbf{c}_1=2 \textbf{a}_2-\textbf{a}_1$ and $\textbf{c}_2= \textbf{a}_2+\textbf{a}_1$. We also wrote there the sublattice index $\alpha=1,\dots,6$ for each hexagon where the fermions sit, based on the inequivalent sites of the variational wave function. The fermions will map to
 \begin{align}
     &I: \quad \psi_{\textbf{r}_1}\rightarrow \psi_{\textbf{T}_1}^5\\
     &  II: \quad \psi_{\textbf{r}_1}\rightarrow \psi_{\textbf{T}_1}^4\\
     &  III: \quad \psi_{\textbf{r}_1}\rightarrow \psi_{\textbf{T}_1}^6
 \end{align}
 \subsection{$\tau=3$ structure factor}
 We also need the translated fermions for $\tau=3,\bm{\tau}=\bm{\delta}_3$ which will be 
 \begin{align}
     &I: \quad \psi_{\textbf{r}_1+\bm{\delta}_3}\rightarrow \psi_{\textbf{T}_1}^2\\
     &  II: \quad \psi_{\textbf{r}_1+\bm{\delta}_3}\rightarrow \psi_{\textbf{T}_1+\textbf{c}_1-\textbf{c}_2}^3\\
     &  III: \quad \psi_{\textbf{r}_1+\bm{\delta}_3}\rightarrow \psi_{\textbf{T}_1}^1
 \end{align}
 We write now the weights as (they have the same arrow direction in our Kasteleyn orientation )
 \begin{align}
     K_{\textbf{r}_1,\textbf{r}_1+\bm{\tau}} \quad \rightarrow \quad 
     &I:\quad  z\\
     &  II: \quad  1\\
     &  III:\quad  z
 \end{align}
 The first part of the connected correlation function factorizes so that we need to consider the on-site correlation function
 \begin{align}
      & s_o(\textbf{q})=\sum_{\textbf{R}_1} e^{i\textbf{q}\cdot \left(-\textbf{R}_1\right)}
     K_{\textbf{r}_1,\textbf{r}_1+\bm{\tau}} \expval{\psi_{\textbf{r}_1} \psi_{\textbf{r}_1+\bm{\tau}} } \\
     & 
     =  \sum^I_{\textbf{R}_1=\textbf{T}_1} \quad +\sum^{II}_{\textbf{R}_1=\textbf{T}_1-\textbf{a}_1} +\sum^{III}_{\textbf{R}_1=\textbf{T}_1-\textbf{a}_2} \\
     & \quad e^{i\textbf{q}\cdot \left(-\textbf{R}_1\right)}
     K_{\textbf{r}_1,\textbf{r}_1+\bm{\tau}} \expval{\psi_{\textbf{r}_1} \psi_{\textbf{r}_1+\bm{\tau}} }\\
     &
     =  \sum^I_{\textbf{R}_1=\textbf{T}_1} e^{-i\textbf{q}\cdot \textbf{R}_1 }   K_{\textbf{r}_1,\textbf{r}_1+\bm{\tau}} \expval{\psi_{\textbf{r}_1} \psi_{\textbf{r}_1+\bm{\tau}} }\\
     &+\sum^{II}_{\textbf{R}_1=\textbf{T}_1-\textbf{a}_1}  e^{-i\textbf{q}\cdot \textbf{R}_1 } K_{\textbf{r}_1,\textbf{r}_1+\bm{\tau}} \expval{\psi_{\textbf{r}_1} \psi_{\textbf{r}_1+\bm{\tau}} } \\
     &+\sum^{III}_{\textbf{R}_1=\textbf{T}_1-\textbf{a}_2} e^{-i\textbf{q}\cdot \textbf{R}_1 }  K_{\textbf{r}_1,\textbf{r}_1+\bm{\tau}} \expval{\psi_{\textbf{r}_1} \psi_{\textbf{r}_1+\bm{\tau}} }\\
     &
     =  \sum^{I}_{\textbf{T}_1} e^{-i\textbf{q}\cdot \textbf{T}_1 }   z \expval{\psi_{\textbf{r}_1} \psi_{\textbf{r}_1+\bm{\tau}} }+\sum^{II}_{\textbf{R}_1=\textbf{T}_1-\textbf{a}_1}  e^{-i\textbf{q}\cdot(\textbf{T}_1-\textbf{a}_1) } \\
     & \expval{\psi_{\textbf{r}_1} \psi_{\textbf{r}_1+\bm{\tau}} } +\sum^{III}_{\textbf{R}_1=\textbf{T}_1-\textbf{a}_2} e^{-i\textbf{q}\cdot (\textbf{T}_1-\textbf{a}_2) } z \expval{\psi_{\textbf{r}_1} \psi_{\textbf{r}_1+\bm{\tau}} }\\
     &
     =  \sum_{\textbf{T}_1} e^{-i\textbf{q}\cdot \textbf{T}_1 }   z \expval{\psi_{\textbf{T}_1}^5 \psi_{\textbf{T}_1}^2  }+\sum^{II}_{\textbf{R}_1=\textbf{T}_1-\textbf{a}_1}  \\
     & \times e^{-i\textbf{q}\cdot(\textbf{T}_1-\textbf{a}_1) } \expval{\psi_{\textbf{T}_1}^4\psi_{\textbf{T}_1+\textbf{c}_1-\textbf{c}_2}^3 }\\
     &+\sum^{III}_{\textbf{R}_1=\textbf{T}_1-\textbf{a}_2} e^{-i\textbf{q}\cdot (\textbf{T}_1-\textbf{a}_2) } z \expval{\psi_{\textbf{T}_1}^6\psi_{\textbf{T}_1}^1}
     \\
     &
     =  \sum_{\textbf{T}_1}   e^{-i\textbf{q}\cdot \textbf{T}_1 } \left( z \expval{\psi_{\textbf{T}_1}^5 \psi_{\textbf{T}_1}^2 } +z e^{i\textbf{q}\cdot \textbf{a}_1}\expval{\psi_{\textbf{T}_1}^6\psi_{\textbf{T}_1}^1} \right.\\
     & \left.+e^{i\textbf{q}\cdot \textbf{a}_2} \expval{\psi_{\textbf{T}_1}^4\psi_{\textbf{T}_1+\textbf{c}_1-\textbf{c}_2}^3 }\right)
 \end{align}
 We note from previous studies that the action in Fourier space and the two point correlator are given by
\begin{align}
    &S= \dfrac{1}{2}\sum_{\textbf{k},\alpha,\beta} h_{\alpha \beta}(\textbf{k})\psi^\alpha_{ \textbf{k}}\psi^\beta_{ -\textbf{k}} \\
    &\expval{\psi^\alpha_{ \textbf{x}} \psi^\beta_{\textbf{y}}}= \dfrac{1}{\tilde{N}}\sum_{\textbf{k}} e^{-i \textbf{k} \cdot(\textbf{x}-\textbf{y})}\left[h^{-1}(\textbf{k})\right]_{ \beta,\alpha}
\end{align}
where the momentum points are given in the new basis $\textbf{k}=\frac{n}{\tilde{L}_x}\textbf{d}_1+\frac{m}{\tilde{L}_y}\textbf{d}_2$, $n,m\in \mathbb{Z}$ so that $\textbf{d}_i \cdot \textbf{c}_j = 2\pi \delta_{ij}$. Solving the linear system of equations we find $\textbf{d}_1= \frac{1}{3}(\textbf{b}_2-\textbf{b}_1),\textbf{d}_2= \frac{1}{3}(2\textbf{b}_1+\textbf{b}_2  ) $ and $\tilde{N}=\tilde{L}_x\tilde{L}_y=N/3$ for the original $N$ site lattice. With the matrix

\begin{align}
&h(\textbf{k})=\left(\begin{array}{cc}
\mathbf{0} & \tilde{\mathbb{W}}(\textbf{k}) \\
-\tilde{\mathbb{W}}^{\dagger}(\textbf{k}) & \mathbf{0}
\end{array}\right),\\
& \tilde{\mathbb{W}}(\textbf{k})=\left(\begin{array}{ccc}
z & -e^{-i k_y} & -z \\
-z & -z & e^{i k_x} \\
-e^{-i k_x+i k_y} & z & -z
\end{array}\right)
\end{align}
Because of the translation invariant form of the correlations, we notice that we can simplify the on-site correlator $s_o(\textbf{q})$
\begin{align}
       & \left( z \expval{\psi_{0}^5 \psi_{0}^2 } +z e^{i\textbf{q}\cdot \textbf{a}_1}\expval{\psi_{0}^6\psi_{0}^1} +e^{i\textbf{q}\cdot \textbf{a}_2} \expval{\psi_{0}^4\psi_{\textbf{c}_1-\textbf{c}_2}^3 } \right) \sum_{\textbf{T}_1}   e^{-i\textbf{q}\cdot \textbf{T}_1 } \notag
       \\ &= \tilde{N} \left( z \expval{\psi_{0}^5 \psi_{0}^2 } +z e^{i\textbf{G}\cdot \textbf{a}_1}\expval{\psi_{0}^6\psi_{0}^1} + e^{i\textbf{G}\cdot \textbf{a}_2}\expval{\psi_{0}^4\psi_{\textbf{c}_1-\textbf{c}_2}^3 }\right) \delta_{\textbf{q},\textbf{G}}
\end{align}
We note that here a $\textbf{G}$ reciprocal lattice vector of the extended unit cell is included since it is defined so that $e^{-i\textbf{q}\cdot \textbf{T}_1 }=1$. We note that $\textbf{K}=\textbf{d}_2$ is a reciprocal lattice vector. We only have a contribution for zero momentum in the thermodynamic limit, consistent with the fact that this term is an on-site correlator. We can come back to the original dimer structure factor and write
\begin{align}
    &\tilde{s}(\textbf{q})= \dfrac{1}{N} \sum_{\textbf{R}_1,\textbf{R}_2} e^{i\textbf{q}\cdot \left(\textbf{R}_2-\textbf{R}_1\right)}
     4K_{\textbf{r}_1,\textbf{r}_1+\bm{\tau}} K_{\textbf{r}_2,\textbf{r}_2+\bm{\tau}}\\
     &\times \expval{\psi_{\textbf{r}_1}\psi_{\textbf{r}_2+\bm{\tau}} }\expval{\psi_{\textbf{r}_1+\bm{\tau}}  \psi_{\textbf{r}_2} } -4 s_o(\textbf{G})  \delta_{\textbf{q},\textbf{G}}\\
     &+4 \dfrac{1}{N}s_o(\textbf{G}) s_o(-\textbf{G}) \delta_{\textbf{q},\textbf{G}}+N \delta_{\textbf{q},\textbf{G}}
\end{align}
We see that the only non-zero momentum (modulo reciprocal lattice vectors) component is given by
\begin{widetext}
\begin{align}
    &\tilde{s}_c(\textbf{q})= \dfrac{1}{N}\sum_{\textbf{R}_1,\textbf{R}_2} e^{i\textbf{q}\cdot \left(\textbf{R}_2-\textbf{R}_1\right)}
     4K_{\textbf{r}_1,\textbf{r}_1+\bm{\tau}} K_{\textbf{r}_2,\textbf{r}_2+\bm{\tau}} \expval{\psi_{\textbf{r}_1}\psi_{\textbf{r}_2+\bm{\tau}} }\expval{\psi_{\textbf{r}_1+\bm{\tau}}  \psi_{\textbf{r}_2} }\\
     &=\dfrac{4}{N}\left(\sum^I_{\textbf{R}_1=\textbf{T}_1} +\sum^{II}_{\textbf{R}_1=\textbf{T}_1-\textbf{a}_1} +\sum^{III}_{\textbf{R}_1=\textbf{T}_1-\textbf{a}_2} \right)\left(\sum^I_{\textbf{R}_2=\textbf{T}_2} +\sum^{II}_{\textbf{R}_2=\textbf{T}_2-\textbf{a}_1} +\sum^{III}_{\textbf{R}_2=\textbf{T}_2-\textbf{a}_2} \right) e^{i\textbf{q}\cdot \left(\textbf{R}_2-\textbf{R}_1\right)}K_{\textbf{r}_1,\textbf{r}_1+\bm{\tau}} K_{\textbf{r}_2,\textbf{r}_2+\bm{\tau}} \expval{\psi_{\textbf{r}_1}\psi_{\textbf{r}_2+\bm{\tau}} }\expval{\psi_{\textbf{r}_1+\bm{\tau}}  \psi_{\textbf{r}_2} }\\
     &=\dfrac{4}{N}  \left(\sum^I_{\textbf{R}_1=\textbf{T}_1}  +\sum^{II}_{\textbf{R}_1=\textbf{T}_1-\textbf{a}_1} +\sum^{III}_{\textbf{R}_1=\textbf{T}_1-\textbf{a}_2} \right)K_{\textbf{r}_1,\textbf{r}_1+\bm{\tau}}\left(\sum^I_{\textbf{R}_2=\textbf{T}_2}  +\sum^{II}_{\textbf{R}_2=\textbf{T}_2-\textbf{a}_1} +\sum^{III}_{\textbf{R}_2=\textbf{T}_2-\textbf{a}_2} \right)K_{\textbf{r}_2,\textbf{r}_2+\bm{\tau}}e^{i\textbf{q}\cdot \left(\textbf{R}_2-\textbf{R}_1\right)}\expval{\psi_{\textbf{r}_1}\psi_{\textbf{r}_2+\bm{\tau}} }\expval{\psi_{\textbf{r}_1+\bm{\tau}}  \psi_{\textbf{r}_2} }\\
     &=\dfrac{4}{N}   \left(\sum^I_{\textbf{R}_1=\textbf{T}_1}z  +\sum^{II}_{\textbf{R}_1=\textbf{T}_1-\textbf{a}_1} +\sum^{III}_{\textbf{R}_1=\textbf{T}_1-\textbf{a}_2} z \right)\left(\sum^I_{\textbf{R}_2=\textbf{T}_2}z +\sum^{II}_{\textbf{R}_2=\textbf{T}_2-\textbf{a}_1} +\sum^{III}_{\textbf{R}_2=\textbf{T}_2-\textbf{a}_2}z \right)e^{i\textbf{q}\cdot \left(\textbf{R}_2-\textbf{R}_1\right)}\expval{\psi_{\textbf{r}_1}\psi_{\textbf{r}_2+\bm{\tau}} }\expval{\psi_{\textbf{r}_1+\bm{\tau}}  \psi_{\textbf{r}_2} }\\
     & = \dfrac{4}{N}   \left(\sum^I_{\textbf{R}_1=\textbf{T}_1}z \sum^I_{\textbf{R}_2=\textbf{T}_2}z  + \sum^I_{\textbf{R}_1= \textbf{T}_1}z\sum^{II}_{\textbf{R}_2=\textbf{T}_2-\textbf{a}_1} +\sum^I_{\textbf{R}_1=\textbf{T}_1}z\sum^{III}_{\textbf{R}_2=\textbf{T}_2-\textbf{a}_2}z  \right. \\
     & \qquad +\sum^{II}_{\textbf{R}_1=\textbf{T}_1-\textbf{a}_1}\sum^I_{\textbf{R}_2=\textbf{T}_2}z+\sum^{II}_{\textbf{R}_1=\textbf{T}_1-\textbf{a}_1}\sum^{II}_{\textbf{R}_2=\textbf{T}_2-\textbf{a}_1} +\sum^{II}_{\textbf{R}_1=\textbf{T}_1-\textbf{a}_1}\sum^{III}_{\textbf{R}_2=\textbf{T}_2-\textbf{a}_2}z \notag\\
     &\left. \qquad + \sum^{III}_{\textbf{R}_1=\textbf{T}_1-\textbf{a}_2} z\sum^I_{\textbf{R}_2=\textbf{T}_2}z +\sum^{III}_{\textbf{R}_1=\textbf{T}_1-\textbf{a}_2} z\sum^{II}_{\textbf{R}_2=\textbf{T}_2-\textbf{a}_1} +\sum^{III}_{\textbf{R}_1=\textbf{T}_1-\textbf{a}_2} z \sum^{III}_{\textbf{R}_2=\textbf{T}_2-\textbf{a}_2}z\right) e^{i\textbf{q}\cdot \left(\textbf{R}_2-\textbf{R}_1\right)}\expval{\psi_{\textbf{r}_1}\psi_{\textbf{r}_2+\bm{\tau}} } \expval{\psi_{\textbf{r}_1+\bm{\tau}}  \psi_{\textbf{r}_2} }\notag
\end{align}
\end{widetext}
We can see a matrix structure arising from the calculation which we make clear now by defining the factor $\phi_\mu(\textbf{q})$,with $m=I,II,III$ or in the sublattice notation of Fig.~\ref{fig:honeycomb} $\mu=5,4,6$ and $\psi_I(\textbf{q})=\phi_5(\textbf{q})=z,\psi_{II}(\textbf{q})=\phi_4(\textbf{q})=e^{i\textbf{q}\cdot \textbf{a}_1},\psi_{III}(\textbf{q})=\phi_6(\textbf{q})=z e^{i\textbf{q}\cdot \textbf{a}_2} $. We also will need the vector $\textbf{u}_\mu=(\textbf{c}_1-\textbf{c}_2)\delta_{\mu,II}$. We also will need the analog of the sublattice when translated by $\tau$, the function $p(\mu)=2,3,1$. We can then write
\begin{align}
    &\tilde{s}_c(\textbf{q})= \dfrac{4}{N} \sum_{\mu,\nu} \sum_{\textbf{T}_1}\sum_{\textbf{T}_2} e^{i\textbf{q}\cdot \left(\textbf{T}_2-\textbf{T}_1\right)}\phi_\mu(\textbf{q})\phi_\nu^*(\textbf{q})\\
    &\quad \times \expval{\psi^\mu_{\textbf{T}_1}\psi^{p(\nu)}_{\textbf{T}_2+\textbf{u}_{\nu} }} \expval{\psi^{p(\mu)}_{\textbf{T}_1+\textbf{u}_{\mu}}\psi^{\nu}_{\textbf{T}_2 } }
\end{align}

\begin{figure}[ht!]
    \centering
    \includegraphics[scale=0.27]{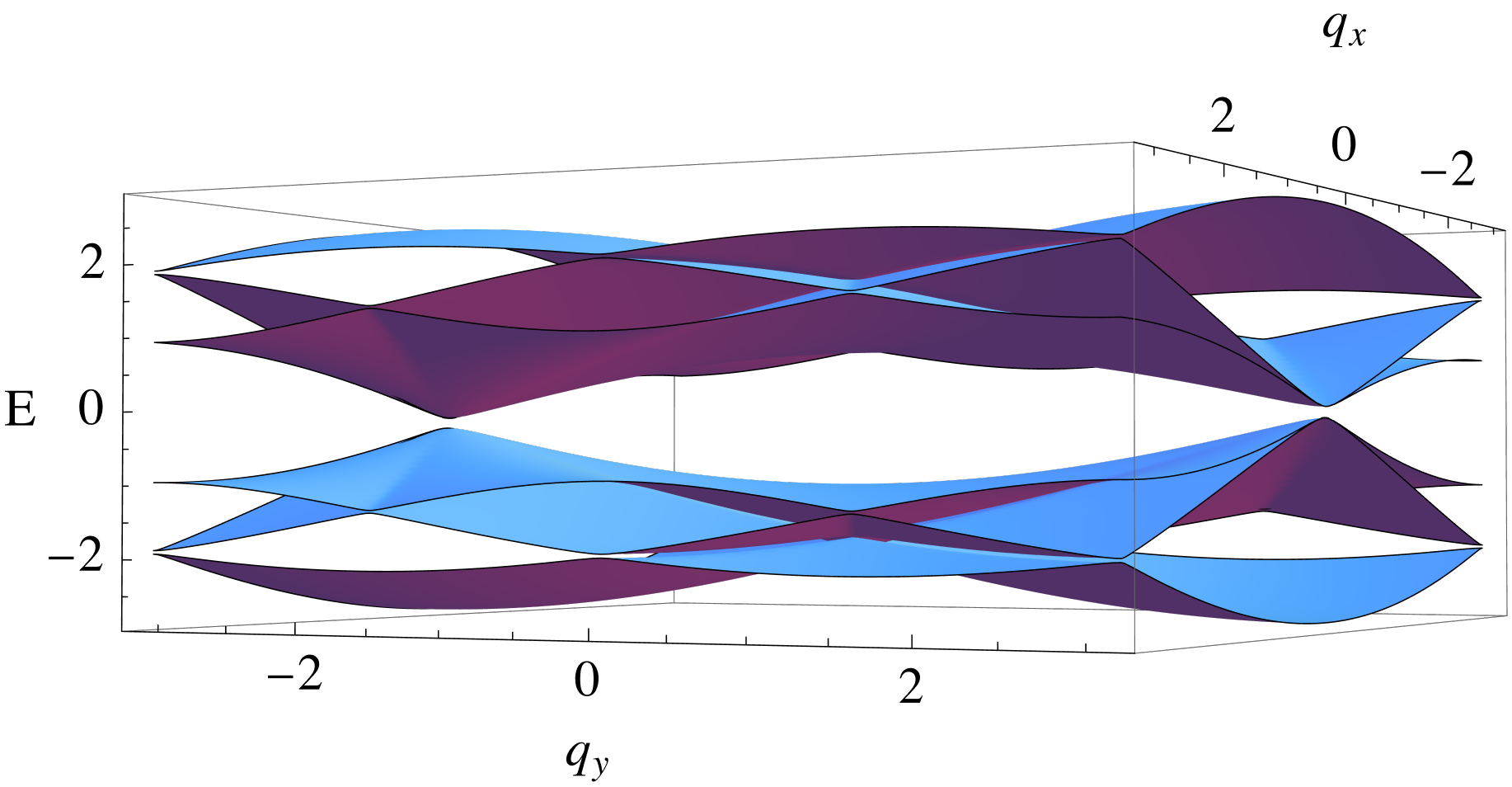}
    \caption{Band structure calculated from diagonalizing $A(\textbf{k})$ with $z=0.9250$ as found in previous studies of the supersolid order.}
    \label{fig:BandsDimer}
\end{figure}

\subsubsection{Polarization bubble}

We note now that we can write the connected structure factor in terms of a static pair fermion bubble, we note that the dimer connected structure factor has an extra term coming from the $\textbf{R}_1=\textbf{R}_2$ correlations that the fermionic mapping fails to calculate. We include this now explicitly
\begin{align}
     \tilde{s}_c(\textbf{q})&= \sum_{\mu,\nu} \Pi^{\mu\nu}(\textbf{q})\phi_\mu(\textbf{q})\phi_\nu^*(\textbf{q}),\\
     s_c(\textbf{q})&=\tilde{s}_c(\textbf{q})+\dfrac{4}{N} \sum_{\textbf{R}_1,\textbf{R}_2}e^{i\textbf{q}\cdot(\textbf{R}_2-\textbf{R}_1)} \expval{n^2_{\tau}(\textbf{R}_1)}\delta_{\textbf{R}_1,\textbf{R}_2}\\
     s_c(\textbf{q})&=\tilde{s}_c(\textbf{q})+\dfrac{4}{N} \sum_{\textbf{R}} \expval{n_{\tau}(\textbf{R})}
\end{align}
where we defined 
\begin{align}
&\Pi^{\mu\nu}(\textbf{q})=\dfrac{4}{N} \sum_{\textbf{T}_1}\sum_{\textbf{T}_2} e^{i\textbf{q}\cdot \left(\textbf{T}_2-\textbf{T}_1\right)}\expval{\psi^\mu_{\textbf{T}_1}\psi^{p(\nu)}_{\textbf{T}_2+\textbf{u}_{\nu} }} \\
&\qquad \times \expval{\psi^{p(\mu)}_{\textbf{T}_1+\textbf{u}_{\mu}}\psi^{\nu}_{\textbf{T}_2 } },
\end{align}
Moreover by using the Fourier transform of the correlations
\begin{align}
     &\Pi^{\mu\nu}(\textbf{q})=\dfrac{4}{N}\sum_{\textbf{T}_1}\sum_{\textbf{T}_2} e^{i\textbf{q}\cdot \left(\textbf{T}_2-\textbf{T}_1\right)} \dfrac{1}{\tilde{N}}\sum_{\textbf{k}} e^{i \textbf{k} \cdot(\textbf{T}_2+\textbf{u}_\nu-\textbf{T}_1)}\notag\\
     &\times \left[h^{-1}(\textbf{k})\right]_{ p(\nu)\mu }\expval{\psi^{p(\mu)}_{\textbf{T}_1+\textbf{u}_{\mu}}\psi^{\nu}_{\textbf{T}_2 } }\\
     &=\dfrac{4}{N}\sum_{\textbf{T}_1}\sum_{\textbf{T}_2} e^{i\textbf{q}\cdot \left(\textbf{T}_2-\textbf{T}_1\right)} \dfrac{1}{\tilde{N}}\sum_{\textbf{k}} e^{i \textbf{k} \cdot(\textbf{T}_2+\textbf{u}_\nu-\textbf{T}_1)}\\
     &\times \left[h^{-1}(\textbf{k})\right]_{ p(\nu)\mu }\dfrac{1}{\tilde{N}}\sum_{\textbf{k}'} e^{i \textbf{k}'\cdot(\textbf{T}_2-\textbf{T}_1-\textbf{u}_\mu)}\left[h^{-1}(\textbf{k}')\right]_{ \nu p(\mu) }\\
     &=\dfrac{4}{N}e^{i\textbf{u}_\mu \cdot (\textbf{q}-\textbf{G})}\sum_{\textbf{k}\in \text{BZ}_\text{T}} e^{i\textbf{k}\cdot(\textbf{u}_\mu+\textbf{u}_\nu)}\notag\\
     &\times h^{-1}_{ p(\nu) \mu}(\textbf{k})h^{-1}_{\nu p(\mu)}(-\textbf{k}-\textbf{q}-\textbf{G}_\text{T}).
\end{align}

\begin{align}
    &\Pi^{\mu\nu}(\textbf{q})=\dfrac{4}{N}e^{i\textbf{u}_\mu \cdot (\textbf{q}-\textbf{G})}\sum_{\textbf{k}\in \text{BZ}_\text{T}} e^{i\textbf{k}\cdot(\textbf{u}_\mu+\textbf{u}_\nu)}\\
    & \qquad \times h^{-1}_{ p(\nu) \mu}(\textbf{k})h^{-1}_{\nu p(\mu)}(-\textbf{k}-\textbf{q}-\textbf{G}_\text{T})
\end{align}
We expressed now $\textbf{q}\rightarrow \textbf{q}+\textbf{G}_\text{T}$, where $\textbf{G}_\text{T}$ is a reciprocal lattice vector of the red hexagon lattice and $\text{BZ}_\text{T}$ is it's first Brillouin Zone so that calling $G(\textbf{k})=h^{-1}(\textbf{k})$ we have
\begin{align}
    \Pi^{\mu\nu}(\textbf{q})=\sum_{\textbf{k}\in \text{BZ}_\text{T}} \hspace{-0.15cm} e^{i\textbf{k}\cdot(\textbf{u}_\mu+\textbf{u}_\nu)+i \textbf{q}\cdot\textbf{u}_\mu }G_{ p(\nu) \mu}(\textbf{k})G_{\nu p(\mu)}(-\textbf{k}-\textbf{q})
\end{align}
We note that $h^T(\textbf{k})=-h(-\textbf{k})$. Let us call then $G(\textbf{k})=h^{-1}(\textbf{k})$, it follows that $G^T(\textbf{k})=-G(-\textbf{k})$, so that $G_{\alpha\beta}=-G_{\beta\alpha}^*$ We now organize the matrix elements of the sublattices appearing before by defining the reduced $\tilde{G}$, we omit momentum index for the moment, matrix

\begin{align}
       &(\tilde{G}_{nm})= (G_{p(\mu)\nu})= \begin{pmatrix}
    G_{p(5)5}  && G_{p(4)5} && G_{p(6)5}\\
    G_{p(5)4} && G_{p(4)4} && G_{p(6)4} \\
    G_{p(5)6} && G_{p(4)6} && G_{p(6)6}
    \end{pmatrix}\\
    &=
    \begin{pmatrix}
    G_{25}  && G_{35} && G_{15}\\
    G_{24} && G_{34} && G_{14} \\
    G_{26} && G_{36} && G_{16}
    \end{pmatrix}
\end{align}
where now we label $n,m=I,II,III$. We then have that sums of products are given by matrix multiplications 
\begin{align} &(G_{\nu p(\mu)})=
    \begin{pmatrix}
    G_{5p(5)}  && G_{5p(4)} && G_{5p(6)}\\
    G_{4p(5)} && G_{4p(4)} && G_{4p(6)} \\
    G_{6p(5)} && G_{6p(4)} && G_{6p(6)}
    \end{pmatrix}\\
    &=
    \begin{pmatrix}
    G_{52}  && G_{53} && G_{51}\\
    G_{42} && G_{43} && G_{41} \\
    G_{62} && G_{63} && G_{61}
    \end{pmatrix}=
     -(\tilde{G}_{nm})^*\\
     & \sum_{\nu} G_{p(5)\nu}G_{p(\nu)5}=G_{25}G_{25}+G_{24}G_{35}+G_{26}G_{15}\\
     &= \sum_{m}\tilde{G}_{Im}\tilde{G}_{mI}
\end{align}
In general, we express the sum as
\begin{align}
    & \sum_{\mu\nu} G_{p(\nu) \mu}(\textbf{k})G_{\nu p(\mu)}(-\textbf{k}-\textbf{q})\\
    &= - \sum_{\mu\nu} G_{p(\mu)\nu }(\textbf{k}+\textbf{q})G_{p(\nu) \mu}(\textbf{k}) \\
    &= -\sum_{nm}\tilde{G}_{nm}(\textbf{k}+\textbf{q})\tilde{G}_{mn}(\textbf{k}),
\end{align}
in our case we have other factors that depend on $\mu,\nu$so the sum is not just a trace. We need to define the object 
\begin{align}
    T_{\mu\nu}(\textbf{k},\textbf{q})=\phi_\mu(\textbf{q})\phi_\nu^*(\textbf{q})e^{i\textbf{u}_\mu \cdot \textbf{q}}e^{i\textbf{k}\cdot(\textbf{u}_\mu+\textbf{u}_\nu)} = (T_{mn})(\textbf{k},\textbf{q})
\end{align}
Then the correlation function becomes a single momentum sum
\begin{align}
     &\tilde{s}_c(\textbf{q})= \sum_{\mu,\nu} \Pi^{\mu\nu}(\textbf{q})\phi_\mu(\textbf{q})\phi_\nu^*(\textbf{q})\\
     & =\dfrac{4}{N}\sum_{\textbf{k}\in \text{BZ}_\text{T}} T_{\mu\nu}(\textbf{k},\textbf{q})G_{p(\mu)\nu }(\textbf{k}+\textbf{q})G_{p(\nu) \mu}(\textbf{k})\\
     =&-\dfrac{4}{3\tilde{N}}\sum_{\textbf{k}\in \text{BZ}_\text{T}} \sum_{nm}T_{mn}(\textbf{k},\textbf{q})\tilde{G}_{nm}(\textbf{k}+\textbf{q})\tilde{G}_{mn}(\textbf{k})
\end{align}
Let us now turn it into an integral and our final result
\begin{align}
     \tilde{s}_c(\textbf{q})=-\dfrac{4}{3(2\pi)^2}\int\displaylimits_{\textbf{k}\in \text{BZ}_\text{T}} \sum_{nm}T_{nm}(\textbf{k},\textbf{q})\tilde{G}_{mn}(\textbf{k}+\textbf{q})\tilde{G}_{nm}(\textbf{k}).
\end{align}
This is the function we integrate and plot in Fig.~\ref{fig:DSF_exp}, with the disconnected part contributing with the delta function peaks shown in white.

\bibliography{manuscript_revised_black.bib}

\end{document}